\documentstyle[12pt]{report}

\begin{document}
\begin{center}
\pagenumbering{roman}
\thispagestyle{empty}
{\Huge \bf Studies in Certain Planar
\vskip 0.3cm
 Field Theories}
\vskip 2.0 true cm
Thesis Submitted for the degree of\\
Doctor of Philosophy (Science)\\
of \\
JADAVPUR  UNIVERSITY
\vskip 3.0 true cm
 2003 \\
{\large \bf TOMY SCARIA}\\
S. N. Bose National Centre for Basic Sciences\\
JD Block, Sector III \\
Salt Lake City \\
Calcutta  700 098\\
India
\end{center}
\newpage
\centerline{\large \bf CERTIFICATE FROM THE SUPERVISORS}
This is to certify that the thesis entitled
{\bf $``$Studies in Certain Planar Field Theories"}
submitted by  {\bf Sri Tomy Scaria} who got his name
registered on {\bf 13.08.2001} for the award of Ph.D.(Science)
degree of Jadavpur University, absolutely based upon his
own work under the supervision of {\bf Dr. Rabin Banerjee and Dr. Biswajit Chakraborty}
and that neither this thesis nor any part of it has been
submitted for any degree/diploma or any other academic
award anywhere before .\\
 
1. {\bf Rabin Banerjee}  \hfill    2.{\bf  Biswajit Chakraborty}\\
    Associate Professor. \hfill     Faculty Fellow \\
    S.N.Bose National Centre \hfill S.N.Bose National Centre \\
    for Basic Sciences,  \hfill     for Basic Sciences\\
 Salt Lake, Calcutta, India. \hfill Salt Lake, Calcutta, India.

\newpage
\thispagestyle{empty}
.
\vskip 6.0cm
\begin{center}
{\Huge \bf  Dedicated}
\vskip 0.5 cm
{\Huge \bf to }
\vskip 0.5cm
{ \Huge \bf   My Parents}
\end{center}
\newpage
\vskip 2.0cm
\begin{center}
{\large \bf ACKNOWLEDGEMENTS}
\end{center} 

 With great pleasure, I express my deep sense of gratitude to my thesis advisors, Dr. Rabin Banerjee
 and Dr. Biswajit Chakraborty for their expert guidance and constant encouragement. I acknowledge with sincere gratitutde, the valuable guidance I received from Dr. Rabin Banerjee and am grateful to him for always being
 there to help me in my work. I record my sincere thanks to Dr. Biswajit Chakraborty for helping me throughout the course of this work and especially for sparing no effort to guide my work even during a period of
 prolonged  illness which caused great physical hardship for him. The timely completion of this thesis is a  result of their unflinching support and I am indebted to them. It is my duty and joy to thank the respective
 families of Dr. Chakraborty and Dr. Banerjee for their hospitality during my visits for professional
 discussions.
       
 I am grateful to late Prof. C. K. Majumdar, Founder-Director of S. N. Bose National Centre for
 Basic Sciences (SNBNCBS), for giving me the opportunity to do research here.
      I thank Prof. S. Dattagupta, Director, SNBNCBS for his support
in my work.
       I am thankful to Prof. A. Mookerjee, Dean of Academic Affairs, SNBNCBS, for the help rendered to me
during my stay here.

  I thank all the academic and administrative staff of SNBNCBS for helping
me in many ways. In particular, I am thankful to the Library staff,
especially Mr. Swapan Ghosh, for the excellent assistance provided to me.
 In this regard, I also thank Mr. Bhupati Naskar who till recently worked in the library.
         
It is my pleasure to thank my friends in Calcutta who with their help and support in both academic
and personal matters made my stay here an experience I cherish much.

        Finally and most importantly, I express my whole hearted gratitude to my family members. It is the  love and
 support of my parents that enabled me to pursue the studies which finally culminated in this thesis. I dedicate this thesis to
 them. I am deeply indebted to my sister and brothers for their enthusiasm and 
constant encouragement in all my efforts. I am greatly obliged to my wife for
her encouragement  and support. 
\newpage
\tableofcontents
\chapter{Introduction and Overview}
\pagenumbering{arabic}
Field theories defined in  2+1 dimensional space-time (planar field theories)
are of importance in theoretical physics in many ways  \cite{witten,carlip,nair,nieto,seigel,jt,forte}.
One of the chief reasons for the  usefulness of planar theories  is that
 they serve as simplified versions of more 
complicated 3+1 dimensional ones that involve formidable mathematical and conceptual difficulties.
In such instances, a study of planar theories may provide  valuable insights and informations 
for developing  possible methods of dealing
with their more realistic 3+1 dimensional counterparts. A famous example for
such a  theory is the quantum gravity which is soluble
in 2+1 dimensions \cite{witten,carlip}. 
Another important
reason for the widespread interest in the study of planar models is that they 
can  describe the physics of 2+1 dimensional systems like surfaces and thin films
in condensed matter physics.   
Elementary particles confined to two spatial dimensions  behave in
distinctly  different  ways compared to the their usual behavior in the familiar 3+1
dimensions \cite{forte}. Such difference in behavior gives rise to interesting observable
phenomenon  thus making the study of planar theories highly interesting 
both from  theoretical and experimental points of view. For example, one such
 important application of planar theories  
 is the well known fractional statistics and  fractional quantum Hall effect \cite{forte,wilczek}.

In  this thesis
we deal with certain planar field theoretical models uncovering various aspects
of these theories and their interrelationships. A major part of the work is
devoted to the study topologically massive gauge theories which possess the interesting
property of gauge invariance co-existing with mass. The Maxwell-Chern-Simons
(MCS) and Einstein-Chern-Simons (ECS) theories  in 2+1 dimensions are  typical
 examples of such theories and have
evinced tremendous interest among theoretical physicists in recent times \cite{djt}.  
While MCS theory is a vector field theory, the ECS theory has a symmetric
second rank tensor as its basic field. We have investigated certain aspects
of  MCS theory and the linearized version of  ECS theory. The massive nongauge
vector theories, namely Proca and Maxwell-Chern-Simons-Proca (MCSP) theories, are
known to be equivalent to  doublets of  MCS theories \cite{djt,bk,bkm}. Also, MCS
theories are equivalent to self and anti-self dual models in 2+1 dimensions
 \cite{dj, br, townsend}.
Similarly, certain studies have suggested the existence of a similar 
connection between linearized versions of ECS theory and the
Einstein-Pauli-Fierz (EPF) theory \cite{djt}. 
We attempt to provide some fresh insight into the interrelationship between
these various theories through the maximally reduced form of the polarization
vectors/tensors of these theories. The polarization vectors and tensors are
of importance in phenomenological calculations \cite{tyutin,tyutin1,melo}.
 Further usefulness of these
polarization vectors/tensors become more transparent when we consider the gauge transformations generated by the translational subgroup of Wigner's little 
group  for massless particles \cite{rb}. It is quite well known that the polarization
tensor of free Maxwell theory in 3+1 dimensions when acted upon by this
translational subgroup undergoes gauge transformations \cite{we,we1,we2,hk,hk1,hk2}. It has also been
recently extended to linearized gravity which has a second rank symmetric 
tensor as the underlying gauge field \cite{ng,sc} and  Kalb-Ramond (KR) theory involving
2-form gauge field \cite{bc1}.
It  has also been shown that the translational group
 $T(3)$, group of translations in 3-dimensional space, in a particular representation acts as generator of gauge transformations
in $B\wedge F$ theory \cite{bc1} which is obtained by topologically coupling  Maxwell field to
the KR field. Analogous to MCS and  ECS theories, $B\wedge F$ model  is a topologically
massive gauge theory, but in 3+1 dimensions \cite{kr,cs,lahiri,lahiri1}. It was later discovered that one
 can in fact  systematically derive this representation of  $T(3)$ which 
generate gauge transformations in a  massive gauge theory in 3+1 dimensions,
using the gauge transformation properties of 4+1 dimensional Maxwell and
massless KR theories by a method known as  $`$dimensional descent' \cite{bc2}.
In the present work,
we not only unravel many new and interesting points pertaining to the
gauge generation by translational groups  in 3+1 dimensional theories, but 
 extend them to planar gauge theories also. We  show that a particular
representation of the 1-dimensional translational group $T(1)$  generate gauge transformations
 in the topologically massive MCS and ECS theories \cite{sc,bc2,bcs1}.
Connection of this representation of $T(1)$ with the representation of 
$T(2)$ (group of  translations in  2-dimensions) that generate
gauge transformations in 
3+1 dimensional massless theories is demonstrated explicitly by dimensional
descent from 3+1 to 2+1 dimensions \cite{sc,bc2}. 

We considered only Abelian  theories in this work  and the
calculation of the polarization vector (or tensor) of these theories is of
crucial importance in our context. In general, polarization vectors/tensors 
capture the Lorentz transformation property of the basic fields when expressed
as a mode expansion. In the usual field theoretical models such as Maxwell theory
or Proca theory, the components of the polarization vectors can
be independent of each other and are real \cite{greiner,felsager}. However, the components of the 
 polarization  vector (tensor) of  the topologically massive MCS (ECS) theory are
necessarily complex and cannot be chosen independently, manifesting the 
internal structure of the theory \cite{sc,bcs}. This is true also for MCSP theory
and a comparison of the 
polarization vectors of MCSP theory with those of a pair of MCS theories with
opposite helicities explicitly shows that the former can be considered to
be equivalent to the MCS doublet \cite{bcs}. This equivalence between a pair of
MCS theories and MCSP theories was earlier studied at the level of the basic fields 
of the models \cite{bk,bkm}. As we shall see later in this report,  the polarization tensor of
ECS theory is the tensor product of the polarization vectors of a pair of
MCS theories with the same helicity. We take recourse to explicit expressions
of the polarization vector or tensor of the theory under consideration in order
to show  how  the appropriate
translational group generate gauge transformation in the theory. For this
purpose we consider only a single mode in the Fourier expansion of the basic
field of the theory and restrict ourselves to a particular reference frame.
We then use the Euler-Lagrange equation to derive the maximally reduced form of the
polarization vector or tensor 
which is devoid of any spurious degrees of freedom and represent only the
physical sector of the theory \cite{felsager}. This procedure is named $`$plane wave method'
and it can be used, with a great economy of effort,  to study  the gauge generating nature of   translational groups
in various theories. Another important feature of this $`$plane wave method' is 
that it yields the mass of the quanta of the  theory under consideration rather
effortlessly. Apart
from the usual massive nongauge theories, this method of extracting the masses
(zero or nonzero)
of the quanta can be applied to ordinary gauge theories having massless
excitations, topologically massive theories having massive quanta and 
nongauge theories elevated to gauge theories by St\"uckelberg mechanism \cite{sc,bc1,bcs,ts}.

In chapter 2, we  establish the equivalence of the MCSP
 model  to a doublet of MCS models  defined in a variety
of covariant gauges. This equivalence is shown to hold at the level of
polarization vectors of the basic fields. The analysis is done
in both Lagrangian and Hamiltonian formalisms and compatible results are
 obtained \cite{bcs}.  A similar equivalence with a doublet of self and anti self dual
models  is briefly discussed.

In chapter 3, we review the role of translational subgroup $T(2)$ of
Wigner's little group for massless particle as a generator of gauge
 transformations in  3+1 dimensional theories.
First it is shown, following Kim et.al. \cite{hk}, how the gauge transformation 
in Maxwell theory is generated by the
defining representation of $T(2)$. 
 It is then shown that
the same representation of $T(2)$  generates gauge transformations in 3+1 dimensional 
linearized gravity \cite{ng,sc} and massless KR theory \cite{bc1}. The gauge transformations generated by
translational  group form only a subset of the full range of gauge 
transformations available to linearized gravity and KR theories \cite{ts}. We also
see in chapter 3 that the reducibility of gauge transformations in KR theory
is clearly manifested in the gauge transformations generated by $T(2)$.
Furthermore, in the case of the topologically massive $B\wedge F$ theory, 
the gauge transformations are generated by $T(3)$ \cite{ts}.

We discuss in chapter 4, the role of $T(1)$ as generator of  gauge
 transformations the topologically massive MCS theory as well the linearized 
ECS theory in 2+1 dimensions. Using plane wave method, we drive the maximally
reduced form of the polarization vector and tensor of these theories and show
that a suitable representation of $T(1)$ generates gauge transformation in
these theories \cite{sc,bcs1}. Polarization vector of the ECS theory is clearly shown to be
the tensor product of the polarization vector of MCS theory with itself \cite{sc}.

The method of dimensional descent \cite{bc2} is reviewed in chapter 5 and the
representation of $T(1)$ that acts as gauge generator for MCS and ECS
theories are derived using this method by starting from gauge transformation
properties of massless gauge theories (Maxwell and linearized gravity) in 3+1
dimensions \cite{sc}.
Finally, the polarization tensor of EPF theory in 2+1
dimensions is shown to split into the polarization tensors of a pair of
ECS theories with opposite helicities suggesting a doublet
structure for EPF theory \cite{sc}.

 One may also
construct massive gauge theories by   converting second class
constrained systems (in the  language of Dirac's theory of constraint dynamics
\cite{dirac1,dirac2}) to first class (gauge) systems using the generalized embedding
prescription of Batalin, Fradkin and Tyutin \cite{bft,bft1,bt}.
On the other hand, in  the Lagrangian framework, one can convert the massive 
nongauge theories to gauge theories
by the  generalized  St\"uckelberg extension mechanism \cite{berg,goto}.
It is pointed out in \cite{bneto} that there exists a one to one correspondence
between this Hamiltonian embedding prescription and the St\"uckelberg extension mechanism based on Lagrangian formalism. 
By such embedding  prescriptions,
one may elevate gauge noninvariant  Proca, massive KR \cite{bb} and EPF theories to gauge theories and  obtain the corresponding
 S\"tuckelberg extended versions of these models.    Chapter 6
is devoted to the study of the connection between gauge generation and 
translational groups in such embedded  massive gauge theories. Though the models considered in
chapter 6 belong to 3+1 dimensional space-time, with suitable modifications 
 these methods and results
can be easily applied also to planar theories. We show that the same 
representation of $T(3)$ that generate gauge transformation in $B\wedge F$
theory also acts as gauge generator in the above mentioned   St\"uckelberg 
extended models \cite{ts}.

A brief description of the major results and conclusions are given in
chapter 7. 

{\bf Notation:} We use subscripts/superscripts in Greek letters 
for denoting  indices in 2+1 dimensional space-time. The letters $a,b,c$ etc.
from the beginning of Latin alphabet are used for indices in 3+1 dimensional 
space-time whereas
$x,y,z$ etc. from the end of the alphabet denote 4+1 dimensions.  
Letters like $i,j,k$ from the middle of Latin alphabet represent spatial components of
vectors/tensors in any dimension. Signature of the metrics used are mostly 
negative.

We adopt the following nomenclature for gauge theories having massive
excitations. We discuss two types of gauge theories where the gauge fields are
massive. When a massless gauge theory is coupled to a topological term,
the theory acquires mass while retaining the gauge symmetry. The MCS, ECS 
\cite{djt} and $B\wedge F$ \cite{cs,lahiri} theories are examples of such gauge
theories where the origin of mass is due to the presence of topological terms
in their actions. These theories are clearly referred to as `topologically
massive gauge theories'. The term `massive gauge theories' are used for
 those gauge theories obtained elevating the massive second class theories to
 first class (gauge) theories by the prevously mentioned embedding prescription
given by Batalin, Fradkin and Tyutin \cite{bft,bft1,bt}. Such massive  gauge theories
 are also called St\"uckelberg extended theories of the corresponding massive nongauge theories because they can also be obtained by generalized St\"uckelberg extension mechanism \cite{berg,goto,bneto}.

This thesis is based on the following publications.

\begin{enumerate}
\item  Polarization vectors and doublet structure in planar
field theory \cite{bcs}\\
R. Banerjee, B. Chakraborty and Tomy Scaria\\
 {\it  Int. J. Mod. Phys.} {A16} (2001) 3967.
 
\item  On the role of Wigner's little group as a generator of gauge transformation in
Maxwell-Chern-Simons theory \cite{bcs1}\\
R. Banerjee, B. Chakraborty and Tomy Scaria \\
{\it  Mod. Phys. Lett. }{\bf A16} (2001) 853. 
\item  Wigner's little group as a gauge generator in linearized gravity
theories \cite{sc}\\
Tomy Scaria and B. Chakraborty \\
{\it Class. Quant. Grav.} {\bf 19} (2002) 4445. 

\item Translational groups as   generators of gauge transformations \cite{ts}\\
Tomy Scaria \\
{\it Phys. Rev.}{\bf D68} (2003) 105013.
\end{enumerate}

\chapter{Polarization Vectors and Doublet Structure in Planar Theories}
The polarization vectors or tensors contained in the mode expansions 
of the basic fields of field theoretical models usually capture the
Lorentz transformation properties of the fields under consideration. 
In some theories the polarization vectors/tensors may carry more information
on the structure of the theory itself. In such cases, a simple analysis of
the polarization vectors of the theories provide valuable information
regarding these theories. In the present chapter, 
we study two such models, namely Maxwell-Chern-Simons theory and
Maxwell-Chern-Simons-Proca theory. The topic of this chapter is the
relationship between these models and we also consider the self and antiself
dual model which have a close connection with these theories.
As mentioned before MCS theory is topologically massive gauge theory where
the gauge invariant mass occurring already at the classical (tree)
level. Now, it is intriguing to note that the MCSP theory
 can be regarded as the embedding of a doublet of
topologically massive gauge theories \cite{bk,bkm}.
Earlier, this was studied at the level of the basic 
fields in the two
theories \cite{bk,bkm}.
Here we shall pursue this mapping at the more fundamental
level involving polarization vectors associated with different modes
of these fields. This is all the more
important since proper evaluation of these vectors is crucial for reduction
formulae and the study of the massless limit of the MCSP theory \cite{tyutin}.
Besides, as stated before,  the transformation properties of the polarization vectors of
various gauge theories under the action of the translational subgroup
of Wigner's little group display the precise gauge symmetry of the theory.
Therefore, it is obvious that polarization vectors play a fundamentally
important role in the momentum space analysis  of these theories.
This motivates us to compute the polarization vectors of these theories
in the Lagrangian and Hamiltonian formulations and make a comparison between them.
A difference in the Hamiltonian approach, in contrast to the Lagrangian
approach, is the need  to introduce a $``$new
set" of polarization vectors for canonically conjugate momentum variables
($\pi^\mu$) along with those of the basic vector fields ($A^\mu$).

\section{Polarization vectors in   
Lagrangian formalism}
\label{pol-lgr}
\subsection{Maxwell-Chern-Simons theory}
\label{mcs1}
\begin{sloppypar}
We first review the calculation of the polarization vectors in the Maxwell-Chern-Simons theory pointing out the differences from  the corresponding 
analysis for the Maxwell theory. Apart from reviewing the standard analysis
\cite{greiner,girotti} where imposition of Lorentz gauge is required,
 an alternative
analysis depending only on the symmetries of the theory  will also be discussed.
The Lorentz gauge condition emerges naturally in the latter method.

The MCS Lagrangian in 2+1 dimensions is given by
\end{sloppypar}
\begin{equation} {\cal L} = -{\frac{1}{4}}{F^{\mu \nu}F_{\mu \nu}} + {\frac{\vartheta}{2}}{\epsilon}^{\mu \nu \lambda}A_{\mu}{\partial}_{\nu}A_{\lambda}.
\label{c1}
\end{equation}
This is the well known topological gauge theory with a single mode of mass
$|\vartheta|$ and spin $\frac{\vartheta}{|\vartheta|}$ \cite{djt}.
The corresponding equation of motion is given by,
\begin{equation}
 -{\partial}^{\nu}F_{\mu \nu} + {\vartheta}\epsilon _{\mu \nu \lambda }{\partial}^{\nu}A^{\lambda} = 0. 
\label{c2}
\end{equation}
Imposing the Lorentz gauge,
\begin{equation}
 {\partial}_{\mu}A^{\mu} = 0
\label{c3}
\end{equation}
the above equation reduces to
\begin{equation}
\left( {\Box}g^{\mu \nu} + {\vartheta \epsilon ^{\mu \lambda \nu }}{\partial}_{\lambda} \right) A_{\nu} = 0. 
\label{c4}
\end{equation}
Substituting the solution
\begin{equation}
 A^{\mu}(x) = {\xi}^{\mu}(k)\exp(ik.x)
\label{c5}
\end{equation}
 for the negative energy component\footnote{Here we simply suppressed the positive energy 
component which is just the complex conjugate of the negative energy component
appearing in (\ref{c5}). Its presence is required to make $A^{\mu}(x)$ real. However, this suppression of the positive frequency part is of no consequence to
our analysis.}
in terms of the polarization vector ${\xi}^{\mu}(k)$
in the above two equations, gives, respectively,
\begin{equation}
 k_{\mu}{\xi}^{\mu} = 0\label{c6}
\end{equation} 
and
\begin{equation}
 {\Sigma}_{(MCS)}^{\mu \nu}{\xi}_{\nu}(k) = 0
\label{c7}
\end{equation}
where,
\begin{equation}
 {\Sigma}_{(MCS)}^{\mu \nu} = -k^2g^{\mu \nu} + i{\vartheta \epsilon ^{\mu \lambda \nu}}k_{\lambda}.\label{c8}
\end{equation}
For a non trivial solution to exist, we must have  
\begin{equation}
 \det {\Sigma}_{(MCS)} = -k^4(k^2 - {\vartheta}^2) = 0.
\label{c9}
\end{equation}
It follows therefore that, we must have either $ k^2 = 0$ or $
  k^2 = {\vartheta}^2 $.
When $k^2 = 0$, the solution is,
\begin{equation}
 {\xi}^{\mu}(k) = k^{\mu}f(k)
\label{c12}
\end{equation}
where $f(k)$ is an arbitrary function. 
Therefore massless excitations are pure gauge artefacts, which may be
ignored. 

Now consider the case $k^2 = {\vartheta}^2$, which implies that the
quanta has mass\footnote{This can also be seen by rewriting the equation of
 motion (\ref{c2}) of MCS theory in terms of the gauge invariant pseudo vector
 dual field $\tilde{F}^\mu \equiv
\frac{1}{2} \epsilon_{\mu\nu\lambda}F_{\nu\lambda}$:
$$ (\Box + \vartheta^2)\tilde{F}^\mu  = 0.$$
This equation clearly shows that MCS theory has massive excitations. Yet another way to arrive this result is by calculating the gauge field propagator whose 
pole gives the mass at $k^2 = \vartheta^2$ \cite{djt}.}
$|\vartheta|$. This enables a passage to the rest frame with  $k^{\mu} = (|\vartheta|,0,0)$. Then 
the equation of motion (\ref{c7}) yields,
\begin{equation}
  -{\vartheta}^2{\xi}_0({\bf 0}) = 0
\label{c13}
\end{equation}
\begin{equation}
 {\vartheta}^2{\xi}_1({\bf 0}) + i{\vartheta}(-|{\vartheta}|){\xi}_2({\bf 0}) = 0 \label{c14}
\end{equation}
\begin{equation}
 {\vartheta}^2{\xi}_2({\bf 0}) + i{\vartheta}|{\vartheta}|{\xi}_1({\bf 0}) = 0\label{c15}
\end{equation}
where $ {\xi}^\mu ({\bf 0}) $ stands for the MCS polarization in the rest frame.
The above set yields,
\begin{equation}
 {\xi}^0({\bf 0}) = 0
\label{c16}
\end{equation}
\begin{equation}
   {\xi}^2({\bf 0}) = -i{\frac{\vartheta}{|\vartheta|}}{\xi}^1({\bf 0}).
\label{c17}
\end{equation}
Therefore in the rest frame the polarization vector is given by, 
\begin{equation}
 {\xi}^{\mu}({\bf 0}) = \left( 0, {\xi}^1({\bf 0}), -i{\frac{\vartheta}{|\vartheta|}}{\xi}^1({\bf 0}) \right)  
\label{c18}
\end{equation}
and is the maximally reduced form of the polarization vector ${\xi}^{\mu} $ in the rest frame representing just the single physical degree of freedom of the MCS theory\footnote{This method of obtaining the maximally reduced form of the polarization vector (or tensor) of a theory by starting from a plane wave solution
of the Euler-Lagrange equation for the basic field is named plane wave method 
and is used extensively in this thesis}.
The above expressions are determined modulo a normalization factor. This can be
fixed from the normalization condition,
\begin{equation}
 \xi^{\ast \mu} ({\bf 0}) \xi_{\mu}({\bf 0}) = -1
\label{c19}
\end{equation}
following essentially from the space-like nature of the vector $\xi^\mu$,
as follows from (\ref{c6}) using the fact that $k^\mu$ is time-like. An
important point of distinction from the Maxwell case is that $\xi^\mu$ has
complex entries while the  $x$ and $y$ components of the polarization
vectors bear a simple ratio between them in the rest frame (\ref{c17}),
so that the number of degrees of freedom reduces to one. Furthermore, the normalization
condition(\ref{c19}) reveals a $U(1)$  invariance in the expression for $\xi^\mu$;
i.e., if $\xi^\mu$ is a solution, then $e^{i\phi} \xi^\mu$ is also a solution.
 This observation will be used later on to show the equivalence among different
forms for $\xi^\mu$.

The normalization condition fixes $|{\xi}^1({\bf 0})|^2 = \frac{1}{2}$. Hence,
\begin{equation}
 {\xi}^{\mu}({\bf 0}) = \frac{1}{\sqrt{2}}\left(0, 1, -i{\frac{\vartheta}{|\vartheta|}}\right).
\label{c20}
\end{equation}

Now we present another  derivation of this result where only the symmetries of the
model are used.  Consider again the equation (\ref{c2}) and assume solutions of the form (\ref{c5}).
Substituting (\ref{c5}) in (\ref{c2}) yields,
\begin{equation}
 \xi_{\nu} k^{\nu}k^{\mu} - k^2 \xi^{\mu} + i\vartheta \epsilon^{\mu \nu \lambda} \xi_{\lambda} k_{\nu}
= 0.
\label{c21}
\end{equation} 
The two  possibilities for $k^2$, corresponding to massless or massive modes  are
$(i)   k^2 = 0$ and $
 (ii) k^2 \not= 0$.
We first take up the case  case (i).
Using $k^2 = 0$ in (\ref{c21}) gives,
\begin{equation}
 k^{\mu}(\xi \cdot k) = -i \vartheta \epsilon^{\mu \nu \lambda} k_{\nu} \xi_{\lambda}
\label{c24}
\end{equation}
Multiplying both sides with $\epsilon_{\mu \alpha \beta} k^{\alpha}$ one arrives at,
 $$0 = i\vartheta k_{\beta}(\xi \cdot k) $$
which implies that the momentum space Lorentz condition 
\begin{equation}
 \xi \cdot k = 0 \label{c25}
\end{equation}
holds automatically. 
Using $k^2 =0$  and (\ref{c25}) in (\ref{c21}), we get $\xi^{\mu} = f(k)k^{\mu}$ which, as mentioned
earlier, shows that massless excitations are pure gauge artefacts.

Next we consider the case  $k^2 \not= 0$, from (\ref{c21}) we have,
\begin{equation}
 \xi_{\mu} = \frac{1}{k^2}\left[ (\xi \cdot k)k_{\mu} + i\vartheta \epsilon_{\mu \nu \lambda} k^{\nu} \xi^{\lambda} \right].
\label{c26}
\end{equation} 
 and we are allowed to go to a rest frame where $k^{\mu} = (m, 0, 0)$ and $k^2 = m^2$.
Let $\xi^{\mu}$ in this frame be given by, 
\begin{equation}
 \xi^{\mu}({\bf 0}) = \left( \xi^0 ({\bf 0}), \xi^1 ({\bf 0}), \xi^2 ({\bf 0})\right).\label{c27}
\end{equation}
Then (\ref{c26}) gives, 
\begin{equation}
 \xi^1({\bf 0}) = \frac{i\vartheta}{m} \xi^2 ({\bf 0}) 
\label{c28}
\end{equation}
and
\begin{equation}
 \xi^2({\bf 0}) = - \frac{i\vartheta}{m} \xi^1 ({\bf 0}).
\label{c29}
\end{equation}
Substituting for $\xi^2({\bf 0})$ from (\ref{c29}) in (\ref{c28}) gives
$$
 \vartheta^2 = m^2 $$
from which it follows,
\begin{equation}
 m = | \vartheta |.
\label{c30}
\end{equation}
From the gauge invariance of the model it  follows that 
${\xi}^0({\bf 0})$ can be set equal to zero.
Therefore from (\ref{c28}), (\ref{c29}) and  (\ref{c30}), we reproduce the 
earlier result  (\ref{c18}).
It is important to note that the result is compatible with the covariance condition (\ref{c6})
 although it was not used explicitly in the analysis.
This is also true for  Maxwell theory where
the polarization vector automatically satisfies this condition;
but there $k^\mu$ corresponds to the massless physical excitations($k^2 = 0$).
On the other hand, the massive excitations in the Maxwell theory
are pure gauge artefacts, as can be
easily seen from (\ref{c21}) by setting $\vartheta = 0$. Thus the roles of
massive and massless excitations in MCS theory is just the opposite of 
Maxwell theory.
It is now straightforward to calculate the polarization vector in a moving
 frame by giving a Lorentz boost \cite{jackson} to the result in the rest frame,
\begin{equation}
 \left( \begin{array}{c}
  {\xi}^0(k) \\ {\xi}^1( k) \\ {\xi}^2(k) \end{array} \right) = \left( \begin{array}{ccc}
   {\gamma} & {\gamma}{\beta}^1 & {\gamma}{\beta}^2	 \\
	{\gamma}{\beta}^1 & 1 + \frac{({\gamma} -1)({\beta}^1)^2}{(\vec{\beta})^2} & \frac{({\gamma} -1){\beta}^1{\beta}^2}{(\vec{\beta})^2} \\
	{\gamma}{\beta}^2 & \frac{({\gamma} -1){\beta}^1{\beta}^2}{(\vec{\beta})^2} & 1 + \frac{({\gamma} -1)({\beta}^2)^2}{(\vec{\beta})^2}
    \end{array} \right)\left( \begin{array}{c}
  {\xi}^0({\bf 0}) \\ {\xi}^1({\bf 0}) \\ {\xi}^2({\bf 0}) \end{array} \right)
\label{c31}
\end{equation}
where $\vec{\beta} = \frac{\bf k}{k^0}$ and $\gamma = \frac{k^0}{|\vartheta|}$.
The ensuing polarization vector is given by,
\begin{equation}
 {\xi}^{\mu}(k) = \left( \frac{{\vec{\xi}({\bf 0})}\cdot{\bf k}}{|\vartheta|},	\vec{\xi}({\bf 0}) + \frac{{\vec{\xi}({\bf 0})}\cdot{\bf k}}{(k^0 + |\vartheta|)|\vartheta|} {\bf k} \right)
\label{c32}
\end{equation}
where $\vec{\xi}({\bf 0})$ stands for the space part of the vector in (\ref{c20}). Thus,
\begin{equation}
 {\xi}^{\mu}(k) = 
\left( \frac{k^1 - i{\frac{\vartheta}{|\vartheta|}}k^2}{{\sqrt{2}}{|\vartheta|}},
\frac{1}{\sqrt{2}} + \frac{k^1 - i{\frac{\vartheta}{|\vartheta|}}k^2}{{\sqrt{2}}{(k^0 + |\vartheta|)|\vartheta|}}k^1, -i\frac{\vartheta}{\sqrt{2}|\vartheta|} + 
\frac{k^1 - i{\frac{\vartheta}{|\vartheta|}
}k^2}{{\sqrt{2}}{(k^0 + |\vartheta|)|\vartheta|}}k^2 \right) 
\label{c33}
\end{equation}
which agrees with the expression given in \cite{girotti} calculated in the Lorentz gauge.

\subsection{Maxwell-Chern-Simons-Proca theory}
\label{mcsp1}
The Maxwell-Chern-Simons-Proca(MCSP) Lagrangian is given by,
\begin{equation}
 {\cal L} = -{\frac{1}{4}}{F^{\mu \nu}F_{\mu \nu}} + {\frac{\theta}{2}}{\epsilon}^{\mu \nu \lambda}A_{\mu}{\partial}_{\nu}A_{\lambda} + {\frac{m^2}{2}}A^{\mu}A_{\mu}.
\label{c34}
\end{equation}
The equation of motion is, 
\begin{equation}
 -{\partial}^{\nu}F_{\mu \nu} + {\theta}\epsilon _{\mu \nu \lambda}{\partial}^{\nu}A^{\lambda} +m^2A_{\mu} = 0 
\label{c35}
\end{equation}
which automatically satisfies the  transversality condition
 ${\partial}_{\mu}A^{\mu} = 0.$
Using this, the equation of motion simplifies to,
\begin{equation}
 \left[({\Box} + m^2)g^{\mu \nu} + {\theta}{\epsilon}^{\mu \lambda \nu}{\partial}_{\lambda} \right]A_{\nu} = 0
\label{c36}
\end{equation}
Substitution of the solution $A^{\mu} = {\varepsilon}^{\mu}(k){\exp}(ik.x)$ yields,
\begin{equation}
 \left[(-k^2 + m^2)g^{\mu \nu} + i{\theta \epsilon ^{\mu \lambda \nu}}k_{\lambda}\right]{\varepsilon}_{\nu} = 0. 
\label{c37}
\end{equation}
From the transversality relation we get $k_{\mu}{\varepsilon}^{\mu} = 0.$
Let us define,
\begin{equation}
 {\Sigma}^{\mu \nu}_{(MCSP)} = (-k^2 + m^2)g^{\mu \nu} + i{\theta \epsilon ^{\mu \lambda \nu}}k_{\lambda} 
\label{c38}
\end{equation}
Then the equation of motion can be written as,
\begin{equation}
 {\Sigma}^{\mu \nu}_{(MCSP)}{\varepsilon}_{\nu} = 0.
\label{c39}
\end{equation}
For ${\varepsilon}_{\nu}$ to have a non trivial solution the determinant of 
${\Sigma}_{(MCSP)}$ should vanish. That is,
\begin{equation}
 (-k^2 + m^2)\left[(-k^2 + m^2)^2 - {\theta}^2k^2\right] = 0.
\label{c40}
\end{equation}
This implies, either, 
\begin{equation}
 -k^2 + m^2 = 0 
\label{c41}
\end{equation}
or
\begin{equation}
 (-k^2 + m^2)^2 - {\theta}^2k^2 = 0.
\label{c42}
\end{equation}
Using (\ref{c41}) in (\ref{c37}), it follows that the solution must have the form, ${\varepsilon}^{\mu}(k) = k^{\mu}f(k)$,
which is however incompatible with the transversality relation and is therefore ignored. The second
case leads to
\begin{equation}
 k^2 = \theta_{\pm}^2 
\label{c43}
\end{equation}
where,
\begin{equation}  \theta_{\pm} = \sqrt{\frac{2m^2 + {\theta}^2 \pm \sqrt{{\theta}^4 + 4m^2{\theta}^2}}{2}} = \sqrt{\frac{\theta^2}{4} + m^2} \pm \frac{\theta}{2}. 
\label{c44}
\end{equation}
Two useful relations follow from this identification,
\begin{equation}
 \theta = {\theta}_+ - {\theta}_-
\label{c45}
\end{equation}
and
\begin{equation}
 m^2 = {\theta}_+{\theta}_-. 
\label{c46}
\end{equation}
We use the notation ${\varepsilon}_{\pm}(k_{\pm})$ for the polarization vectors corresponding to $k^2 = {\theta_{\pm}}^2$ and let ${\varepsilon}_{\pm}({\bf 0})$ denote the polarization vectors in the rest frame.
Taking the rest frames to be the ones in which $ k^\mu = (k^0_\pm,0,0)^T = (|{\theta}_{\pm}|,0,0)^T$ we have from the equation of motion (\ref{c37}),
$$
 (m^2 - {{\theta}_{\pm}}^2){\varepsilon}_{\pm 0}({\bf 0}) = 0, $$
$$ -(m^2 - {{\theta}_{\pm}}^2){\varepsilon}_{\pm 1}({\bf 0}) - i{\theta}{\theta}_{\pm}{\varepsilon}_{\pm 2}({\bf 0}) = 0 $$
$$ -(m^2 - {{\theta}_{\pm}}^2){\varepsilon}_{\pm 2}({\bf 0}) + i{\theta}{\theta}_{\pm}{\varepsilon}_{\pm 1}({\bf 0}) = 0 $$  
where ${\varepsilon}_{\pm}^{\mu} = ({\varepsilon}_{\pm}^0, {\varepsilon}_{\pm}^1, {\varepsilon}_{\pm}^2).$
From the above set of equations we arrive at, 
\begin{equation}
 {\varepsilon}_{\pm 0}({\bf 0}) = 0  
\label{c47}
\end{equation}
\begin{equation}
 {\varepsilon}_{\pm 2}({\bf 0}) = \frac{i{\theta}{\theta}_{\pm}}{m^2 - {{\theta}_{\pm}}^2} {\varepsilon}_{\pm 1}({\bf 0}) = \mp i{\varepsilon}_{\pm 1}({\bf 0})
\label{c48}
\end{equation}
where the connection among various parameters has been used.
Using a  normalization condition analogous to (\ref{c19})
\begin{equation}
 {\varepsilon}_{\pm}^{\ast \mu}({\bf 0}){\varepsilon}_{\pm \mu}({\bf 0}) = -1
\label{c49}
\end{equation}
gives,
\begin{equation}
 |{\varepsilon}_{\pm 1}({\bf 0})|^2 =\frac{1}{2}.
\label{c50}
\end{equation}
Hence, 
\begin{equation}
 {\varepsilon}_{\pm}^{\mu}({\bf 0}) = \frac{1}{\sqrt{2}}\left(0, 1, \mp i\right). 
\label{c51}
\end{equation}
The transversality condition $k_{\mu}{\varepsilon}^{\mu} = 0 $ is preserved
which acts as a consistency check.  
\bigskip
The polarization vectors in a moving frame corresponding to the two massive modes with masses ${\theta}_{\pm}$ are easily found, as before, by giving a
Lorentz boost,
\begin{equation}
 {\varepsilon}_{\pm}^{\mu}(k_{\pm}) =
 \left(\frac{k^1 \mp i
k^2}{{\sqrt{2}
}{{\theta}_{\pm}}}, \frac{1}{\sqrt{2}} + \frac{k^1 \mp i
k^2}{{\sqrt{2}}{(k^0_{\pm} + {\theta}_{\pm}){\theta}_{\pm}}}k^1, \mp \frac{i}{\sqrt{2}} +\frac{k^1 \mp ik^2}{{\sqrt{2}}{(k^0_{\pm} + {\theta}_{\pm}){\theta}_{\pm}}}k^2 \right)
\label{c52}
\end{equation}
The pair of polarization vectors are related by the parity transformation in
two space dimensions $k^1 \rightarrow k^1, k^2 \rightarrow  -k^2$ augmented by $ k^0 _+ \rightarrow k^0 _-$(which also implies   $\theta_+ \rightarrow \theta_-$),
$$
 {\varepsilon}_+^0(k^0_+, k^1, k^2) = {\varepsilon}_-^0(k^0_- \rightarrow k^0_+, k^1 \rightarrow k^1, k^2 \rightarrow  -k^2) $$
\begin{equation}
 {\varepsilon}_+^1(k^0_+, k^1, k^2) = {\varepsilon}_-^1(k^0_- \rightarrow k^0_+, k^1 \rightarrow k^1
, k^2 \rightarrow  -k^2) 
\label{c53}
\end{equation}
$$ {\varepsilon}_+^2(k^0_+, k^1, k^2) =  -{\varepsilon}_-^2(k^0_- \rightarrow k^0_+, k^1 \rightarrow k^1
, k^2 \rightarrow  -k^2). $$
Also, the pair is related by complex conjugation,
\begin{equation}
 {\varepsilon}_+^{\mu}(k_+) = {\varepsilon}_-^{\ast \mu}(k_-).
\label{c54}
\end{equation}
where it is implied that this operation flips the parameter $\theta_+ \rightarrow \theta_-$.
Now it may be pointed out that the polarization vectors satisfy the conditions,
$$ {\varepsilon}_\pm ^{\mu}({\bf 0}){\varepsilon}_{\pm \mu}({\bf 0}) = 0. $$
These polarization vectors are therefore light-like.
Here however, it is possible to interpret these conditions as a
consequence of the usual orthogonality relations,
$$ {\varepsilon}_+^{\ast \mu}({\bf 0}){\varepsilon}_{- \mu}({\bf 0}) = 0. $$
together with the parity law (\ref{c54}).
These observations suggest an inbuilt doublet structure in the MCSP model.
The embedded doublet structure, related by the augmented parity transformations, in the
MCSP theory will be further explored in the next subsection.
\section{Application of polarization vectors}
\label{appl}
\subsection{$U(1)$ invariance and doublet structure}
\label{alt}
The above methods of calculating the polarization vectors depend on the
 existence of a rest frame. The results obtained in this frame are Lorentz
boosted to an arbitrary moving frame. There is another approach which directly
yields the polarization vectors from a solution of the free field equations
of motion. We now discuss this and compare with the previous analysis.

Let us consider the MCS theory (with $\vartheta > 0$). Since it has a
 single physical mode
of mass $\vartheta$, it is possible to write a general expression for 
the polarization vector, satisfying the Lorentz gauge condition (\ref{c6})
and the equation of motion (\ref{c7}),
\begin{equation}
 \tilde{\xi}^{\mu}(k) = N \left(k^{\mu} - g^{\mu 0}\frac{{\vartheta}^2}{\omega} - i \frac{\vartheta}{\omega} \epsilon^{\mu \alpha 0} k_{\alpha} \right) 
\label{nc1}
  \end{equation}
with, $\omega = k_0 = \sqrt{{\vartheta}^2 + |{\bf k}|^2}$ and $N$ is the
normalization. This is fixed from the condition ($\tilde{\xi}^{\ast \mu} \tilde{\xi}_{\mu} = -1)$,
$$ N = \frac{1}{\sqrt{2}} \frac{\omega}{\vartheta |{\bf k}|} $$
This expression for the polarization vector was given in \cite{tyutin}. Though
(\ref{nc1}) appears quite different from the previous result (\ref{c33})
they only differ by a $U(1)$ phase factor. To see this, we express (\ref{nc1}) in component form as follows.
\begin{equation}
 \tilde{\xi}^{\mu}(k) = \frac{1}{\sqrt{2}\vartheta} \left(|{\bf k}|, \frac{\omega}{|{\bf k}|}
(k_1 + i \frac{\vartheta}{\omega}k_2), \frac{\omega}{|{\bf k}|}(k_2 -i\frac{\vartheta}{\omega}k_1)\right)
\label{nc2}
\end{equation}
By introducing a phase angle $\phi$ we can write the spatial components of $k^\mu$ as
\begin{eqnarray}
 k^1 = |{\bf k}| \cos \phi \label{ncc3} \\
k^2 = |{\bf k}| \sin \phi
\label{nc3}
\end{eqnarray}
Using (\ref{ncc3}, \ref{nc3}), since $\frac{\omega}{\vartheta} = 1+ \frac{|{\bf k}|^2}{ (\omega + \vartheta) \vartheta}$, (\ref{nc2}) can be rewritten as 
\begin{eqnarray}
 \tilde{\xi}^{\mu}(k) = \frac{1}{\sqrt{2}}\left[\frac{|{\bf k}|}{\vartheta}, \left(1+ \frac{|{\bf k}|^2}{ (\omega + \vartheta) \vartheta}\right)\cos \phi + i\sin \phi,
\left( 1+ \frac{|{\bf k}|^2}{ (\omega + \vartheta) \vartheta}\right)\sin \phi -i \cos \phi
 \right]
\nonumber \\
 = \frac{1}{\sqrt{2}} e^{i\phi} \left[e^{-i\phi} \frac{|{\bf k}|}{\vartheta}, 1+ \frac{|{\bf k}|^2}{ (\omega + \vartheta) \vartheta} e^{-i\phi} \cos \phi, -i +
\frac{|{\bf k}|^2}{ (\omega + \vartheta) \vartheta} e^{-i\phi} \sin \phi \right]
\nonumber
\end{eqnarray}
$$ =  e^{i\phi} \xi^{\mu}(k).$$
(We  have used (\ref{ncc3}, \ref{nc3}) again in the last step.)
Up to a $U(1)$ phase, this exactly coincides with (\ref{c33}) thereby proving
the equivalence of the two results. However, one may notice that the representation (\ref{nc1}) does not have smooth rest frame limit whereas for  (\ref{c33}) 
this limit is
a well behaved one.

Following identical techniques the polarization vectors for MCSP theory turn
out as,
\begin{equation}
 \tilde{\varepsilon}_{\pm \mu} =
\frac{1}{\sqrt{2}}\frac{\omega_{\pm}}{|{\bf k}|\sqrt{\omega_{\pm}^2 - |{\bf k}|^2}} \left(k_{\pm \mu} - g_{\mu 0}\frac{\omega_{\pm}^2 - |{\bf k}|^2}{\omega_{\pm}} - i \frac{\omega_{\pm}^2 - |{\bf k}|^2 - m^2}{\theta \omega_\pm}
\epsilon_{\mu \alpha 0} k^{\alpha}\right)
\label{nc4}
\end{equation}
where, $\omega_\pm = \sqrt{\theta_{\pm}^2 + |{\bf k}|^2}$.
Once again this does not have a smooth transition to the rest frame. But
we can show its equivalence with the expressions (\ref{c52}) by adopting 
the previous
procedure. Expressions similar to (\ref{nc1}) and (\ref{nc4}) were reported earlier in \cite{tyutin}.

Different representations of the polarization vectors find uses in different contexts.  For instance when MCS theory is coupled to fermions (MCS-QED),  the infra-red 
singularities of the 2+1 dimensional model leads to gauge dependence of the
one-loop fermion mass-shell \cite{djt}.  The expressions (\ref{nc1}) and (\ref{nc4}) are used in
\cite{tyutin} for
 analyzing this fermion mass variance in the  MCS-QED mentioned above.
 On the other hand, the
 expressions given in (\ref{c33}) and (\ref{c52}) reveal the presence of a doublet
structure in the MCSP model.
Specifically, the pair of polarization vectors $ {\varepsilon}_{\pm}^{\mu}$, 
corresponding to the distinct massive modes ${\theta}_{\pm}$, can be exactly
identified with the polarization vectors for a doublet of MCS models,
\begin{equation}
 {\cal L}_+ = -{\frac{1}{4}}{F^{\mu \nu}(A)F_{\mu \nu}(A)} + {\frac{{\theta
}_+}{2}}{\epsilon}^{\mu \nu \lambda}A_{\mu}{\partial}_{\nu}A_{\lambda}
\label{nc5}
\end{equation}
\begin{equation}
 {\cal L}_- = -{\frac{1}{4}}{F^{\mu \nu}(B)F_{\mu \nu}(B)} - {\frac{{\theta
}_-}{2}}{\epsilon}^{\mu \nu \lambda}B_{\mu}{\partial}_{\nu}B_{\lambda}. 
\label{ncc5}
\end{equation}
The necessary symmetry features are preserved provided both ${\theta}_{\pm}
> 0 $ or ${\theta}_{\pm} < 0 $. It then follows from  (\ref{c33}) that the
polarization vectors of the MCS doublet exactly match with (\ref{c52}). The two
massive modes $\theta_{\pm}$ of the doublet are exactly identified with
the pair found in the MCSP model.

Yet another way of understanding the doublet structure is to look at the
$m^2 \rightarrow 0$ limit of the MCSP model (\ref{c34}), which then reduces to 
the MCS model. From (\ref{c45}) and (\ref{c46}) we see that this limit corresponds to two
possibilities;
\begin{eqnarray}
 (i) \hskip 1.0 cm \theta_+ \rightarrow 0; \theta \rightarrow -\theta_- \\
 (ii) \hskip 1.0 cm \theta_- \rightarrow 0; \theta \rightarrow \theta_+
\label{nc6}
\end{eqnarray}
These two cases $(\theta \rightarrow \pm \theta_\pm)$ exactly correspond 
to the MCS doublet (\ref{nc5}) and (\ref{ncc5}). Likewise the polarization vectors(\ref{c52}) also
map to the corresponding doublet structure. Note that this expression is 
divergent for $\theta_+ \rightarrow 0$ or $\theta_- \rightarrow 0$, but this
mode does not corresponds to the physical scattering amplitude when fermions
are coupled \cite{tyutin}.

It is worthwhile to mention that the limit $m^2 \rightarrow 0$ takes a second
class system  to a first class one. From the view point of a constrained system,
 such a limit is generally not smooth. However, the polarization vector shows
a perfectly smooth transition. We might also recall that the $m^2 \rightarrow 0$ limit in the second class Proca model, to pass to the Maxwell theory, is 
problematic due to the change in the nature of the constraints. This is also
manifested in the structure of the polarization vectors. Setting the $m^2 \rightarrow 0$ limit in  
the relevant expressions for the Proca model does not yield the Maxwell
theory polarization vector. In this way, therefore, the massless limit in the MCSP theory is quite distinctive. Since a pair of MCS theories get mapped to the
MCSP theory, such a smooth transition exists.

 It is interesting to compare the above discussed relation between the
second class MCSP theory and a doublet of first class  MCS theories to the 
previously (chapter 1) mentioned embedding prescription due to Batalin, Fradkin
and Tyutin \cite{bft,bft1,bt}. In their  prescription, a second class system is
embedded in a first class system. In contrast,  in the present context a
pair of first class (MCS) theories gets embedded in a second class (MCSP) 
theory. Therefore, the mapping between MCSP is model and the doublet of
MCS models is, in some sense, the opposite of the embedding procedure described in \cite{bft,bft1,bt}.

Another point worth mentioning is that, if $\theta_+ = \theta_-$, then 
$\theta = 0$  from (\ref{c45}). This means that a doublet of MCS theories having
 the same topological mass but with opposite helicities, maps to a Proca model, i.e., massive 
Maxwell model, with mass $m^2 = \theta_+^2 = \theta_-^2$. In this case parity is a symmetry
which is also seen from the generalized transformations (\ref{c53}). This mapping was
 discussed earlier \cite{bw,deser} in terms of the basic fields of the respective
models.

\subsection{Mapping with a  doublet of self dual models}
\label{dual}
We have shown how a doublet of  MCS theories can be
 mapped to a  MCSP theory
using the explicit expressions for the polarization vectors of the theories.
This subsection is devoted to a discussion of this seemingly  paradoxical
mapping between a doublet of  gauge  (MCS) theories
 and a non-gauge (MCSP)
 theory. We provide the justification for such a mapping and 
 elucidate its  physical interpretation and  other related issues.

In order to explain the various subtleties regarding the mapping mentioned
above, we begin by noting the well established equivalence between a 
self-dual model and the MCS theory \cite{dj, br, townsend}.
The Lagrangian of the self-dual model is given by
\begin{equation}
 {\cal L}_{SD} = {\frac{1}{2}} f^{\mu}  f_{\mu} - {\frac{1}{2M}} {\epsilon}^{\mu \nu \lambda} f_{\mu}{\partial}_{\nu}f_{\lambda} 
\label{c55}
\end{equation}
 An obvious  difference between the two theories is that, whereas the MCS
theory is manifestly a gauge theory, possessing only first class constraints, 
the self dual model is a non-gauge theory and has second class constraints.
 It has been shown  \cite{br}, using both operator and path integral 
techniques,  that
the gauge invariant sector of the MCS theory given by $\cal L_+$ (\ref{nc5})  gets mapped to the
self dual model{\footnote{This situation is just analogous to the well known
equivalence between gauge non-invariant nonlinear sigma model(NLSM) and 
$CP^1$ model which is a $U(1)$ gauge theory. Here the mapping between the
NLSM fields $n^a$ satisfying $n^a n^a = 1 (a = 1, 2, 3)$ and the gauge-variant
 $CP^1$ field doublet $ Z = \left( \begin{array} {c} z_1 \\ z_2
\end{array}\right) $ satisfying $Z^{\dagger} Z = 1$ is given by the
Hopf map $n^a = Z^{\dagger} \sigma^a Z$ with $\sigma^a$ being  Pauli matrices.}}.
   Specifically,  the fundamental 
 field $f^{\mu}$ of the self dual model and the dual field  $\tilde{F}^{\mu} =
\frac{1}{2}\epsilon^{\mu \nu \lambda} F_{\nu \lambda}$ of the MCS theory get identified, so
that $f^{\mu} = \tilde{F}^{\mu} $. Likewise the mass parameters in the two
theories are also equated ($\theta_+ = M$). 

In the present context the connection between the self dual and MCS theories
will be discussed in terms of the polarization vectors. Indeed it can
be verified 
 explicitly
that the  polarization vector of the self dual model matches with the
 physical polarization vector [(\ref{c20}) and (\ref{c32})] of MCS theory.
The equation of motion following from (\ref{c55}) is 
\begin{equation}
 f^{\mu} - {\frac{1}{M}} {\epsilon}^{\mu \nu \lambda} {\partial}_{\nu}f_{\lambda} = 0.
\label{c56}
\end{equation}
As was done before, we consider a solution of the form
\begin{equation}
  f^{\mu} = \xi^{\mu}(k) e^{ik.x}
\label{c57}
\end{equation}
where $\xi^{\mu}(k)$ stands for the polarization vector of the self dual theory.
Substitution of (\ref{c57}) in (\ref{c56}) yields the equation
\begin{equation}
 \left[g^{\mu \lambda} -\frac{ i}{M} \epsilon^{\mu \nu \lambda} k_{\nu} \right] \xi_{\lambda}(k) = 0
\label{c58}
\end{equation}
which will have a nontrivial solution only if 
\begin{equation}
 det \left[g^{\mu \lambda} -\frac{ i}{M} \epsilon^{\mu \nu \lambda} k_{\nu} \right] = 0
\label{c59}
\end{equation}
Now, (\ref{c59}) leads to the condition  $k^2 = M^2$ which in turn implies that
the excitations are massive and that there exists a rest frame for the quanta.
Proceeding exactly as was done in the case of MCS and MCSP theory before,
one can find the rest frame polarization vector of the self dual model as
\begin{equation}
 \xi^{\mu}({\bf 0}) = \frac{1}{\sqrt 2}\left(0, 1, -i\right)
\label{c60}
\end{equation} 
To make a comparison, we now calculate the polarization vector
 $\tilde{\xi}^{\mu}(k)$ for the dual field $\tilde{F}^{\mu} =
 \tilde{\xi}^{\mu}(k) e^{ik.x} $ in the MCS theory given by the Lagrangian $\cal L_+$ (\ref{nc5}).
This is done
by using the structure (\ref{c5}) for the $A_{\mu}$ field and passing to
 the momentum frame.
$$
  \tilde{F}^0 = \epsilon_{ij}\partial_i A_j = i {\bf k} \times {\vec{\xi}} e^{ik.x} $$
$$  \tilde{F}^1 = \epsilon^{1 \nu \lambda} \partial_{\nu} A_{\lambda} = i(k_2 \xi_0 - k_0 \xi_2)e^{ik.x}      $$
$$ \tilde{F}^2 =  \epsilon^{2 \nu \lambda} \partial_{\nu} A_{\lambda} = i(k_0 \xi_1 - k_1 \xi_0)e^{ik.x} $$
In the rest frame (where $ k^{\mu} = (\theta_+, 0, 0))$ the above set of equations
 reduces to
\bigskip
$$  \tilde{F}^0 = 0,  \tilde{F}^1 = \theta_+e^{ik.x} ,  \tilde{F}^2 = -i \theta_+ e^{ik.x}  $$
where use has been made of the explicit form  for $\xi$ given in (\ref{c20}).
Therefore,  the polarization vector $\tilde{\eta}^{\mu}({\bf 0}) $ in the rest frame is given by
\begin{equation}
 \tilde{\xi}^{\mu}({\bf 0}) =  \theta_+(0, 1, -i)   
\label{c61}
\end{equation}
We thus find that the polarization vector for $f^{\mu}$ matches (up to a
normalization factor) with that of
$\tilde{F}^{\mu}$ thereby providing an alternative interpretation of the
equivalence between the self dual and MCS theories. Moreover since the 
polarization vector for $\tilde{F}^{\mu} $ matches with that of $A^{\mu}$
calculated in the covariant gauge, this shows the equivalence of $f^{\mu}$
with $A^{\mu}$ taken in the covariant gauge.
Clearly, the polarization vectors of $f^{\mu}$ and $\tilde{F}^{\mu}$ will match
 in a moving frame also.
Similarly one can show that the polarization vector of the anti self dual
model whose Lagrangian is given by
\begin{equation}
 {\cal L}_{ASD} = {\frac{1}{2}} f^{\mu}  f_{\mu} + {\frac{1}{2M}} {\epsilon}^{\mu \nu \lambda} f_{\mu}{\partial}_{\nu}f_{\lambda} \label{c62}
\end{equation}
corresponds to that of the dual field in the MCS model given by $\cal L_-$.

The expressions for the polarization vectors in the self and anti self
dual models obviously agree with the doublet structure found in the MCSP model.
 This shows the mapping of the self dual and anti self dual models with the
MCSP theory\footnote{Such a mapping was also analysed in \cite{bkm}.}.
A pair of gauge noninvariant models is mapped to a composite gauge noninvariant
model. Since the self (or anti self) dual model is equivalent to the gauge 
invariant sector of the MCS theory, this clarifies the equivalence of the
MCS doublet with the MCSP model. This equivalence should be interpreted 
as the mapping of the gauge invariant sector of the MCS doublet with the
 MCSP theory. Furthermore,since the mapping involves only the gauge invariant
sector, the purported equivalence will hold in any gauge. Here it has been
explicitly shown for a covariant gauge. For a proof in a noncovariant gauge
like the Coulomb gauge we take recourse to an indirect argument. The
equivalence of the self dual model with the MCS theory in different gauges
including the Coulomb gauge  has been shown in \cite{br}. Since the mapping of
 the self dual and anti self dual pair with the MCSP has been illustrated, this
shows that an analogous mapping must also hold for the MCS doublet in the
Coulomb gauge.

One may also note that it is 
not possible to consider the two excitations in MCSP theory as arising from
two scalar bosons. It was explicitly shown in \cite{bkm} that  the spins of the
two  excitations of MCSP theory are $\pm 1$.
Furthermore,  group theoretical analysis of 2+1 dimensional 
theories
 shows that  massive 
excitations arising due to the presence of parity violating terms(eg. the Chern-Simons term)  have spin  $\pm 1$ \cite{binegar}. For
MCS theory it was shown that the spin of the massive excitation in 
$\cal L_+ $ is +1 while that in $\cal L_- $ the spin is -1 \cite{djt}.
This is also consistent with the mapping analysed here. 

\section{Polarization vectors in   Hamiltonian formalism}
\label{pol-hmt}
The analysis of the polarization vectors of MCS theory in the Lagrangian formalism presented
above is restricted to a single covariant gauge - the Lorentz gauge.
However it is possible to obtain the polarization vectors with 
other gauge choices also. For example,
Devecchi et. al.\cite{girotti} studied the case of polarization vector
of MCS theory in the Coulomb gauge which turned out to be of a  different
structure in comparison to (\ref{c33}). The MCSP theory not being a gauge theory,
the form (\ref{c52}) of the polarization vectors of the theory  is unique and
corresponds to the fact that the Lorentz condition $\partial_{\mu} A^{\mu} = 0$
is automatically satisfied here, unlike MCS theory where it is imposed
by hand. 
In this context one might wonder
if it is possible to  establish the doublet structure of MCSP theory
in any other Lorentz covariant gauge. Furthermore,  Lagrangian framework is a manifestly
covariant formalism while the Hamiltonian formulation does not  possess
this covariance because of its very nature of singling out time. Also, because of the
presence of  momenta $\pi^\mu(x)$ conjugate to the field variables $A_\mu(x)$
in the Hamiltonian formalism, one has to introduce additional polarization
vectors ($\pi^\mu(x)$) for  to implement the mode expansion of the momenta. Hence it is not clear 
if the results obtained in the Lagrangian and the Hamiltonian formalisms
are mutually compatible. To settle these issues we compute the polarization vectors in the Hamiltonian formalism 
based on Dirac's constrained algorithm. Different covariant
 gauge conditions will be considered. 
\subsection{Maxwell-Chern-Simons theory}
\label{mcs2}
In this subsection we again consider the MCS field in the Lorentz gauge, but
with the difference that the Lorentz gauge condition is now  imposed at 
the level
of the Lagrangian of the model itself. Consider the Lagrangian,
\begin{equation}
 {\cal L}_{\alpha} = -{\frac{1}{4}}{F^{\mu \nu}F_{\mu \nu}} + {\frac{\vartheta}{2}}
{\epsilon}^{\mu \nu \lambda}A_{\mu}{\partial}_{\nu}A_{\lambda} - \frac{1}{2 \alpha}
(\partial \cdot A)^2  
\label{c63}
\end{equation}
which is obtained from (\ref{c1}) by adding an extra gauge fixing term
$- \frac{1}{2 \alpha}(\partial \cdot A)^2$. (In this subsection $\alpha$ represents the gauge parameter). For simplicity, the parameter $\vartheta$ is taken 
positive.
If the vector field $A^{\mu}$ satisfies the Lorentz constraint (\ref{c3}), the 
Lagrangian ${\cal L}_{\alpha}$  is equivalent to the MCS Lagrangian given by (\ref{c1}).
The value of the gauge parameter $\alpha$ being
arbitrary, we make the choice $\alpha = 1$(Feynman gauge). With this choice, after an integration
by parts in the action, ${\cal L}_1$ transforms to,
$$ {\cal L}_1 = -\frac{1}{2} {\partial}_{\mu} A_{\nu} \partial^{\mu} A^{\nu} + \frac{1}{2}\partial_{\mu}\left[ A_{\nu}(\partial^{\nu} A^{\mu}) - (\partial_{\nu} A^{\mu})A^{\mu}\right] + {\frac{\vartheta}{2}} {\epsilon}^{\mu \nu \lambda}A_{\mu}{\partial}_{\nu}A_{\lambda}. $$
Ignoring the total divergence term we write,
\begin{equation}
{\cal L}_1 = -\frac{1}{2} {\partial}_{\mu} A_{\nu} \partial^{\mu} A^{\nu} + {\frac{\vartheta}{2}} {\epsilon}^{\mu \nu \lambda}A_{\mu}{\partial}_{\nu}A_{\lambda}.
 \label{c64}
\end{equation} 
 The conjugate momenta are given by,
\begin{equation}
 \pi^{\mu} = \frac{{\partial}{\cal L}_1}{{\partial}\dot{A}_{\mu}} = (-\dot{A}^0 , -\dot{A}^i + \vartheta \epsilon^{i j} A_j)
\label{c65}
\end{equation}
The Hamiltonian corresponding to (\ref{c64}) is given by 
$$
 H_1 = \int d^2 x \left[-{\frac{1}{2}} \pi^{\mu}\pi_{\mu} + {\frac{1}{2}}\partial^kA^\nu \partial_k A_\nu \right]  $$
\begin{equation}
 + \int d^2 x\left[
 - \frac{\vartheta}{2} \epsilon^{i j}\left( A_i \pi_j + A_0 \partial_i A_j + A_i \partial_j A_0 \right) + {\frac{1}{8}}\vartheta^2 {\bf A}^2 \right]
\label{cc65}
\end{equation}
The  Hamilton's equations following from
$$\dot{A}^{\mu} = \{ A^{\mu}, H_1 \}$$
 and 
$$\dot{\pi}^{\mu} =\{ {\pi}^{\mu} , H_1 \}$$
  are explicitly given as follows.
\begin{eqnarray}
\dot{A}^0 = -\pi^0 \label{c66-1} \\
 \dot{A}^1 = -\pi^1 - \frac{\vartheta}{2} A^2 \label{c66-2} \\
 \dot{A}^2 = -\pi^2 + \frac{\vartheta}{2} A^1 \label{c66-3}\\
 \dot{\pi}^0 = -\nabla^2 A^0 + \vartheta \epsilon^{i j}\partial_i A_j \label{c66-4} \\
 \dot{\pi}^1 = -\nabla^2 A^1 - \vartheta \partial^2 A^0 + \frac{\vartheta^2}{4}A^1 -\frac{\vartheta}{2}\pi^2 \label{c66-5} \\
 \dot{\pi}^2 = -\nabla^2 A^2 + \vartheta \partial^1 A^0 + \frac{\vartheta^2}{4}A^2 + \frac{\vartheta}{2}\pi^1 \label{c66-6}
\end{eqnarray}
Since our aim is to obtain the explicit form of the polarization vectors of the
field, we consider solutions of the form,
\begin{eqnarray}
 A^{\mu} = \sum_{\lambda = 1}^2\xi^{\mu}(\lambda, {\bf k})a_{{\bf k} \lambda} \exp[ik \cdot x] + c.c  \label{c67-1}\\
 \pi^{\mu} = \sum_{\lambda = 1}^2 \xi^{\mu}(\lambda, {\bf k})b_{{\bf k} \lambda} \exp[ik \cdot x] + c.c
\label{c67}
\end{eqnarray}
Note that the polarization vectors $\xi^{\mu}(k)$ used in the previous section have been expanded in terms of their basis vectors $\xi^{\mu}(\lambda, {\bf k})$. Since $\xi^{\mu}(k)$ is
space-like there are utmost two  linearly independent vectors $\xi^{\mu}(\lambda, {\bf k})$ characterized
by $\lambda$, which are used in the expansion of $A^{\mu}$. The justification for
using only the  two space-like basis vectors $\xi^{\mu}(\lambda, {\bf k})$ for the mode expansion of $\pi^{\mu}$ is that the  defining relation (\ref{c65}) of $\pi^{\mu}$ involves components
of $A^{\mu}$ (which can be written in terms of space-like polarization vectors) and their time derivatives. Therefore, $\pi^{\mu}$ also can be written in terms
of the same space-like polarization vectors.
The above solutions when substituted in (\ref{c66-1}) and (\ref{c66-4}) give, in the 
rest frame $(k_0, 0, 0)$ of the quanta of excitations,
\bigskip
$$ (ik_0 a_1 + b_1)\xi^0 (1,{\bf 0}) + (ik_0 a_2 + b_2)\xi^0 (2,{\bf 0}) = 0 $$
and
$$ (ik_0 b_1)\xi^0 (1,{\bf 0}) + (ik_0 b_2)\xi^0 (2,{\bf 0}) = 0. $$
In these equations, the symbol ${\bf k}$ in $a_{{\bf k} \lambda}$ and $b_{{\bf k} \lambda} $ has been suppressed.
It can be seen that the determinant $(a_1b_2 - a_2b_1)$ of the coefficients
does not vanish, since in that case $A^{\mu}$ would be proportional to $\pi^{\mu}$. Hence the only solutions to the above set of two equations are the trivial ones.
That is
\begin{equation}
 \xi^0(\lambda, {\bf 0}) = 0 \label{c66}
\label{c68}
\end{equation}
Similar substitution of (\ref{c67-1},\ref{c67}) in (\ref{c66-2}), (\ref{c66-3})
(\ref{c66-5}) and (\ref{c66-6}) gives, in the rest
frame,
\begin{equation}
 {\Sigma}_{MCS(H)} \bar{\xi}(0) = 0 
\label{c69}
\end{equation}
where
\begin{equation}
 {\Sigma}_{MCS(H)} = \left( \begin{array}{cccc}
  ik_0 a_1 + b_1 & \frac{\vartheta}{2}a_1 & ik_0 a_2 + b_2  & \frac{\vartheta}{2}a_2 \\
  -\frac{\vartheta}{2}a_1 & ik_0 a_1 + b_1 & -\frac{\vartheta}{2}a_2 & ik_0 a_2 + b_2  \\
  ik_0 b_1 - \frac{\vartheta^2}{4}a_1 & \frac{\vartheta}{2}b_1 & ik_0 b_2 - \frac{\vartheta^2}{4}a_2 & \frac{\vartheta}{2}b_2 \\
 -\frac{\vartheta}{2}b_1 & ik_0 b_1 - \frac{\vartheta^2}{4}a_1 & -\frac{\vartheta}{2}b_2 & ik_0 b_2 - \frac{\vartheta^2}{4}a_2 \end{array} \right)
\label{c70}
\end{equation}
and 
\begin{equation}
\bar{\xi}(0) = \left( \begin{array}{c}
{\xi}^1(1,{\bf 0}) \\
{\xi}^2(1,{\bf 0}) \\
{\xi}^1(2,{\bf 0}) \\
{\xi}^2(2,{\bf 0})
\end{array} \right)
\label{cc70}
\end{equation}
The solutions of (\ref{c69}) are given by, 
\begin{equation}
   {\xi}^2(\lambda, {\bf 0}) = -i{\xi}^1(\lambda, {\bf 0}) 
\label{c71}
\end{equation}
For $\vartheta > 0,$ (\ref{c68}) and (\ref{c71}) agree with the result (\ref{c16}) and (\ref{c17}) obtained in the Lagrangian approach. The agreement clearly will be preserved
for the polarizations vectors in a moving frame also.

Now we show that the above result is a special case of a more general one in which we introduce a 
Nakanishi-Lautrup auxiliary field $\cal B$ in the MCS Lagrangian and linearize the gauge fixing
term \cite{gitman}.
In this covariant gauge formalism the Lagrangian (\ref{c63}) is expressed as,
\begin{equation}
 {\cal {L}}_{\alpha} = -{\frac{1}{4}}{F^{\mu \nu}F_{\mu \nu}} + {\frac{\vartheta}{2}}{\epsilon}^{\mu \nu \lambda}A_{\mu}{\partial}_{\nu}A_{\lambda} +
{\cal {B}} \partial_{\mu}A^{\mu} +
 {\frac{{\alpha}}{2}} {{\cal {B}}^2}
\label{c72}
\end{equation}
Notice that ${\cal L}_{\alpha}$ does not 
contain $\dot{\cal B}$ and that it is linear in $\dot{A}^0$. The Euler-Lagrange equations of motion which follow from (\ref{c72}) are,
\begin{equation}
 \Box A^{\mu} - \partial^{\mu}(\partial_{\nu} A^{\nu} + {\cal{B}}) +
\vartheta \epsilon^{\mu \nu \lambda} \partial_{\nu} A_{\lambda} = 0 
\label{c73}
\end{equation}
and
\begin{equation}
 {\cal {B}} = - \frac{1}{\alpha} \partial_{\nu}A^{\nu}.
\label{c74}
\end{equation}
Note that with the choice ${\alpha} = 1 $ and eliminating $\cal B $ using the
equation of motion (\ref{c74}), one can get ${\cal L}_1$. The momenta conjugate to the fields $A^0, A^i$ and $\cal B$ are, respectively, given by
\bigskip
$$ \pi_0 = {\cal{B}}  $$
$$ \pi_i = \partial_i A_0 - \dot{A}_i + \frac{\vartheta}{2} \epsilon_{i j} A^j 
$$
$$ \pi_{{\cal{B}}} = 0 $$
The Hamiltonian obtained from ${\cal {L}}_{\alpha}$ is
\bigskip
$$ H_{\alpha} = \frac{1}{2}\int d^2 x \left[(\pi_i)^2 + \vartheta \epsilon^{i j}\pi^i A^j + \frac{\vartheta^2}{4} + \frac{1}{2}(F^{i j})^2\right] $$
\begin{equation}
 + \int d^2 x \left[A^0 (\partial^i \pi^i - \frac{\vartheta}{2} \epsilon^{i j}
\partial^i A^j) - {\cal{B}}(\partial^i A^i) - \frac{1}{\alpha} {\cal {B}}^2\right]  
\label{c75}
\end{equation}
The constraints of the model are,
\begin{equation}
 \Phi_1 = \pi_0 - {\cal{B}} \approx 0
\label{c76}
\end{equation}
and
\begin{equation}
 \Phi_2 = \pi_{\cal{B}} \approx 0 
\label{c77}
\end{equation}
which form a second class pair. Setting the second class  constraints strongly equal
to zero, one obtains $\pi_0 = {\cal{B}}$, using which one can eliminate the 
auxiliary field ${\cal{B}}$ from the Hamiltonian:
$$
 H_{\alpha} = \frac{1}{2}\int d^2 x \left[(\pi_i)^2 + \vartheta \epsilon^{i j}\pi^i
A^j + \frac{\vartheta^2}{4} + \frac{1}{2}(F^{i j})^2\right] $$
\begin{equation}
 + \int d^2 x \left[A^0 (\partial^i \pi^i - \frac{\vartheta}{2} \epsilon^{i j}
\partial^i A^j) - \pi^0 (\partial^i A^i) - \frac{1}{\alpha} (\pi^0)^2\right]
\label{c78}
\end{equation}
The Poisson brackets between the fields $A^{\mu}$ and their conjugate momenta 
$\pi_{\nu}$ are given by,
\begin{equation}
 \{ A^{\mu}({\bf x}, t), \pi_{\nu}({\bf y}, t) \} = {g^{\mu}}_{\nu} \delta
({\bf x} - {\bf y}). 
\label{c79}
\end{equation}
It should be mentioned that the Dirac brackets in the ($A^{\mu}, \pi_{\nu}$)
sector are identical to their Poisson brackets.
Hence the Hamilton's equations for $A^{\mu}$ and $\pi^{\mu}$ are obtained from
$$ \dot{A}^{\mu} = \{A^{\mu}, H_{\alpha} \}, \hskip 0.5cm 
 \dot{\pi}^{\mu} = \{ \pi^{\mu},  H_{\alpha} \} $$
and are the following.
\begin{eqnarray}
 \dot{A}^0 = \partial^i A^i - {\alpha} \pi^0 \label{80-1} \\
 \dot{\pi}^0 = \frac{\vartheta}{2} \epsilon^{i j} \pi^i A^j -
 \partial^i \pi^i \label{80-2}\\
 \dot{A}^i = - \pi^i - \frac{\vartheta}{2} \epsilon^{i j} A^j + \partial^i A^0 \label{80-3}
\\
 \dot{\pi}^i = \frac{\vartheta^2}{4}A^i + \partial^j F^{i j} - \frac{\vartheta}{2}\epsilon^{i j} \left(\partial^i A^0 + \pi^j\right) - \partial^i \pi^0
\label{c80}
\end{eqnarray}
Substitution of the expansions (\ref{c67-1}) and (\ref{c67}) in (\ref{80-1}) and (\ref{80-2}), in the rest frame
$(k_0, 0, 0)$ leads to,
\bigskip
$$ \left[ \xi^0(1, {\bf 0})a_1 + \xi^0(2, {\bf 0})a_2 \right]ik_0 + \left[ \xi^0(1, {\bf 0})b_1 +
\xi^0(2, {\bf 0})b_2 \right]\alpha = 0 $$
and
$$ \left[ \xi^0(1, {\bf 0})b_1 + \xi^0(2, {\bf 0})b_2 \right]ik_0  = 0 $$ 
If $\alpha \neq \infty$ we can rewrite the above set as
$$\left( \begin{array}{cc}
a_1 & a_2 \\
b_1 & b_2 
\end{array} \right)\left( \begin{array}{c}
 \xi^0(1, {\bf 0}) \\
 \xi^0(2, {\bf 0})
\end{array} \right) = 0 $$
The above equation does not have any nontrivial solution as $a_1b_2 - a_2b_1 \neq 0$ for
the same reason mentioned earlier in the case of $\alpha = 1$. Hence,   
$$ \xi^0(\lambda, {\bf 0}) = 0 $$
if $\alpha$ is finite. When $\alpha = \infty$ there is no unambiguous solution.
This case corresponds to the fact that when $\alpha = \infty$, the gauge
fixing term in $\cal{L}_\alpha$ (\ref{c63}) vanishes. The expansions (\ref{c67-1}), (\ref{c67})
when substituted in (\ref{80-3},\ref{c80}) yields,
$$ \xi^2(\lambda, {\bf 0}) = -i\xi^1(\lambda, {\bf 0}) $$ 
These results are independent of the gauge parameter and naturally agree with
the previous $\alpha = 1$ calculation  and are identical to the results
(\ref{c16}, \ref{c17}) obtained in the Lagrangian framework. Clearly the results will
agree in a boosted frame also. Therefore one can conclude that the form (\ref{c33},)
of the physical polarization vector of MCS theory holds for the different
Lorentz covariant gauge conditions considered here.

\subsection{Maxwell-Chern-Simons-Proca theory}
Taking the MCSP Lagrangian (\ref{c34}) the canonical momenta are defined as,
\begin{equation}
 \pi^i = \frac{{\partial}{\cal L}}{{\partial}\dot{A}_i} = -\left(F^{0 i} + \frac{\theta}{2}\epsilon^{i j}A^{j}\right)
\label{c81}
\end{equation}
and 
\begin{equation}
 \pi^0 \approx 0 
\label{c82}
\end{equation}
is the primary constraint.
The canonical Hamiltonian is, 
 $$ H_{MCSP} = {\frac{1}{2}}\int d^2 x \left[{{\pi}_i}^2 + \frac{1}{2}{F_{i j}}^2 + \left(\frac{{\theta}^2}{4} + m^2\right){A_i}^2 -
\theta {\epsilon_{i j}}A_i \pi_j + m^2 {A_0}^2\right] $$
\begin{equation}
 + \int d^2 x{A_0}\Omega 
\label{c83}
\end{equation}
where,
\begin{equation}
 \Omega = \partial_i\pi_i - \frac{\theta}{2}\epsilon_{i j}\partial_i A_j -m^2A_0 \approx 0
\label{cc84}
\end{equation}
is the secondary constraint. Using (\ref{cc84}) to eliminate $A_0$ from (\ref{c83}), we obtain the reduced Hamiltonian,
$$
 H_R = {\frac{1}{2}}\int d^2 x \left[{{\pi}_i}^2 + \left(\frac{1}{2} + \frac{{\theta}^2}{8m^2}\right){F_{i j}}^2 +
 (\frac{{\theta}^2}{4} + m^2){A_i}^2 - \theta\epsilon_{i j}A_i\pi_j\right]  $$
\begin{equation}
  + {\frac{1}{2m^2}}\int d^2 x \left[(\partial_i \pi_i)^2 - \theta \partial_i \pi_i \epsilon_{l m} \partial_l A_m \right].
\label{c85}
\end{equation}
The only  non vanishing bracket between the phase space variables is,
\begin{equation}
 \{A^i({\bf x},t), \pi^j({\bf y},t)\} = - \delta^{i j}\delta({\bf x} - {\bf y})  
\label{c86}
\end{equation}
Therefore the Hamilton's equations are given by
\begin{equation}
 \dot{A}^i = \{ A^i, H_R \}
             = -\pi^i + \frac{1}{m^2}\partial^i\partial^j\pi^j -  \frac{\theta}{2}\epsilon^{i j}A^j 
- \frac{\theta}{2m^2}\epsilon^{l m}\partial^i\partial^lA^m \label{c87}
\end{equation}
 and
\begin{equation}
 \dot{\pi}^i = \{ \pi^i, H_R \}
 = \left(1 + \frac{\theta^2}{4m^2}\right)\left[\partial^i\partial^jA^j - \partial^j\partial^jA^i + m^2A^i\right]
-\frac{\theta}{2}\epsilon^{i j}\left[\pi^j + \frac{1}{m^2}\partial^j\partial^k\pi^k\right]
\label{c88}
\end{equation}
We consider solutions (in terms of the polarization vectors $\varepsilon({\bf k}, \lambda))$ of the form,
\begin{eqnarray}
 A^i = \sum_{\lambda = 1}^2 \varepsilon^i(\lambda, {\bf k})a_{{\bf k} \lambda} \exp[ik \cdot x] + c.c \\
 \pi^i = \sum_{\lambda = 1}^2 \varepsilon^i(\lambda, {\bf k})b_{{\bf k} \lambda} \exp[ik \cdot x] + c.c
\label{c89}
\end{eqnarray}
Substitution of the above solutions in the Hamilton's equations (\ref{c87}) and (\ref{c88})
yields, respectively,
\bigskip
$$ \sum_{\lambda = 1}^2 -ik_0\varepsilon^i(\lambda, {\bf k})a_{{\bf k} \lambda} = \sum_{\lambda = 1}^2  \{[\varepsilon^i(\lambda, {\bf k}) +
{\frac{1}{m^2}}k^ik^j\varepsilon^j(\lambda, {\bf k})]b_{{\bf k} \lambda}   $$
\begin{equation}
 + {\frac{\theta}{2}}
[\epsilon^{i j}\varepsilon^j(\lambda, {\bf k}) - {\frac{1}{m^2}}\epsilon_{l m}k^ik^l\varepsilon^m(\lambda, {\bf k})]
a_{{\bf k} \lambda} \} \label{c90}
\end{equation}
and 
\bigskip
$$ \sum_{\lambda = 1}^2 -ik_0\varepsilon^i(\lambda, {\bf k})b_{{\bf k} \lambda} = \sum_{\lambda = 1}^2 \{[k^ik^j
\varepsilon^{j}(\lambda, {\bf k}) - k^jk^j\varepsilon^{i}(\lambda, {\bf k}) - m^2 \varepsilon^{i}(\lambda, {\bf k})]
(1 + \frac{\theta^2}{4m^2})a_{{\bf k} \lambda} $$
\begin{equation}
 + \frac{\theta}{2}\epsilon^{i j}[\varepsilon^{j}(\lambda, {\bf k}) - {\frac{1}{m^2}}k^jk^k\varepsilon^{k}(\lambda,
 {\bf k})]b_{{\bf k} \lambda} \}
\label{c91}
\end{equation}
In the rest frame $(k^0, 0, 0)$, the above set of four equations can be written in the matrix form as, 
\begin{equation}
 {\Sigma}_{MCSP(H)}\bar{\varepsilon} = 0  
\label{c92}
\end{equation}
where
$$ {\Sigma}_{MCSP(H)} = $$ 
\begin{equation} \left( \begin{array}{cccc}
  b_1 + ik_0a_1 & \frac{\theta}{2}a_1 & b_2 + ik_0a_2 & \frac{\theta}{2}a_2  \\
  -\frac{\theta}{2}a_1 & b_1 + ik_0a_1 & -\frac{\theta}{2}a_2 & b_2 + ik_0a_2 \\
  -Dm^2a_1 + ik_0b_1 & \frac{\theta}{2}b_1 &  -Dm^2a_2 + ik_0b_2 & \frac{\theta}{2}b_2 \\
  -\frac{\theta}{2}b_1 & -Dm^2a_1 + ik_0b_1 & -\frac{\theta}{2}b_2 & -Dm^2a_2 + ik_0b_2
\end{array} \right)  \label{c93}
\end{equation}
and 
\begin{equation}
 \bar{\varepsilon} = \left( \begin{array}{c}
{\varepsilon}^1(1,{\bf 0}) \\
{\varepsilon}^2(1,{\bf 0}) \\
{\varepsilon}^1(2,{\bf 0}) \\
{\varepsilon}^2(2,{\bf 0}) 
\end{array} \right)
\label{c94}
\end{equation}
with $D= (1 + \frac{\theta^2}{4m^2}) $.
For a nontrivial solution of (\ref{c92}), $ \det{\Sigma}_{MCSP(H)} = 0. $ This condition, after a straightforward
algebra, reduces to 
\begin{equation}
 (a_1b_2 - a_2b_1)^2\left[k_0^4 -2k_0^2(\frac{\theta^2}{2} + m^2) + m^4\right] = 0 
\label{c95}
\end{equation}
from which it follows that
\begin{equation}
 \left[ k_0^4 -2k_0^2(\frac{\theta^2}{2} + m^2) + m^4 \right] = 0
\label{c96}
\end{equation}
That is, 
\begin{equation}
 k_0 = \sqrt{\frac{\theta^2}{4} + m^2} \pm \frac{\theta}{2} = \theta_\pm 
\label{c97}
\end{equation}
Replacing $k_0$ with $\theta_+$ in (\ref{c92}) we obtain, after suitable
manipulations, the following relationship between the components of
$\vec{\varepsilon}(\lambda, {\bf 0})$;
\bigskip
$$ \varepsilon^2(\lambda, {\bf 0}) = -i\varepsilon^1(\lambda, {\bf 0}) $$
For $k_0 = \theta_- $ the corresponding relationship is given by
$$ \varepsilon^2(\lambda, {\bf 0}) = +i\varepsilon^1(\lambda, {\bf 0}). $$
Denoting  the rest frame polarization vectors corresponding
to $ k_0 = \theta_\pm $ by $\varepsilon_\pm({\bf 0})$, the above two
expressions can be written as,
\begin{equation}
 \varepsilon^2_\pm({\bf 0}) = \mp i \varepsilon^1_\pm({\bf 0}).
\label{c98}
\end{equation}
which agrees with the relationship obtained from the  Lagrangian
formalism. The polarization vectors in moving frames are obtained
by  boosting the rest frame vectors appropriately, and the result
obviously agrees with that obtained earlier within the Lagrangian framework.

\section{Summary}
A detailed analysis of the polarization vectors in planar field theories
involving both a topological mass and a usual mass has been done. The
structure of these vectors is crucial for the reduction formulae, the
study of unitarity in topologically massive gauge theories\cite{tyutin}, as
well as for considering the massless limit of these theories augmented by a normal mass term.
Our general approach using either Lagrangian or Hamiltonian techniques,
has shown a $U(1)$ invariance (in the k-space) in the form of the
polarization vectors. This is quite distinct from the usual Abelian
invariance associated with gauge theories. The $U(1)$ invariance
reported here is connected with the presence of the Chern-Simons(CS)
term and has nothing to do with the presence or absence of gauge freedom
in the action. The CS term leads to complex entries in the polarization 
vector, thereby manifesting a $U(1)$ invariance. This can be contrasted with
the pure Maxwell theory where all entries are real so that there is no
$U(1)$ invariance of this type. It was seen that in MCS theories, the massive modes were
physical while the massless ones could be gauged away and hence were
unphysical. This is the exact counterpart of the Maxwell theory where
the roles of the massive and massless modes are reversed.
The structures found here naturally revealed
a mapping between the Maxwell-Chern-Simons-Proca  model and a doublet
of Maxwell-Chern-Simons theories with opposite helicities. This was also
helpful in studying the massless limit of the MCSP model.
A mapping between the MCSP model and  a pair of self and anti self dual
models is  shown, once again on the basis of the polarization vectors.
The Hamiltonian analysis of MCS theory with various covariant gauge conditions
revealed that the structure of the physical polarization vector for the theory
is the same for all these gauge choices. For MCSP theory also both Lagrangian
and Hamiltonian analyses produced identical results. This enables one to 
conclude that the mapping between MCSP theory to a doublet of MCS theories
is manifested at the level of polarization vectors  of the basic fields for
arbitrary gauge parameter.


\chapter{Translational groups as generators of gauge transformations}
Wigner's little group is quite familiar to physicists mainly
because of  its role in the classification of elementary particles.
Wigner introduced the concept
of little group in a seminal paper \cite{w} published in 1939 and showed
how particles can be classified on the basis of their spin/ helicity quantum
numbers using the little group. Several decades later Wigner along with Kim
showed that the little group relates the internal symmetries of massive
 and massless particles \cite{wigner1,wigner2,kim}.   A comparatively less
known facet of the little group, namely  its role as a generator of gauge
transformations in various Abelian gauge theories, was also unraveled in the mean time. Historically, the first attempts to study this aspect of little group
were in the contexts
of free Maxwell theory  \cite{we,hk,hk1} and linearized Einstein gravity \cite{ng}.
These studies revealed  that the defining representation of 3+1 dimensional
 Wigner's  little group  for
massless particles (which is isomorphic to $E(2)$ - Euclidean group in two
dimensions)\footnote{A  brief review of the essential properties of
this little group is given in appendix A.}, or more precisely its translational subgroup $T(2)$, 
acts as generators of gauge
transformations  in 3+1 dimensional Maxwell theory and linearized gravity. 
Recently, this gauge generating property of the little group received more
attention \cite{bc1, bc2, bcs1}.
 Moreover, it was shown that the same
translational group was found to be generating gauge transformation in 
the 3+1 dimensional Kalb-Ramond (KR) theory involving a massless 2-form gauge field \cite{bc1}. (A further  study on the gauge generating property of Wigner's little group in the context of massless KR
theory  can be found in \cite{malik} where its connection with BRST cohomology
is explored.)
On the other hand, the same study \cite{bc1} showed that
  for the $B\wedge F$ theory which is  obtained by
topologically coupling Maxwell field to a KR field, the generator of gauge
transformation is a particular representation of the three dimensional
translational group $T(3)$. Note that $B\wedge F$ theory is a topologically massive
gauge theory where  gauge invariance co-exist with mass \cite{kr,cs,lahiri}. Thus we see that
different translational groups in their appropriate representations generate
gauge transformations in various Abelian gauge theories in 3+1 dimensions.
In this chapter we provide a review of the gauge generating property of
translational group $T(2)$
 and $T(3)$ 
with necessary details. All the
theories considered in this chapter belong to 3+1 dimensional space-time.
In later chapters, we borrow some of the techniques developed in the context
of  these
3+1 dimensional theories for similar investigations in planar theories.

\section{Translational subgroup of Wigner's little group}
\label{tswlg}
A brief review of the essential aspects of Wigner's little group for massless
particles 
\cite{we,hk,wkt}
which are relevant to the present study is provided in appendix A.
Wigner's little group is defined as the subgroup of homogeneous Lorentz group that
preserves the energy-momentum vector of a particle. For a massive particle,
it is trivial to see that the little group is given by $SO(3)$ as one makes the transition to rest frame. As discussed in appendix A,
an element of the little group that
preserves the four-momentum $k^a = (\omega, 0, 0, \omega )^T$ of a massless
particle moving in the $z$-direction  is given by
\begin{equation}
W(p, q;  \phi) = W(p, q) R(\phi)
\label{3-2}
\end{equation}
where
\begin{equation}
W(p, q) \equiv W(p, q;  0) = \left( \begin{array}{cccc}
1+ \frac{p^2 + q^2 }{2} & p & q  & -\frac{p^2 + q^2 }{2} \\
p & 1 & 0 & -p \\
q & 0 & 1 & -q \\
\frac{p^2 + q^2 }{2} &  p & q  & 1 -\frac{p^2 + q^2 }{2}
\end{array}\right)
\label{3-3}
\end{equation}
is  a particular representation of the  translational subgroup  $T(2)$ of
the little group and $ R(\phi)$ represents a $SO(2)$ rotation about the $z$-axis.
Note that the representation $W(p, q)$ satisfies the  relation $ W(p, q)W(\bar{p},\bar{q}) = W(p+\bar{p}, q+\bar{q})$.

\section{Maxwell theory}
\label{tm}
We now discuss how this representation of $T(2)$
generates gauge transformation in Maxwell theory which has the well known
Lagrangian,
\begin{equation}
{\cal L} = -\frac{1}{4}F_{ab }F^{ab }.
\label{3-4-1}
\end{equation}
Maxwell theory is invariant under the  gauge transformation
\begin{equation}
A_{a}(x) \rightarrow
A^{\prime}_{a}(x) =  A_{a}(x) + \partial_{a}\tilde{f}(x)
\label{3-4-2}
\end{equation}
where $\tilde{f}(x)$ is an arbitrary scalar function.
The Lagrangian (\ref{3-4-1}) leads to following  equation of motion:
\begin{equation}
\partial_a F^{ab} = 0.
\label{3-4-3}
\end{equation}
Denoting the polarization vector of a photon by $\varepsilon^{a}(k)$, the  gauge field $A^a (x)$  can be written as
\begin{equation}
A^{a}(x) = \varepsilon^{a}(k) e^{ik \cdot x}.
\label{3-4-4}
\end{equation}
In terms of the
polarization vector, the gauge transformation (\ref{3-4-2}) is expressed  as
\begin{equation}
\varepsilon_{a}(k) \rightarrow \varepsilon^{\prime}_{a}(k) =  \varepsilon_{a}(k) + f(k)k_{a}.
\label{3-4-5}
\end{equation}
where $f(k)$ is proportional to the Fourier component of  $\tilde{f}(x)$. 
The equation of motion, in terms of the polarization vector,  will now be given
by
\begin{equation}
k^2 \varepsilon^{a} - k^{a} k_{b} \varepsilon^{b} = 0
\label{3-4-6}.
\end{equation}
The massive excitations corresponding to $k^2 \neq 0$ leads to a solution
$\varepsilon^{a} \propto k^{a}$ which can therefore be gauged away by a
suitable choice of $f(k)$ in (\ref{3-4-5}). For
massless excitations  ($k^2 = 0$), the Lorentz condition $
k_a \varepsilon^{a} =0$
 follows immediately
 from (\ref{3-4-6}).
For a  photon of energy $\omega$ propagating in the $z$-direction (i.e., $k^{a} = (\omega, 0, 0, \omega)^T$, it follows from (\ref{3-4-6}) that the corresponding polarization tensor   $\varepsilon^{\mu}(k)$ takes
the form $(\varepsilon^{0}, \varepsilon^1, \varepsilon^2, \varepsilon^{0})$
which can be reduced to the
maximally reduced form
\begin{equation}
\varepsilon^{a}(k) =
(0, \varepsilon^1, \varepsilon^2, 0)^T
\label{3-4}
\end{equation}
by a suitable  gauge transformation (\ref{3-4-5}) with $f(k) = \frac{-\varepsilon^0}{\omega}$. Note that the maximally reduced
form
 displays just the two transverse physical degrees of freedom
$\varepsilon^1$ and $\varepsilon^2$.
Under the action (\ref{3-3}) of the translational group $T(2)$  this polarization
vector transforms  as follows:
\begin{equation}
\varepsilon^{a} \rightarrow \varepsilon^{\prime a} = {W^{a}}_{b}(p, q) \varepsilon^{b} = \varepsilon^{a} +
 \left( \frac{p\varepsilon^1 + q\varepsilon^2}{\omega}\right)k^{a}~.
\label{3-5}
\end{equation}
Clearly, this can be identified as a gauge transformation of
the form (\ref{3-4-5}) by
choosing
\begin{equation}
f(k) = \frac{p\varepsilon^1 + q\varepsilon^2}{\omega}.
\label{naya}
\end{equation}
 Hence one 
 says that the translational subgroup
of Wigner's little group for massless particles acts as a gauge generator in 
Maxwell theory.

\section{Linearized gravity}
\label{tlg}
The pure
Einstein-Hilbert action in 3+1 dimensions is given by
\begin{equation}
I^E = -\int d^4 x  {\cal L}^E,  ~~~~~ {\cal L}^E = \sqrt{g} R = \sqrt{g} g^{ab}R_{ab}
\label{32-1}
\end{equation}
where ${\cal L}^E$ is the Einstein Lagrangian and $R_{ab}$ is the Ricci
tensor.
  In the linearized approximation the metric $g_{ab}$ is
assumed to be close to the  flat background part $\eta^{ab}$ and therefore
\begin{equation}
g_{ab} = \eta_{ab} + h_{ab}
\label{32-2}
\end{equation}
where $h_{ab}$ is the deviation such that $|h_{ab}|<< 1 $. When the
deviation is small one considers only terms up to first order
in $h_{ab}$. The raising and lowering of indices is done using $\eta^{ab}$
and $\eta_{ab}$ respectively.

The linearized version of
Einstein-Hilbert Lagrangian  is
\begin{equation}
{\cal L}_L^E = \frac{1}{2}h_{ab} \left[ R^{ab}_L - \frac{1}{2} \eta^{ab} R_L\right].
\label{32-3}
\end{equation}
Here $ R^{ab}_L$ is the linearized Ricci tensor given by
\begin{equation}
 R^{ab}_L = \frac{1}{2}(- \Box h^{ab} + \partial^a \partial_c  h^{c b} + \partial^b\partial_c  h^{c a} - \partial^a \partial^b h)
\label{32-4}
\end{equation}
with $h = {h^c}_c$. Similarly $R_L = {R^c_L}_c$.
As mentioned before, the translational subgroup of Wigner's little group for massless particles
generates gauge transformations in linearized Einstein gravity also \cite{ng, sc}. However, the gauge
transformations generated by the translational group constitute only a subset of the entire set of gauge
transformations available in linearized gravity. Here we give a detailed review of this partial gauge
generation in linearized gravity by the translational group $T(2)$.

Gravity (linearized) in $d$ space-time dimensions is described by a symmetric  second rank tensor gauge field and  has $\frac{1}{2}d(d-3)$ degrees of
freedom\footnote{The degree of freedom counting can be done by following
Weinberg \cite{weinberg1}.
To start with, note that a symmetric second rank tensor in $d$ dimensions has
$\frac{1}{2}d(d+1)$ independent components. Analogous to
the Lorentz gauge condition ($\partial^\mu A_\mu = 0$) of Maxwell theory,
in general relativity we have the harmonic gauge condition
$g^{\mu \nu} {\Gamma^\lambda}_{\mu \nu} = 0$ which amounts to $d$ constraints
on the components of
$g_{\mu \nu}$. These along with the $d$ independent components of the   gauge parameter (which by itself is a $d$-vector now;
see the ensuing discussion below, particularly (\ref{32-6})), in the linearized
version of the theory, reduces the number of independent components of the
tensor
field to $ \frac{1}{2}d(d+1) - 2d = \frac{1}{2}d(d-3)$.}. Therefore general
relativity in 3+1 dimensions has two degrees of freedom.
The field
equations for $ h^{ab}$ following from the Lagrangian (\ref{32-3}) is given by
\begin{equation}
- \Box h^{ab} + \partial^a \partial_c  h^{c b} + \partial^b\partial_c  h^{ca}
 - \partial^a \partial^b h + \eta^{ab}(\Box h -\partial_c \partial_d h^{cd}) = 0.
\label{32-5}
\end{equation}
The above equation is invariant under the following gauge transformation:
\begin{equation}
h^{ab} \rightarrow h'^{ab} = h^{ab} + \partial^a \zeta^b (x) + \partial^b  \zeta^a (x)
\label{32-6}
\end{equation}
Here $ \zeta^a(x)$ are completely arbitrary except that they
 are considered to be small.
Following the plane wave method,  we now adopt  the ansatz
\begin{equation}
h^{ab} = \chi^{ab} (k) e^{ik.x} + c.c.
\label{32-7}
\end{equation}
where $\chi^{ab} (k)$ is the symmetric  polarization tensor. 
With the choice
\begin{equation}
 \zeta^a(x) = -i\zeta^a(k) e^{ik.x} + c.c.
\label{32-8}
\end{equation}
the gauge transformation in $h^{ab}$ can be written in terms of the
polarization tensor as
\begin{equation}
 \chi^{ab} (k)  \rightarrow {\chi}'^{ab} (k) = \chi^{ab} (k) + k^a \zeta^b (k) +  k^b \zeta^a (k).
\label{32-9}
\end{equation}
Just as in the Maxwell case, hereafter we  will consider only  the negative
frequency  part for simplicity.
Substituting the ansatz in the equation of motion yields
\begin{equation}
k^2\chi^{ab} - k^a k_c \chi^{cb} -
k^b k_c \chi^{ca }
+  k^a k^b \chi
+\eta^{ab}(-k^2 \chi +k_c k_d \chi^{cd}) =
0.
\label{32-10}
\end{equation}

As was done  in the previous cases, we separately consider the two possibilities $k^2 \ne 0$ and $k^2 = 0$. Choosing the massive ( $k^2 \ne 0$) case first,
we contract the equation of motion (\ref{32-10}) with $\eta^{ab}$ to obtain
\begin{equation}
k^2\chi - k_c k_d \chi^{cd} = 0~;~~~~ \chi = {\chi^a}_b.
\label{32-11}
\end{equation}
A general solution to this equation is given by
\begin{equation}
\chi^{ab}(k) = k^{a} f^{b}(k) +  k^{b} f^{a}(k)
\label{32-12}
\end{equation}
with $f^{a}(k)$ being arbitrary functions of $k$. Therefore, it can be
 easily seen that this solution can be $`$gauged away' by appropriate choice of the variables
$\zeta^a (k)$ in (\ref{32-9}) as this corresponds to pure gauge. Thus, analogous to Maxwell theory, the
massive excitations of linearized Einstein gravity  are gauge artefacts.

For massless ($k^2 = 0$) excitations, the equation of motion (\ref{32-10}) reduces
 to
\begin{equation}
- k^a k_c \chi^{cb} -
k^b k_c \chi^{ca }
+  k^ak^b \chi
+\eta^{ab}k_c k_d \chi^{cd} = 0.
\label{32-13}
\end{equation}
In  a frame  of reference where $k^{a} = (\omega , 0, 0, \omega )^T$
the above equation can be written as
\begin{equation}
-\omega [k^{a} (\chi^{0b} - \chi^{3 b}) +
k^{b} (\chi^{0a} - \chi^{3 a})] + k^{a} k^{b}  \chi +\omega^2 \eta^{ab}(\chi^{00} + \chi^{33} - 2 \chi^{03}) = 0.
\label{32-14}
\end{equation}
Various components of the above equation together with the symmetricity of
$\chi^{ab}$ leads to the reduction in the number of independent
components of $\chi^{ab}$. In (\ref{32-14}), ${a = b = 0}$ leads
to $ \chi^{11} = -\chi^{22}$; ${a =0, b = i}$ to
$ \chi^{0i} = \chi^{3i}$
 and ${a = b = i}$ to $ \chi^{00}
 = \chi^{33}$. Therefore, one can write the polarization tensor $\chi^{ab}$ in a reduced form as
\begin{equation}
\{\chi^{ab}\} = \left(
\begin{array}{cccc}
 \chi^{00} & \chi^{01} &  \chi^{02}& \chi^{00} \\
\chi^{01} & \chi^{11} &  \chi^{12} & \chi^{01} \\
 \chi^{02} & \chi^{12} & -\chi^{11} &  \chi^{02} \\
 \chi^{00} & \chi^{01} &  \chi^{02}& \chi^{00}
\end{array}
\right).
\label{32-15}
\end{equation}
Now if we make a momentum space gauge transformation  with the choice $ \zeta^0 = \zeta^3 = -\frac{\chi^{00}}{2\omega},
\zeta^2  =-\frac{\chi^{01}}{\omega}, \zeta^2  =-
\frac{\chi^{02}}{\omega}$, the polarization tensor $ \{\chi^{ab}\} $ can be written in the maximally reduced form  
 as follows:
\begin{equation}
  \{\chi^{ab}\} = \left(
\begin{array}{cccc}
0 & 0 & 0 & 0\\
0 & \chi^{11} & \chi^{12} & 0 \\
0 & \chi^{12}  & -\chi^{11} & 0 \\
0 & 0 & 0 & 0
\end{array}
\right).
\label{32-16}
\end{equation}
Here $\chi^{11}$ and $\chi^{12}$   represent the two physical degrees of freedom
for the theory. (This form of the polarization tensor of linearized Einstein
gravity in 3+1 dimensions is derived in \cite{weinberg1} following a different approach.)
In this form (\ref{32-16}),  the polarization tensor $\{\chi^{ab}\}$ satisfies the harmonic gauge condition,
\begin{equation}
k_a \chi^{ab} = \frac{1}{2} k^b \chi^{a}_{a}
\label{32-17}
\end{equation}in momentum space,
automatically.
Using the maximally reduced form (\ref{32-16}) of the polarization tensor, it
is now
straightforward to show that the group $W(p, q) $ in (\ref{3-3}) generate
gauge transformations in linearized Einstein gravity also. For this purpose
consider the action of $W(p, q)$ on $\{\chi^{ab}\}$ in (\ref{32-16}),
$$\{\chi^{ab}\} \rightarrow \{{\chi}'^{ab}\}= W(p, q) \{\chi^{ab}\} W^T(p, q)$$
\begin{equation}
 = \{\chi^{ab}\} +\left(
\begin{array}{cccc}
\left( \begin{array}{c} (p^2 - q^2)\chi^{11}\\ + 2pq\chi^{12}) \end{array}\right)  & (p\chi^{11} + q\chi^{12}) & (p\chi^{12} -q\chi^{11}) & \left( \begin{array}{c} (p^2 - q^2)\chi^{11}\\ + 2pq\chi^{12}\end{array}\right) \\
(p\chi^{11} + q\chi^{12}) &0 & 0 & (p\chi^{11} + q\chi^{12}) \\
(p\chi^{12} -q\chi^{11})  & 0 & 0 & (p\chi^{12} -q\chi^{11}) \\
\left(\begin{array}{c}(p^2 - q^2)\chi^{11} \\+ 2pq\chi^{12}\end{array}\right) &( p\chi^{11} + q\chi^{12}) &( p\chi^{12} -q\chi^{11}) & \left(\begin{array}{c}(p^2 - q^2)\chi^{11} \\ + 2pq\chi^{12}\end{array}\right)
\end{array}\right).
\label{32-18}
\end{equation}
The above transformation can be cast in the form of a gauge transformation (\ref{32-9}) with the
following choice for  the arbitrary functions $\zeta^a(k)$:
\begin{equation}
\zeta^0 = \zeta^3 = \frac{(p^2 - q^2)\chi^{11} + 2pq\chi^{12}}{\omega},
~~~
\zeta^1 = \frac{p\chi^{11} + q\chi^{12}}{\omega},~~~
\zeta^2 = \frac{p\chi^{12} -q\chi^{11}}{\omega}.
\label{32-19}
\end{equation}
However, since $k^a = (\omega , 0,0, \omega )^T$,  a general gauge
transformation for this polarization tensor (\ref{32-16}) has the form
$$\{\chi^{ab} \} \rightarrow \{\chi'^{ab}\}= \{\chi^{ab} \} + \{ k^a \zeta^b \} +
\{k^b \zeta^a\} $$
\begin{equation}
 = \{\chi^{ab}\} +\omega\left(
\begin{array}{cccc}
2 \zeta^0   & \zeta^1 & \zeta^2  & (\zeta^0 + \zeta^3) \\
\zeta^1 &0 & 0 & \zeta^1 \\
\zeta^2  & 0 & 0 & \zeta^2 \\
(\zeta^0 + \zeta^3) &\zeta^1 &\zeta^2 & 2 \zeta^3
\end{array}\right).
\label{32-20}
\end{equation}
Upon comparing the above form of general gauge transformation with the one
generated by $W(p,q)$ given in (\ref{32-18}), it becomes clear that the latter is
only a special case of the former as the relation $\zeta^0 = \zeta^3 $ in
(\ref{32-19})  restricts the number of independent components of the arbitrary
vector  $ \zeta^a$. Therefore, the translational subgroup $T(2)$ of
Wigner's little  group for massless particles generates only a subset of the
full set of gauge transformations in linearized gravity. The reason for
this partial gauge transformation is as follows. We must notice that our
starting point of gauge generation by $W(p,q)$ in linearized gravity is the
maximally reduced polarization tensor (\ref{32-16}) which contains just the
physical sector of the theory (in the reference frame where $k^\mu =
(\omega , 0,0, \omega )^T$) and is devoid of any arbitrary variables to
begin with.
Hence in the gauge generation by $W(p,q)$, we must rely entirely on the two
parameters $p$ and $q$ of the translational group to manufacture the
gauge equivalence class of the state corresponding the polarization tensor
(\ref{32-16}). However, the gauge freedom in linearized gravity is represented by
the arbitrary vector variable $ \zeta^a$ having four components. Naturally,
in the gauge generation by $W(p,q)$ in linearized gravity, only two of the four
components of $ \zeta^a$  remain independent (as is evident from(\ref{32-19}))  when expressed in terms of the two parameters $(p,q)$ of the
translational group and therefore the gauge generation is only partial.
It was noted in \cite{ng} that the gauge generation by the little group in
linearized gravity is subject to the $`$Lorentz condition' $k_a \zeta^a (k) =
0$. This also can be seen from the first relation $\zeta^0 = \zeta^3$ in
(\ref{32-19}). Thus, we  have unraveled all the constraints
behind the partial gauge generation by Wigner's little group in linearized
gravity. In contrast, we may note that the gauge freedom in free Maxwell theory is represented
by a single arbitrary scalar variable $f(k)$ (\ref{naya}) which can be expressed
(without any restrictions) in
terms of the two parameters of  $W(p,q)$ in the gauge generation by little
group as is evident from (\ref{3-5}).
Hence translational subgroup of Wigner's little group  generates the full set of gauge
transformations in Maxwell theory.

\section{Massless Kalb-Ramond theory}
\label{tkr}
By similar methods it can be shown that gauge
transformations are generated by $T(2)$ in the 3+1 dimensional
Kalb-Ramond(KR) theory \cite{bc1} which has a second rank antisymmetric tensor
as its basic field.  The KR theory is described by the Lagrangian 
\begin{equation}
{\cal{L}} = \frac{1}{12}H_{abc}H^{abc}; ~~~~H_{abc} =
\partial_a B_{bc} + \partial_b B_{ca} + \partial_c B_{ab}
\label{3-6}
\end{equation}
where $B_{ab}$ is a 2-form gauge field:
\begin{equation}
B_{ab}  = - B_{ba}.
\label{3-8}
\end{equation}
The equation of motion is
\begin{equation}
\partial_a H^{abc} = 0.
\label{3-9}
\end{equation}
The KR theory is invariant under the gauge transformation
\begin{equation}
B_{ab}(x) \rightarrow   B'_{ab}(x) = B_{ab}(x) + \partial_a F_b
(x) - \partial_b F_a (x)
\label{3-10}
\end{equation}
where $F_a (x)$ are arbitrary functions.
However, these gauge transformations are not all independent.
One can see that under the transformation
\begin{equation}
F_a (x)
 \rightarrow F'_a (x) = F_a (x) + \partial_a \beta (x)
\label{3-11}
\end{equation}
 (where $\beta (x)$ is an arbitrary scalar function) the gauge transformation (\ref{3-10}) remains invariant.
In particular, if $F_a = \partial_a \Lambda$,  the gauge transformation vanish
trivially.
This is known as  the $`$gauge invariance of gauge transformations' and is  a typical
property of reducible gauge theories
where the generators
of gauge transformation
are not all independent \cite{gomis}. Since the components of the arbitrary
field $F_a (x)$ (which represents the gauge freedom in the KR theory)
are not all independent, there exists some superfluity in the
 gauge transformation (\ref{3-10}). 

In order to obtain the maximally reduced form of the 
 antisymmetric polarization tensor $\varepsilon_{ab}(k)$ of the massless
KR theory, as was done in the previous models, we employ the plane wave method using the ansatz 
\begin{equation}
B_{ab}(x) = \varepsilon_{ab}(k)e^{i k \cdot x}.
\label{3-12}
\end{equation}
In terms of the polarization tensor $\varepsilon^{ab}$, the gauge
transformation (\ref{3-10}) can be written in the momentum space as
\begin{equation}
\varepsilon_{ab}(k) \rightarrow \varepsilon^{\prime}_{ab}(k) = \varepsilon_{ab} (k)+ i(k_{a}f_{b}(k) - k_{b}f_{a}(k))
\label{3-13}
\end{equation}
while the counterpart of (\ref{3-11}) is given by
\begin{equation}
f_a(k) \rightarrow f'_a(k) + ik_a \tilde{\beta}
\label{3-13-13}
\end{equation}
(where we have written $F_a(x) = f_{a}(k) e^{ik\cdot x}$ and $\beta (x) =\tilde{\beta} (k) e^{ik.x}$) 
and the equation of motion (\ref{3-9}) as
\begin{equation}
k_{a}[k^{a} \varepsilon^{bc} + k^{b} \varepsilon^{ca} +
k^{c} \varepsilon^{ab}] =0.
\label{3-14}
\end{equation}
For massive excitations (i.e., when  massive $k^2 \ne 0$), we have
\begin{equation}
\varepsilon^{bc}(k) = \frac{1}{k^2}[k^{b}(k_{a}\varepsilon^{ac}) - k^{c}(k_{a}\varepsilon^{ab})]
\label{3-15}
\end{equation}
Using (\ref{3-13}), this can be gauged away by choosing
\begin{equation}
f^{c}(k) = \frac{i}{k^2}k_{a}\varepsilon^{ac}
\label{3-16}
\end{equation}
We thus find that massive excitations of KR theory are gauge artefacts.

For massless excitations ($k^2 = 0$), the momentum-space equation of motion
 (\ref{3-14}) reduces to 
\begin{equation}
k_{a}\varepsilon^{ab} = 0
\label{3-17}
\end{equation}
which is equivalent to the $``$Lorentz condition" $\partial_{a}B^{ab}
= 0$. Using this condition, the six independent components of the
antisymmetric polarization matrix $ \{\varepsilon^{ab}\}$ can be reduced
further. In the reference frame where the light-like vector $k^a$
takes the form $k^a = (\omega, 0, 0, \omega)^T$, the condition (\ref{3-17})
can be written as 
\begin{equation}
\{\varepsilon^{ab}\} \cdot p = \left( \begin{array}{cccc}
0 & \varepsilon^{01} & \varepsilon^{02} & \varepsilon^{03} \\
-\varepsilon^{01} & 0 & \varepsilon^{12} & \varepsilon^{13}  \\
-\varepsilon^{02} & -\varepsilon^{12} & 0 & \varepsilon^{23}  \\
-\varepsilon^{03} & -\varepsilon^{13}  & -\varepsilon^{23} & 0
\end{array} \right) \left( \begin{array}{c}
\omega \\
0 \\
0 \\
\omega
\end{array} \right) = 0.
\label{3-18}
\end{equation}
The above equation can easily be simplified to
\begin{equation}
\varepsilon^{03} = 0,~~~ \varepsilon^{01} = \varepsilon^{13},~~~ \varepsilon^{02} = \varepsilon^{23}
\label{3-19}
\end{equation}
so that the polarization tensor $\varepsilon^{ab}$ can now be written as
\begin{equation}
\{\varepsilon^{ab}\} =
\left( \begin{array}{cccc}
0 & \varepsilon^{01} & \varepsilon^{02} & 0 \\
-\varepsilon^{01} & 0 & \varepsilon^{12} & \varepsilon^{01} \\
-\varepsilon^{02} & -\varepsilon^{12} & 0 & \varepsilon^{02} \\
0 & -\varepsilon^{01} & -\varepsilon^{02} & 0
\end{array} \right).
\label{3-20}
\end{equation}
With the gauge choice $f^1 = \frac{i}{\omega} \varepsilon^{01}$
and
 $f^2 = \frac{i}{\omega} \varepsilon^{02}$, if we now  make a gauge
transformation (\ref{3-13}), the above form of the polarization tensor
yields the following maximally reduced form:
\begin{equation}
\{\varepsilon^{ab}\} = \varepsilon^{12}
  \left( \begin{array}{cccc}
0 & 0 & 0 & 0 \\
0 & 0 & 1 & 0 \\
0 & -1 & 0 & 0 \\
0 & 0 & 0 & 0
 \end{array} \right)
\label{3-21}
\end{equation}
The transformation of this maximally reduced polarization tensor (\ref{3-21})
of the massless KR theory
under the translational subgroup $W(p, q)$
(\ref{3-3}) of Wigner's little group, can be written as
\begin{equation}
\{\varepsilon ^{ab} \}\rightarrow \{\varepsilon'^{ab}\}  = W(p,q) \{\varepsilon^{\mu \nu}\}  W^T (p,q) =
 \{\varepsilon^{ab} \}+ \varepsilon^{12}\left( \begin{array}{cccc}
0 & -q & p & 0 \\
q & 0& 0 & q \\
-p & 0 & 0 &  -p \\
0 & -q & p & 0
\end{array} \right)
\label{3-22}
\end{equation}
This can be cast in the
form of  (\ref{3-13}) with
\begin{equation}
f^1 = \frac{-q\varepsilon^{12}}{i\omega},~~~
f^2 =  \frac{p\varepsilon^{12}}{i\omega}, ~~~
f^3 = f^0.
\label{3-23}
\end{equation}
Hence we can say that defining representation $W(p, q)$ of the translational 
subgroup $T(2)$ of Wigner's little group for massless particles generate gauge
transformations in massless KR theory also. However, as in the case of linearized gravity, on account of the
requirement $f^3 = f^0$ the gauge transformations generated by the
translational group $T(2)$ fails to include the entire set of gauge transformations
in KR theory. The general form of gauge
transformation (\ref{3-13}) in the matrix form is
\begin{equation}
\{\varepsilon^{ab} \} \rightarrow \{\varepsilon'^{ab} \} = \{\varepsilon^{ab}\}  + \omega \left( \begin{array}{cccc}
0 & f^1 & f^2 & f^0 - f^3 \\
-f^1 & 0& 0 &  -f^1\\
-f^2 & 0 & 0 &  - f^2 \\
 f^3  -  f^0 & f^1 &  f^2 & 0
\end{array} \right)
\label{3-24}
\end{equation}
which makes it quite explicit that the transformation (\ref{3-22})
 does not exhaust (\ref{3-24}), but is only a special case  (where $f^0 = f^3$) of it.
Here again, for the case of gauge transformation (\ref{3-22}) generated by the translational
group $W(p,q)$, the arbitrary vector function $f^a(k)$ which correspond
to the gauge freedom of KR theory satisfy the $`$Lorentz condition'
$k_a f^a(k) = 0$ since $k^a = (\omega , 0,0, \omega )^T $ corresponds to a
KR quantum propagating in the $z$-direction.

Similar to the gauge generation  (\ref{32-18}) in linearized gravity, the
transformation (\ref{3-22}) is an attempt to generate the gauge equivalence class
of the completely physical (maximally reduced) polarization tensor (\ref{3-21})
of KR theory using only the two parameters of the translational group $W(p,q)$
while the full gauge freedom of the theory is represented by the four  arbitrary
components of the vector  $f^a$. Here again, the components $f^1$ and $f^2$ of
$f^\mu$ can be expressed in terms of the parameters $p, q$ of the
 translational group $W(p,q)$ and they remain independent of each other as
can be seen from (\ref{3-23}). However, unlike in the case of linearized
gravity, the other two components ($f^0, f^3$)
 are
independent of the parameters (and of the components of  maximally reduced polarization
 tensor) and are left completely undetermined subject only
to the  constraint $f^0 = f^3$.
Thus, in the gauge generation by $W(p,q)$ in KR theory, corresponding to any given pair
($f^1, f^2$) there exists a continuum of allowed choices for $f^0 (= f^3)$
representative of the invariance of gauge transformations (\ref{3-13})
under (\ref{3-13-13}). Therefore, the partial gauge generation by $W(p,q)$
in massless KR theory clearly exhibits the reducibility of its gauge transformations.
The  reducibility of the gauge transformation (\ref{3-10}) is manifested
in the special choice
(\ref{3-23}) which makes the transformation (\ref{3-22}) of
 the maximally reduced polarization tensor $\varepsilon^{ab}$ effected by
$W(p,q)$, a gauge transformation of the KR theory. This may be compared to
the gauge generation (\ref{32-18}) in linearized gravity by $W(p,q)$ where all the components of the arbitrary vector variable $\zeta^\mu$  are expressed in terms of the
parameters $(p,q)$ (see (\ref{32-19})) hence indicating the absence of any reducibility
in the gauge transformation of the theory.

Notice that the transformation (\ref{3-11}) is of same form as the
gauge transformation (\ref{3-4-2}) of Maxwell theory where the generator of
gauge transformations is $W(p,q)$. Hence, one may consider that the $`$gauge
transformation (\ref{3-11}) of gauge transformations' in KR theory
as being generated by a translational subgroup $W(p,q)$ of little group for
massless particles. Therefore, in KR theory which is a 2-form gauge theory,
  two independent elements of the translational group $W(p,q)$ are involved
in generating gauge transformations, one for the underlying 2-form field
$B_{\mu\nu}$ and the other for the field $F_\mu$.
In the gauge generation for massless theories by the translational group
$W(p,q)$, we therefore perceive an appealing
hierarchical structure, namely  in a $n$-form theory, $n$  elements of the
translational group $W(p,q)$ being involved in gauge
generation.

\section{$B\wedge F$ theory}
\label{bf}
$B\wedge F$ theory \cite{kr,cs,lahiri,lahiri1} is obtained by topologically coupling the $B_{ab}$ field of
 Kalb-Ramond theory (\ref{3-6}) with the Maxwell field $A_{a}$ and is
described by the Lagrangian
\begin{equation}
{\cal L} = -\frac{1}{4}F_{ab}F^{ab} + \frac{1}{12}H_{abc}H^{abc} - \frac{m}{6}\epsilon^{abcd} H_{abc} A_{d}.
\label{d1}
\end{equation}
The   equations for motion for
$A_{a}(x)$ and $B_{ab}(x)$ fields are given by
\begin{equation}
\partial_{a}F^{ad} - \frac{m}{6}\epsilon^{ab c d}H_{abc} = 0
\label{d2}
\end{equation}
and
\begin{equation}
\partial_{a}H^{abc} = \frac{1}{2}m \epsilon^{deb c}F_{de}.
 \label{d3}
\end{equation}
The gauge transformations of the fields $A_{a}(x)$ and $B_{ab}(x)$ are respectively
of the same form as  (\ref{3-4-2}) and (\ref{3-10}). Just like the massless 
KR theory discussed in section \ref{tkr}, the gauge transformation of the
$B_{ab}$ is reducible for $B\wedge F$ theory also. Substituting for $A_{a}$ and
$B_{ab}$   respectively from (\ref{3-4-4}) and (\ref{3-12}), we obtain the momentum space version of the above  equations of motion in terms of the polarization vector $\varepsilon_{a}$ and $\varepsilon_{ab}$ as follows;
\begin{eqnarray}
k^2\varepsilon^{d} - k^{d}k_{a}\varepsilon^{a} +
\frac{i}{2}m \epsilon^{abcd}k_{a}\varepsilon_{bc} = 0
\label{d4}\\
k^2\varepsilon^{ab} + \varepsilon^{ca}k_{c}k^{b} -
\varepsilon^{cb}k_{c}k^{a} + imk_{d}\varepsilon_{e}
\epsilon^{de ab} = 0.
\label{d5}
\end{eqnarray}
The momentum space gauge transformations of the fields are obviously of the
form (\ref{3-4-5}) and (\ref{3-13}). 
Considering the massless case $k^2 = 0$, we see using (\ref{d4}) that
\begin{equation}
\epsilon_{dabc}k^{d}k_{e}\varepsilon^{e} =
im(k_a \varepsilon_{bc} - k_{b}\varepsilon_{ac} +
k_c\varepsilon_{ab})
\label{d6}
\end{equation}
Contracting  with $k^{b}$ on either side of (\ref{d6}) yields,
\begin{equation}
k_{a}k^{b}\varepsilon_{bc} + k_{c}k^{b}\varepsilon_{ab} = 0.
\label{d7}
\end{equation}
Using (\ref{d7}) and the masslessness condition ($k^2 = 0$), one can
immediately see using (\ref{d5}) that
\begin{equation}
k_{d}\varepsilon_{a}\epsilon^{dabc}= 0
\label{d8}
\end{equation}
so that any general solution of $\varepsilon_{a}$ can now be written
as,
 \begin{equation}
\varepsilon_{a} = f(k)k_{a}
\label{d9}
\end{equation}
for some function $f(k)$. Therefore, using (\ref{3-4-5}), one can  easily see that
massless excitations, if any, are gauge artefacts. This is in contrast with
the Maxwell and KR models considered earlier, where the massive excitations are
gauge artefacts. Next let us consider the massive case
($k^2 = M^2$). Going to the rest frame with $k^{a} = (M,0,0,0)^T$, one can relate the spatial components of $\varepsilon^{a}$ and
$\varepsilon^{ab}$ by making use of (\ref{d4}) and (\ref{d5}) to
get the following coupled equations
 \begin{equation}
\varepsilon^i = -\frac{im}{2M}\epsilon^{0ijk}\varepsilon_{jk}
\label{d10}
\end{equation}
\begin{equation}
\varepsilon^{ij} =  -\frac{im}{M}\epsilon^{0ijk}\varepsilon_{k}
\label{d11}
\end{equation}
whereas $\varepsilon^0$ and $\varepsilon^{0i}$ remain arbitrary
which can be trivially gauged away by making use of the gauge transformations
(\ref{3-4-5}) and (\ref{3-13}) and the  form  $k^{a} = (M,  0, 0,0)^T$ for
 the four-momentum in the rest frame. On the other hand, the mutual compatibility of the pair of equations (\ref{d10}) and (\ref{d11}) implies
 that  we must have
\begin{equation}
M^2 = m^2.
\label{d12}
\end{equation}
This indicates that the strength $`m$' (taking $m$ to be positive)  of B$\wedge$F
term in (\ref{d1}) can be identified as the mass of the quanta in B$\wedge$F
model. With this, (\ref{d10}) and (\ref{d11}) simplify further and
one can write $\varepsilon^{ab}$ and $\varepsilon^{a}$ in terms of the
three independent parameters,
\begin{equation}
\{\varepsilon^{ab}\} = \left( \begin{array}{cccc}
0 & 0 & 0 & 0 \\
0 & 0 & c & -b \\
0 & -c & 0 & a \\
0 & b & -a & 0
\end{array} \right),
~~~~~\varepsilon^{a} = -i \left( \begin{array}{c}
0 \\
a \\
b \\
c
\end{array} \right).
\label{d14}
\end{equation}
displaying a dual structure.
Another way of understanding the degree of freedom count is to recall that
the B$\wedge$F Lagrangian (\ref{d1}) can be regarded either as a massive
Maxwell (i.e., Proca) theory or a massive KR theory \cite{bb,ms1,ms2}. This can be achieved by
eliminating once the KR field or, alternatively, the vector field from the
coupled set of equations (\ref{d2}, \ref{d3}). Both these theories have
three massive degrees of freedom. It is interesting  to note that the polarization tensor/vector  in  (\ref{d14}), satisfy an
 orthogonality relation,
 \begin{equation}
\varepsilon^{ab} \varepsilon_{b} = 0.
\label{d144}
\end{equation}

By a straightforward calculation involving the explicit forms of the 
polarization tensor $\varepsilon^{ab}$  and vector $\varepsilon^{a}$
 (\ref{d14}),   one can see that $W(p,q)$ fails to be a generator in $B\wedge F$ theory. Therefore, it appears that the translational 
$T(2)$ in the representation $W(p,q)$ , in contrast to the Maxwell and KR examples, is not a generator
of gauge transformation in the B$\wedge$F theory. So, what would be 
 generator of gauge transformations in the topologically massive B$\wedge$F theory? In order to answer this question consider the matrix,
\begin{equation}
D(p,q,r) = \left( \begin{array}{cccc}
1 & p & q & r \\
0 & 1 & 0 & 0 \\
0 & 0 & 1 & 0 \\
0 & 0 & 0 & 1
\end{array} \right)
\label{d15}
\end{equation}
involving three real parameters $p,q,r$. This generates
gauge transformations (\ref{3-4-5}) and (\ref{3-13})  acting on the polarization tensor  and polarization vector 
(\ref{d14}):
\begin{equation}
\varepsilon^{a} \rightarrow \varepsilon'^a = {D^{a}}_{b}(p,q,r)\varepsilon^{b} = \varepsilon^{a} - \frac{i}{m}(p a +
q b + r c)k^{\mu},
\label{d16}
\end{equation}
$$ \{\varepsilon^{ab}\} \rightarrow \{\varepsilon'^{ab}\} = D(p,q,r) \{\varepsilon^{ab}\} D^T(p,q,r) $$
\begin{equation}
 =\{ \varepsilon^{ab}\} +
\left( \begin{array}{cccc}
0 & (r b- q c) & (p c - r a) & (q a - p b) \\
-(r b- q c) & 0 & 0 & 0 \\
-(p c - r a) & 0 & 0 & 0 \\
-(q a - p b) & 0 & 0 & 0
\end{array} \right)
\label{d17}
\end{equation}
as both (\ref{d16}) and (\ref{d17}) can be easily cast into the form  (\ref{3-4-5}) and (\ref{3-13}) with
the proper  choices of $f(k)$ and $f^a(k)$ given by
\begin{equation}
f(k) = \frac{pa + qb +rc}{i m}
\label{d18}
\end{equation}
\begin{equation}
f^1(k) = \frac{rb -qc }{im},~~~
f^2(k) =  \frac{pc - ra}{im}, ~~~
f^3(k) = \frac{q a - p b}{im}.
\label{d19}
\end{equation}
while the component $f^0(k)$  of $f^a(k)$ remain completely undetermined manifesting
the reducible nature of the gauge transformation of the $B^{ab}$ field as explained in the previous section in the context of KR theory. However, unlike KR theory, the gauge transformations generated by
$D(p,q,r)$ (\ref{d15})exhaust the entire set of gauge transformations in $B\wedge F$ theory. This is because  the components of $f^a(k)$ as given in (\ref{d19})
 are independent of one another:
the three spatial components $f^1,f^2,f^3$ are expressed in terms of the three
parameters $(p,q,r)$ while $f^0$ is left completely undetermined. This is unlike
massless KR theory where the gauge transformations generated by $D(p,q,r)$
is restricted by the condition $f^0 = f^3$.

We can now identify the group in which $D(p,q,r)$ belongs.
One can easily show that
\begin{equation}
D(p,q,r) \cdot D(p',q',r') = D(p+p',q+q',r+r')
\label{eq-51}
\end{equation}
and
\begin{equation}
[T_1, T_2] = [T_1, T_3] = [T_2, T_3] = 0
\label{eq-52}
\end{equation}
where
\begin{equation}
T_1 = \frac{\partial D(p, 0, 0)}{\partial p};~~~ T_2= \frac{\partial D(0, q, 0)}{\partial q};~~~ T_3 = \frac{\partial D(0, 0, r)}{\partial r}
\label{mutually}
\end{equation}
can be thought of as three mutually commuting $``$translational" generators.
The group can therefore be identified with $T(3)$ - the invariant subgroup
of E(3)  or ISO(3) \cite{wkt}. Although this gauge generating representation of
$D(p,q,r)$ has been obtained here in a somewhat empirical manner, it can be derived systematically from Wigner's little group in a space-time of one higher
dimension, i.e. in 4+1 dimensions through dimensional descent.  We shall
elaborate on this in a subsequent chapter.

\section{Summary}
In this chapter we have reviewed the gauge generating nature of the
translational subgroup $T(2)$ of Wigner's little group for massless particle.
We have seen that the representation of $T(2)$ inherited from the defining representation
of Wigner's little group  generates gauge
transformations in several massless Abelian gauge theories in 3+1 dimensions.
Our illustrative examples consisted of free Maxwell theory, Kalb-Ramond theory
and linearized gravity. In the case of Maxwell theory $T(2)$ is found
to generate the entire spectrum of gauge transformations while in KR and 
linearized gravity, it generates only a subset of the whole range of available
gauge transformations. In Kalb-Ramond theory theory, the reducibility of the
gauge transformation is clearly manifested in the gauge generation by $T(2)$.
When it comes to the topologically massive $B\wedge F$ theory, 
 one has to go beyond the Wigner's little group and it is the
 translational group $T(3)$ that generates the full set of gauge transformation in this theory. The 
generation  by $T(3)$ also explicitly manifest the reducibility of the
gauge transformation in the antisymmetric tensor field in $B\wedge F$ theory.

\chapter{Translational groups as gauge generators in planar theories} 
We have seen in the previous chapter that the translational subgroup
of Wigner's little group generate gauge transformations in 3+1 dimensional
massless gauge theories. However, the 2+1 dimensional little group
for massless particles has only one parameter and is isomorphic to 
the  translational group $T(1)$ in 1-dimension. Nevertheless, this little group
will generate gauge transformation in 
Maxwell theory in 2+1 dimensions\footnote{The linearized gravity and
massless KR theories do not have any propagating degree of freedom in
2+1 dimensions.}. As we know, there exist topologically massive gauge theories
in 2+1 dimensions, namely the MCS and ECS theories. The question now is
if the translational group $T(1)$  generate gauge
transformations in these theories as well. The present chapter addresses
this question. We first discuss the 2+1 dimensional little group briefly
and then go on to study the relationship between the little group and gauge
transformations in the topologically massive theories.

\section{Wigner's little group in 2+1 dimensions}
\label{pwl}
Following identical techniques as in 3+1 dimensions, one can derive the
2+1 dimensional Wigner's little group that preserves the momentum 3-vector
$k^\mu = (\omega,0,\omega)^T$ for a massless particle \cite{bcs1} as
\begin{equation}
\{  {W^{\mu}}_{\nu}\} (p) = \left( \begin{array}{ccc}

  1 + \frac{p^2}{2} & a p & -\frac{p^2}{2} \\
  p & a & -p \\
  \frac{p^2}{2} & ap & 1- \frac{p^2}{2}
\end{array} \right)
\label{eq-10}
\end{equation} 
where $a= \pm1$.
For $a = +1$, one can easily show that, 
\begin{equation}
 W(p)\cdot W(p') = W(p + p')
\label{eq-11}
\end{equation} 
and therefore the little group represented by (\ref{eq-10}) is isomorphic to
$\cal R $,  the additive group of real numbers.
It is well known that the  Wigner's little group for massless particles (in 2+1 dimensions) is isomorphic to ${\cal R} \times {\cal Z}_{2}$
\cite{binegar}. The $ {\cal Z}_{2}$ factor is     
required to take into account of the fact that the value of $a$ is restricted to $\pm1$.

The generator G in the representation $W(p)$ in (\ref{eq-10}) is clearly given (with $a = +1$)  by,
\begin{equation}
 G = \frac{\partial W}{\partial p}\mid_{p = 0} =
\left( \begin{array}{ccc}
       0 & 1 & 0 \\
       1 & 0 & -1 \\
       0 & 1 & 0
\end{array} \right)
\label{eq-12}
\end{equation} 
satisfying,
\begin{equation}
 G^2 = \left( \begin{array}{ccc}
          1 & 0 & -1 \\
          0 & 0 & 0 \\
          1 & 0 & -1 
\end{array} \right); ~~~  G^3 = 0
\label{eq-13}
\end{equation}
so that $W(p)$ can be re-expressed as, 
\begin{equation}
 W(p) = e^{p G} = 1 + p G + \frac{1}{2}p^2 G^2 
\label{eq-14}
\end{equation}

One can thus construct various representations of Wigner's little group
for massless particles in 2+1 dimensions just as in the usual 3+1 dimensions.
The important point is that although the various representations are, by
definition, isomorphic to each other, not all of them belong to the Lorentz
group. 
In the following section we
construct  another representation of the little group in 2+1 dimensions,
not inherited from $SO(1,2)$, which will be shown to act as a gauge generator
in topologically massive MCS theory.

The little group for the massive particle,
which in this case can be  trivially seen to be $O(2)$, however, does not have any
role as a generator of gauge transformation. This does not mean however that
they are completely unrelated. In fact,
one can write $W(p)$ for $|p| < 1$, as a product of three matrices;
\begin{equation}
 W(p) = B^{-1}_y(p) R(p) B_x(p) 
\label{eq-1.7}
\end{equation}
where
\begin{equation}
 B^{-1}_y(p) =  \left( \begin{array}{ccc}
 \frac{2 - p^2}{2\gamma} & 0 & -\frac{p^2}{2\gamma} \\
 0 & 1 & 0 \\
 -\frac{p^2}{2\gamma} & 0 & \frac{2 - p^2}{2\gamma}
\end{array} \right);  R(p) = \left( \begin{array}{ccc}
 1 & 0 & 0 \\
 0 & \gamma & -p \\
 0 & p & \gamma 
\end{array} \right);
 B_x(p) = \left( \begin{array}{ccc}
\frac{1}{\gamma} & \frac{p}{\gamma} & 0 \\
 \frac{p}{\gamma} & \frac{1}{\gamma} & 0 \\
0 & 0 & 1
\end{array} \right)
\label{eq-1.8}
\end{equation} 
with $ \gamma = \sqrt{1 - p^2}$. These matrices are themselves the elements
of the Lorentz group $SO(1,2)$; $ B_x$ represents a boost along the $x$-direction,
$R$ represents a  spatial rotation in the $x-y$ plane and $B^{-1}_y$
represents a boost along the negative $y$-direction. Appropriate 
transformations in this order can preserve the energy-momentum 3-vector
of a massless particle moving in the $y$-direction. Here $R$ clearly corresponds to the
little group of a massive particle.
Thus (\ref{eq-1.7})  relates the elements of the connected parts of identity element
of  the little group of massless particles with massive ones as long as
$|p| < 1 $. But this does not provide the natural homomorphism existing
between ${\cal R} $ (the additive group of real numbers) with $SO(2)$.

\section{Little group as gauge generator in Maxwell-Chern-Simons theory}
It is easy to see that the little group $W(p)$ in (\ref{eq-10}) with $a = +1$,
generate gauge transformations in 2+1 dimensional Maxwell theory. For that 
consider a photon of energy $\omega$ moving in the $y$-direction and polarized
in the $x$-direction  so that the potential $3$-vector takes the form 
\begin{equation}
A^\mu(x) = \xi^\mu_x \exp (-ik. x) = \xi^\mu_x \exp(-i\omega( t- y)), 
\label{eq-11-1}
\end{equation}
where \begin{equation} \xi^\mu_x  = \left( \begin{array}{c} 0\\1\\0 \end{array}
\right)
\label{eq-111}\end{equation}
is the polarization vector and the subscript denotes that this vector is in the
$x$-direction. Under the action of $W(p)$ (\ref{eq-10}), $\xi^\mu_x $ undergoes
the transformation 
\begin{equation}\xi^\mu_x \rightarrow \xi'^\mu_x \equiv {W^\mu}_\nu \xi^\nu_x
= \xi^\mu_x +\frac{p}{\omega}k^\mu.\label{eq-112}\end{equation}
This can be identified as the gauge transformation as the corresponding gauge
field undergoes the transformation 
\begin{equation}
A^\mu(x)  \rightarrow  A'^\mu(x) = A^\mu(x) + \partial^\mu \left( \frac{ip}{\omega} e^{-i\omega(t-y)}\right).
\label{eq-11-2}\end{equation}

In contrast to Maxwell theory, MCS excitations are massive as we have seen 
in section \ref{mcs1} and  the  rest frame polarization vector $\xi^\mu ({\bf 0})$ takes the simple
form (\ref{c20}) in the rest frame. That is, 
\begin{equation}
\xi^\mu ({\bf 0}) = \frac{1}{\sqrt{2}}\left( \begin{array}{c}
                                0 \\
                               1 \\
                               -i\frac{\vartheta}{|\vartheta|}
\end{array} \right). 
\label{eq-1.3}
\end{equation}
where $\vartheta$ is the CS parameter.
Note that it has complex entries having both $x$ and
$y$ components unlike the Maxwell photon polarization $\xi^{\mu}_x $.
In fact in their Coulomb gauge analysis Devecchi et. al.\cite{girotti} have pointed out 
that the spin $(\frac{\vartheta}{|\vartheta|})$ of the MCS quanta stems from this
particular complex structure  of the polarization vector. 

We shall now investigate whether this same little group can generate similar
gauge transformation on the MCS polarization vector (\ref{eq-1.3}). To that
end, let us apply $W(p)$ on $\xi^\mu ({\bf 0})$
(\ref{eq-1.3}). Without loss of generality, henceforth we shall consider $\vartheta < 0$ case only. We find that it undergoes the following transformation,
\begin{equation}
 \xi^\mu ({\bf 0}) \rightarrow \xi^{\prime \mu} ({\bf 0}) \equiv {W^{\mu}}_{\nu}(p)\xi^\nu ({\bf 0}) =   \frac{1}{\sqrt{2}}\left( \begin{array}{c}
   p - \frac{i}{2}p^2 \\
    1 - ip \\
   p+ i(1- \frac{p^2}{2})
\end{array} \right)
\label{eq-1.4}
\end{equation}
Clearly this cannot be cast in the form of (\ref{eq-112}).
One cannot therefore
interpret this transformation as a gauge transformation. However, taking  
advantage of the fact that this little group involves a single parameter
only, we can easily construct a (non-unique) representation which does the
required job. This is given by,
\begin{equation}
 D(p) = \left( \begin{array}{ccc}
       1 & p & -ip \\
       0 & 1 & 0 \\
       0 & 0 & 1 
\end{array} \right); ~~~p \in {\cal R}
\label{eq-1.5} 
\end{equation}
so that in place of (\ref{eq-1.4}) one has the desirable form in the sense that it can now be 
put in the form of (\ref{eq-112});
\begin{equation}
 \xi^\mu ({\bf 0}) \rightarrow \xi^{\prime \mu}({\bf 0}) \equiv
 {{D}^{\mu}}_{\nu}  \xi^\nu ({\bf 0}) = \xi^\mu ({\bf 0}) + \frac{\sqrt{2}p}{|\vartheta|} k^\mu  
\label{eq-1.6}
\end{equation}
where $k^\mu = (|\vartheta|, 0, 0)^T$ is the energy-momentum
3-vector of a MCS particle in the rest frame.
This shows that $D(p)$ acts a generator of gauge transformation in
the MCS theory. 

By denoting the rest frame polarization vectors of a doublet ${\cal L}_\pm$ 
((\ref{nc5})and  ( \ref{ncc5})) of MCS theories with helicities $\pm 1$ by $\xi^\mu_\pm$ and the corresponding
elements of  2+1 dimensional little group that generate their gauge 
transformation by $D_\pm(p_\pm)$ respectively, we may write
\begin{equation}  
\xi^\mu_\pm \rightarrow D_\pm(p_\pm) \xi^\mu_\pm  = \xi^\mu_\pm + \frac{\sqrt{2}p_\pm}{|\vartheta|} k^\mu
\label{eeq-1.6}
\end{equation}
where
\begin{equation}
D_\pm(p_\pm) = \left( \begin{array}{ccc}
       1 & p_\pm & \pm ip_\pm \\
       0 & 1 & 0 \\
       0 & 0 & 1
\end{array} \right), ~~~ ~~~ \xi^\mu_\pm  = 
                                   \frac{1}{\sqrt{2}}\left( \begin{array}{c}
0\\
1\\
\mp i\end{array} \right).
\label{eq-1.6.1}
\end{equation}
with $p_\pm$ representing the parameters of the little group elements $D_\pm$.
In the next chapter we show how we can derive these representations of Wigner's
little group in 2+1 dimensions from the gauge generating representation $W(p,q)$ of $T(2)$ for Maxwell theory in 3+1 dimensions by a method called dimensional
descent. 

Now, certain comments on some subtle points regarding  representation $D(p)$ (\ref{eq-1.5})
 are in order. Although $D(p)$ is not an element of the Lorentz
group, it is perfectly admissible to regard it as a 
representation\footnote{Note that  this representation is different from the
 defining representation, the latter can only
be obtained as a subgroup of the Lorentz group $SO(1,2)$.} of
the little group for a massless particle in 2+1 dimensions. This is because  it satisfies  $D(p) \cdot D(p') =
 D(p +p') $, so that   there exists a natural isomorphism
between $W(p)$  (\ref{eq-10}) and $D(p)$ (\ref{eq-1.5}). This is analogous to 3+1 dimensional case of 
 Wigner's little group (\ref{3-2}) for  massless particles. The
  Lie algebra of this  little group (which acts as gauge generators \cite{hk,hk1,hk2,bc1,bcs1,sc})  is isomorphic to the algebra of the Euclidean group $E(2)$ as explained in appendix A. However, notice that the algebra of the little
group, being a combination of boost and rotation generators (\ref{aaab22}), is a subalgebra of the homogeneous Lorentz algebra whereas  the defining representation of $E(2)$ algebra (comprising of two
translational generators 
  and a rotational generator in a plane), is
a subalgebra of Poincare algebra but not of the homogeneous Lorentz algebra.

Coming back to the issue of similarities and dissimilarities between  the
polarization vectors of pure Maxwell theory and that of MCS theory, note that a
photon state is entirely characterized by (\ref{eq-111}) where both the $``$spatial" transversality condition, ${\bf {k. \vec{\xi}_x}}=0$ and the temporal gauge condition $ \xi_x^0 = 0$ are trivially satisfied. Therefore the gauge field
configuration (\ref{eq-11-1}) corresponds to the radiation gauge.
Clearly the same gauge condition will no longer be valid under a Lorentz boost.
However, we shall show now  that the radiation gauge condition can 
still be satisfied, provided the gauge field undergoes an appropriate
gauge transformation preceding the Lorentz boost. Considering the Maxwell case first, the gauge transformed field configuration $A^{\prime \mu}(x)$ corresponding to a photon polarized along the $x$-direction and propagating along the $y$-direction 
can be written as,
\begin{equation}
 A^{\prime \mu}(x) = A^{\mu}(x) + \partial^\mu \tilde{p}(x) = A^{\mu}(x) +p k^\mu e^{-i k \cdot x}
\label{eq-1.9}
\end{equation}
where the  scalar function $\tilde{p}(x)$is  taken  to be of the form $ip(k)$ . A Lorentz
boost of velocity $v = \tanh \phi $ for example, in the $x$-direction yields,
\begin{equation}
 \tilde{A}^{\prime \mu} = \left( \begin{array}{ccc}
 \cosh \phi & \sinh \phi & 0 \\
 \sinh \phi & \cosh \phi & 0 \\
 0 & 0 & 1 
\end{array} \right) \left( \begin{array}{c} p \\ 1 \\ p
\end{array} \right) e^{-ik^{\prime} \cdot x^{\prime}} 
 = \left( \begin{array}{c} p \cosh \phi + \sinh \phi \\ p \sinh \phi + \cosh \phi \\
p \end{array} \right) e^{-ik^{\prime} \cdot x^{\prime}}
\label{eq-1.10}
\end{equation}
where $k^{\prime \mu} $ is the appropriate energy-momentum 3-vector in the
new boosted co-ordinate frame $x^{\prime \mu}$ and is given by, 
\begin{equation}
 k^{\prime \mu} = \omega \left( \begin{array}{c}
  \cosh \phi \\
  \sinh \phi \\
    1
\end{array} \right)
\label{eq-1.11}
\end{equation} 
Preservation of spatial transversality condition implies that we must have,
\begin{equation}
 {\tilde{\bf A}}^{\prime}(x^{\prime}) \cdot {\bf k}^{\prime} = \omega
\left( \begin{array}{c}
\cosh \phi + p \sinh \phi \\
p
\end{array} \right)^{T}
\left( \begin{array}{c}
\sinh \phi \\
1 
\end{array} \right) = 0
\label{eq-1.12}
\end{equation}
Solving for $p$, one gets, 
\begin{equation}
 p = -\tanh \phi = -v
\label{eq-1.13a}
\end{equation} 
This solution, when substituted back in (\ref{eq-1.10}) yields,
\begin{equation}
 \tilde{A}^{\prime 0} (x^{\prime}) = 0
\label{eq-1.13b}
\end{equation}
which is nothing but the temporal gauge condition. Thus with an appropriate
gauge transformation preceding a Lorentz boost, the radiation gauge condition
can be satisfied. But, as we shall see now,  the same is not true for 
MCS Theory.
Upon a gauge transformation,
the polarization vector (\ref{eq-1.3}), in the rest frame, becomes
\begin{equation}
 \xi^\mu (x) \rightarrow \tilde{\xi}^\mu (x) = D
(p) \xi^\mu (x) = 
 \frac{1}{\sqrt{2}} \left( \begin{array}{ccc}
 1 & p & -ip \\
 0 & 1 & 0 \\
 0 & 0 & 1 
\end{array} \right) \left( \begin{array}{c}
0 \\
1 \\
i
\end{array} \right)
 = \frac{1}{\sqrt{2}}\left( \begin{array}{c}
2 p \\
1 \\
i
\end {array} \right) 
\label{eq-1.14}
\end{equation} 
Then a Lorentz boost like (\ref{eq-1.10}) along $x$-axis, for example transforms this to,
\begin{equation}
 \tilde{\xi}^\mu \rightarrow {\tilde{\xi}}^{\prime \mu} = \frac{1}{\sqrt{2}}
\left( \begin{array}{c}
\sinh \phi + 2 p \cosh \phi \\
\cosh \phi + 2 p \sinh \phi \\
i
\end{array} \right)
\label{eq-1.15}
\end{equation}
Simultaneously, the $k^\mu = (1, 0, 0)^T$,
associated to the rest frame, transforms  to, 
\begin{equation}
 k^\mu \rightarrow k^{\prime \mu} = \left( \begin{array}{c}
\cosh \phi \\
\sinh \phi \\
0
\end{array} \right) 
\label{eq-1.16}
\end{equation}
Demanding that the spatial transverality condition is
satisfied in an arbitrary boosted frame (i.e., $\phi \neq 0$) and using (\ref{eq-1.15}) and 
(\ref{eq-1.16}) we get,
\begin{equation}
 \left( \begin{array}{c}
\sinh \phi \\
0
\end{array} \right)^T \left( \begin{array}{c}
\cosh \phi + 2 p \sinh \phi \\
i
\end{array} \right) = 0,
\label{eq-1.17}
\end{equation}
which when solved for the gauge transformation parameter $p$  in terms of the 
boost parameter $\phi$, yields,
\begin{equation}
 p = - \frac{1}{2 \tanh \phi} = -\frac{1}{2v}.
\label{eq-1.18} 
\end{equation}
So just like in the Maxwell case the spatial transversality condition can be 
maintained in any boosted frame, provided the boost is preceded by a suitable gauge transformation. However, in contrast to Maxwell case (\ref{eq-1.13b}),  the temporal gauge
condition  ($A^0= 0$) is not satisfied simultaneously since for 
$p$ satisfying (\ref{eq-1.18})
\begin{equation}
 {\tilde{\xi}}^{\prime 0} = -\frac{1}{\sqrt{2}\sinh \phi}.
\label{eq-1.19}  
\end{equation}
Nevertheless, $ \tilde{\xi}^{\prime 0}$
(\ref{eq-1.19}) can be made to vanish in the infinite momentum frame
(in the limit $\phi \rightarrow \infty$), i.e., when $p 
\rightarrow -\frac{1}{2} $ and $\tanh \phi \rightarrow 1$.

For a boost along the $x$-direction, one can write
\begin{equation}
 \tanh \phi = \frac{k^1}{k^0}.
\label{eq-1.20}
\end{equation}
Using (\ref{eq-1.18}),  (\ref{eq-1.20}) and the mass-shell condition $k^2 = \vartheta^2$, one can simplify (\ref{eq-1.15}) to get the polarization vector as,
\begin{equation}
 \tilde{\xi}^{\prime \mu} = \frac{1}{\sqrt{2}}\left( \begin{array}{c}
\frac{\vartheta}{k^1} \\
0 \\
i
\end{array} \right)
\label{eq-1.21}
\end{equation}
At this stage,  we can see clearly that although the spatial transversality 
(${\bf k}. \vec{\xi} =0$)
holds trivially in the rest frame,  the temporal gauge condition is not
well defined (since $\tilde{\xi}^{\prime 0}\rightarrow \infty $ as $k^1 \rightarrow 0$). This is expected from the simple consideration that a Lorentz
transformation (to the rest frame) alone should not also lead to a complete
gauge fixing. Note that, unlike MCS theory, a rest frame is not available in Maxwell theory.

\section{Linearized Einstein-Chern-Simons theory}
As explained in section \ref{tlg}, pure gravity in 2+1 dimensions is a null
theory in the sense that it does not have a propagating degree of freedom. However, 2+1 dimensional gravity coupled to a non-Abelian Chern-Simons
 topological term, with gauge group being the Lorentz group itself,
possesses a single propagating massive degree of freedom \cite{djt}. Just like the MCS theory,
the gauge invariance coexists with mass in the linearized version of this theory too where the gauge group reduces to Abelian group $T(1)$. In this section we
study the role of  translational group in generating gauge transformations
in the linearized version of gravity coupled to Chern-Simons term 2+1 dimensions \cite{sc}.

The full action of the topologically massive gravity in 2+1 dimensions is 
\begin{equation}
I^{ECS} = I^E + I^{CS}
\label{57+1}
\end{equation} 
where the 2+1 dimensional Einstein action here is
\begin{equation} 
I^E = \int d^3 x \sqrt{g} R
\label{57+2}
\end{equation}
and the Chern-Simons action $I^{CS}$ is given by 
\begin{equation}
I^{CS} = -\frac{1}{4\mu} \int d^3 x \epsilon^{\mu\nu\lambda}[R_{\mu\nu\alpha\beta } {\omega_\lambda}^{\alpha\beta} +
\frac{2}{3}{\omega_{\mu\beta}}^\gamma {\omega_{\nu \gamma}}^\alpha{\omega_{\lambda \alpha}}^\beta].
\label{57+3}
\end{equation}
The   $\omega_{\mu \alpha \beta}$ are the components of the spin connection one-form and 
are related to the curvature two-form by the
 second Cartan's equation of structure (${R^\alpha}_\beta = d{\omega^\alpha}_\beta  + 
{\omega^\alpha}_\gamma \wedge {\omega^\gamma}_\beta $). Note that the sign in front of the Einstein action is now opposite to
the conventional one (\ref{32-1}) and is required to make the full theory free of ghosts \cite{djt}. 
The linearization, $g_{\mu\nu} = \eta_{\mu\nu} + h_{\mu\nu}$, of the ECS theory (\ref{57+1}) results in the
Abelian theory \cite{djt,dargam} given by  
\begin{equation}
I^{ECS}_L =\int d^3 x {\cal L}_L^{ECS} 
\label{57+4}
\end{equation}
where
\begin{equation} 
{\cal L}_L^{ECS} = {\cal L}_L^E + {\cal L}_L^{CS}.
\label{58}
\end{equation}
Here
\begin{equation}
{\cal L}_L^E = -\frac{1}{2}h_{\mu\nu} \left[ R^{\mu\nu}_L - \frac{1}{2} \eta^{\mu\nu} R_L\right]
\label{59}
\end{equation}
now is the Lagrangian for linearized version of pure gravity in 2+1 dimensions
and is the same as (\ref{32-3}) except that it has the opposite sign and the indices, in this case, vary over $0, 1, 2$. Similarly,
\begin{equation}
{\cal L}_L^{CS} = -\frac{1}{2\mu} \epsilon_{\alpha\beta\gamma} 
\left[ R^{\beta\delta}_L - \frac{1}{2} \eta^{\beta\delta} R_L\right]
\partial^\alpha h^\gamma_\delta
\label{60}
\end{equation}
is the linearized Chern-Simons term with the  Chern-Simons parameter  $\mu$.
Under the gauge transformation
$h_{\mu\nu} \rightarrow h'_{\mu\nu} = h_{\mu\nu} + \partial_\mu \tilde{\zeta}_\nu (x) + \partial_\nu \tilde{\zeta}_\mu (x)$, the
Chern-Simons part ${\cal L}_L^{CS}$ changes by a total derivative:
\begin{equation}
\delta {\cal L}^{CS}_L = \frac{1}{\mu}\epsilon_{\mu\nu\lambda} \partial_\delta \left(
R^{\nu\delta}_L \partial^\mu \tilde{\zeta}^\lambda\right)
\label{61}
\end{equation}
The equation of motion corresponding to ${\cal L}_L^{ECS}$ is given by \cite{djt,grignani}
$$ \Box h^{\mu\nu} -\partial^\mu \partial_\gamma h^{\gamma\nu} - \partial^\nu \partial_\gamma h^{\gamma\mu}
+ \partial^\mu\partial^\nu h - \eta^{\mu\nu} (\Box h - \partial_\gamma\partial_\delta h^{\gamma\delta})
$$
\begin{equation}
-\frac{1}{2\mu} \epsilon^{\mu\gamma\delta} \partial_\gamma (\Box h^\nu_\delta
- \partial_\tau \partial^\mu h^\tau_\delta) - 
\frac{1}{2\mu} \epsilon^{\nu\gamma\delta} \partial_\gamma (\Box h^\mu_\delta
- \partial_\tau \partial^\mu h^\tau_\delta) 
= 0.
\label{62}
\end{equation}
With the ansatz  $h^{\mu\nu} = \chi^{\mu\nu} (k) e^{ik.x} $,  
 the above equation of motion can be written
in terms of the symmetric polarization tensor $\chi_{\mu\nu}(k)$ and the 3-momentum
$k^\mu$ as follows
$$
-k^2\chi^{\mu\nu} + k^\mu k_\gamma \chi^{\gamma\nu} +
k^\nu k_\gamma \chi^{\gamma\mu } -  k^\mu k^\nu \chi
-\eta^{\mu\nu}(-k^2 \chi +k_\gamma k_\delta \chi^{\gamma\delta})
$$
\begin{equation}
 -\frac{i}{2\mu}\left[ \epsilon^{\mu\gamma\delta} k_\gamma (-k^2 \chi^\nu_\delta +
k_\tau k^\nu \chi^\tau_\delta)~ +~ \epsilon^{\nu\gamma\delta}
k_\gamma (-k^2   \chi^\mu_\delta
+ k_\tau k^\mu \chi^\tau_\delta)\right] = 0.
\label{n63}
\end{equation}
Analogous to (\ref{32-9}), the expression for the gauge transformation for
ECS theory in terms of its polarization tensors $\chi^{\mu\nu}(k)$ is given by
\begin{equation} 
\chi^{\mu\nu}(k) \rightarrow \chi'^{\mu\nu}(k) = \chi^{\mu\nu}(k) + k^\mu \zeta^\nu(k) + k^\nu \zeta^\mu(k)
\label{new63}
\end{equation}
where $\zeta_\mu(k)$ are small arbitrary functions of $k$.  
Depending on whether the excitations are  massless or massive, we have  two
options for $k^2$:\\
\begin{center} (i) $k^2 = 0$ or (ii) $k^2 \neq 0$. \\ \end{center}
{\it case} (i): $k^2 = 0$ \\
Contracting the (\ref{n63}) with $\eta_{\mu\nu}$ gives
\begin{equation}
k_\mu k_\nu \chi^{\mu\nu} = 0.
\label{64}
\end{equation}
A general solution to this equation consistent with the equation of motion (\ref{n63}) is 
\begin{equation}
\chi^{\mu\nu} = k^\mu f^\nu(k) + k^\nu f^\mu(k)
\label{65}
\end{equation}
where $f^\mu(k)$ are arbitrary functions of $k$. However, with $f^\mu = -\zeta^\mu$ we can
$`$gauge away' these solutions. Therefore, massless excitations of ECS theory
are pure gauge artefacts.
We now proceed to the other option:\\
{\it case} (ii) $k^2 \neq 0$ \\
Let $k^2 = m^2$. On contraction with $\eta_{\mu\nu}$ (\ref{n63}) gives
\begin{equation}
k_\mu k_\nu \chi^{\mu\nu} = m^2\chi
\label{66}
\end{equation}
where $\chi = {\chi^\mu}_\mu$.
With $k^{\alpha} = (m, 0, 0)^T$, this yields
\begin{equation}
\chi_{11} + \chi_{22} = 0.
\label{67}
\end{equation}
By considering the spatial part of (\ref{n63} ) one can show that the mass $m$ of the excitations can be identified with the
Chern-Simons parameter $\mu$ as follows. The spatial part of (\ref{n63} )
is
$$
-m^2\chi^{ij} + k^i k_\mu \chi^{\mu j} +
k^j k_\mu \chi^{\mu i } -  k^i k^j \chi
-\eta^{ij}(-k^2 \chi +k_\mu k_\nu \chi^{\mu\nu})
$$
\begin{equation}
 -\frac{i}{2\mu}\left[ \epsilon^{i \gamma\delta} k_\gamma (-k^2 \chi^j_\delta +
k_\tau k^j \chi^\tau_\delta)~ +~ \epsilon^{j \gamma\delta}
k_\gamma (-k^2   \chi^i_\delta
+ k_\tau k^i \chi^\tau_\delta)\right] = 0.
\label{68}
\end{equation}
In this equation $i, j$ takes values $1$ and  $2$.
On passing to the rest frame the above equation simplifies to
\begin{equation}
-\chi^{ij} +\eta^{ij} {\chi^k}_k  -\frac{im}{2\mu} \left[
\epsilon^{ik}\chi^j_k + \epsilon^{jk}\chi^i_k\right] = 0
\label{69}
\end{equation}
from which we obtain (for  $i=j=1$ and $i=j=2$  respectively)
\begin{equation}
\chi^{11} = +\frac{im}{\mu}\chi^{12}~; ~~~~~\chi^{22}=
-\frac{im}{\mu} \chi^{12}. 
\label{70}
\end{equation}
With $i=1$ and $j=2$ we have
\begin{equation}
\chi^{12} = +\frac{im}{\mu}\chi^{22}.
\label{71}
\end{equation}
This relation together with (\ref{70}) implies that $m^2 = \mu^2$.
The remaining components can be made to vanish by a suitable gauge choice.
 Finally, for the Chern-Simons parameter $\mu > 0 $ , the polarization tensor 
{\bf $\chi_+$} of the gravity coupled to Chern-Simons
theory in 2+1 dimensions  in the rest frame can be written as 
\begin{equation}
{\bf \chi_+ } \equiv \{\chi_+^{\mu\nu}\}= \left(
\begin{array}{ccc}
0 & 0 & 0   \\
0 & 1 & -i \\
0 & -i & -1 
\end{array}
\right)\tau
\label{72}
\end{equation}
where $\tau$ is an arbitrary real parameter. Notice that the ECS theory
has only a single degree 
of freedom
corresponding to the  parameter $\tau$. Similarly, the rest frame
 polarization tensor for
an ECS theory having the Chern-Simons parameter $\mu < 0$ is
\begin{equation}
{\bf \chi_-} \equiv \{\chi_-^{\mu\nu}\}= \left(
\begin{array}{ccc}
0 & 0 & 0   \\
0 & 1 & i \\
0 & i & -1
\end{array}
\right)\tau.
\label{73}
\end{equation}  

It is important to note that these rest frame polarization tensors {\bf $\chi_\pm$}
of
ECS theories (with $ \tau = \frac{1}{2}$) can be obtained as
direct products of the rest frame polarization vectors $\xi_\pm^\mu$ (\ref{eq-1.6.1})
of MCS theories. i.e.,
\begin{equation}
\chi^{\mu\nu}_\pm  = \xi^{\mu}_\pm \xi^{\nu}_\pm. 
\label{74}
\end{equation}
This suggests that we adopt   orthonormality conditions for {\bf $\chi_\pm$} which  are similar to the ones
(\ref{c19}) used for $\xi^\mu$. Hence we require
\begin{equation}
tr\left( (\chi_+)^{\dagger}(\chi_-)\right) = 0~; ~~~~~    tr\left( (\chi_\pm)^{\dagger}(\chi_\pm)\right) = 1.
\label{75}
\end{equation}
Therefore,  we have the following
maximally reduced form for the polarization 
tensors of a pair of  ECS theories with opposite helicities
\begin{equation} 
\chi_\pm = \frac{1}{2}\left(
\begin{array}{ccc}
0 & 0 & 0   \\
0 & 1 & \mp i \\
0 & \mp i & -1
\end{array}
\right).
\label{76}
\end{equation}
Note that these polarization matrices of ECS theories are traceless and 
singular.
We are now equipped to study the role
played by the
translational group in
generating the gauge transformation in this theory. The representations of
$T(1)$ that generates gauge transformation in pair of MCS theories with
opposite helicities is given by $D_\pm (p)$ (\ref{eq-1.6.1}). On account of the
relation (\ref{74}) it is expected that
the same representations will generate gauge transformations in ECS theories
also. Indeed one can easily see that $D_\pm(p_\pm) $ are the gauge generators in
ECS theories:
\begin{equation}
\chi_\pm \rightarrow \chi_\pm '=  D_\pm (p_\pm) \chi_\pm  D^T_\pm (p_\pm ) = \chi_\pm
+  \left(\begin{array}{ccc}
2p_\pm^2 & p_\pm & \mp ip_\pm  \\
 p_\pm & 0 & 0 \\
 \mp ip_\pm & 0 & 0
\end{array}
\right).
\label{79+0}
\end{equation}
This transformation  can be cast  in the form of the  gauge transformation 
(\ref{new63}) with the following choice of $\zeta$'s;
\begin{equation} 
\zeta_0 = \frac{p_\pm^2}{|\mu |}, ~~~ \zeta_1 = \frac{p_\pm}{|\mu |},~~~ \zeta_2 = \frac{\mp ip_\pm}{ |\mu |}.\label{79+1}
\end{equation}  
One can obtain the moving frame expression for  polarization tensors
{\bf $\chi_\pm(k)$} from the above 
rest frame results by applying appropriate Lorentz boost  as follows:
$${\bf \chi_\pm}(k) = \Lambda^T(k) {\bf \chi_\pm}(0)  \Lambda(k)$$
\begin{equation}
 = \frac{1}{2\mu^2}\left(\begin{array}{ccc}
k^2_0 - \mu^2 & k^0 k^1 \mp i|\mu| k^2 &  k^0 k^2 \pm i|\mu| k^1 \\
k^0 k^1 \mp i|\mu| k^2 & \frac{(k^0 k^1 \mp i|\mu| k^2)^2}{k^2 _0 - \mu^2} &
\frac{(k^0 k^1 \mp i|\mu| k^2)( k^0 k^2 \pm i|\mu| k^1)}{k^2 _0 - \mu^2}  \\
k^0 k^2 \pm i|\mu| k^1 & \frac{(k^0 k^1 \mp i|\mu| k^2)( k^0 k^2 \pm i|\mu| k^1)}{k^2 _0 - \mu^2} & \frac{(k^0 k^2 \pm i|\mu| k^1)^2}{k^2 _0 - \mu^2}
\end{array}
\right) e^{\pm 2i \phi(k)}
\label{77}
\end{equation} 
where $ \phi(k) = \arctan (\frac{k^2}{k^1})$ and  the momentum space boost matrix \cite{jackson}
\begin{equation}
\Lambda (k)~~ =~~ \left( \begin{array}{ccc}
   {\gamma} & {\gamma}{\beta}^1 & {\gamma}{\beta}^2      \\
        {\gamma}{\beta}^1 & 1 + \frac{({\gamma} -1)({\beta}^1)^2}{(\vec{\beta})^2} & \frac{({\gamma} -1){\beta}^1{\beta}^2}{(\vec{\beta})^2} \\
         {\gamma}{\beta}^2 & \frac{({\gamma} -1){\beta}^1{\beta}^2}{(\vec{\beta})^2} & 1 + \frac{({\gamma} -1)({\beta}^2)^2}{(\vec{\beta})^2}
    \end{array} \right)
\label{78}
\end{equation}
with $\vec{\beta} = \frac{\bf k}{k^0}$ and $\gamma = \frac{k^0}{|\mu|}$ as given in (\ref{c31}). Results identical to (\ref{74}) and (\ref{77}) were obtained in other contexts
 \cite{grignani} by different
methods. Here, we have shown how these results can be obtained in a simpler and
straightforward manner just by considering the momentum space expression of the
equation of motion in the rest frame using the plane wave method with
 a subsequent boost transformation.
Obviously, the relation (\ref{74}) holds true in the moving frame also.

\section{Summary}
Though, the defining representation of Wigner's little group for massless
particles in 2+1 dimension generate gauge transformations for Maxwell theory,
the same representation does not generate gauge transformations in
the topologically massive Maxwell-Chern-Simons and Einstein-Chern-Simons
theories. However, using the fact that this little group has only one parameter,  we have obtained a different representation of the
little group that generate gauge transformations in these theories. 
The similarities and dissimilarities between the Maxwell and Maxwell-Chern-Simons theories in the context of gauge fixing (spatial transversality and temporal gauge) are also analyzed. Detailed analysis of the polarization tensor
of Einstein-Chern-Simons theory is carried out and the polarization tensor 
is found to be a
tensor product of a pair of polarization vectors of Maxwell-Chern-Simons
theory with the same helicity. This is quite natural since Maxwell-Chern-Simons
(spin $\pm 1$) and Einstein-Chern-Simons (spin $\pm 2$) theories correspond to different spin representations of the 2+1 dimensional Poincare algebra\footnote{Note that spin is a scalar in 2+1 dimensions.} \cite{dunne} and these spin 
representations are related by a tensor product \cite{djt,grignani}.

\chapter{Dimensional descent}
As introduced in \cite{bc2}, dimensional descent is a method by which one can obtain the
 energy-momentum
vector, polarization vector/tensor and the gauge generating representation of the
 translational
subgroup of Wigner's little group etc,  in a  topologically massive gauge theory
living in a
certain space-time
dimension from similar results for massless gauge theories inhabiting a space-time of one higher dimension. In this sense,
dimensional descent is a unification scheme for the results presented in the previous chapters
regarding various gauge theories. In the present chapter, we discuss
dimensional descent as applied to topologically massive gauge theories 
\cite{sc,bc2}.
 Dimensional descent is also applicable to  massive gauge theories \cite{ts} obtained by
the previously mentioned embedding procedure of Batalin, Fradkin and Tyutin and
we provide a short account of dimensional descent for such theories in the 
following chapter. It may be noted  in this connection that the
relationship between massless gauge theories in a given space-time dimension
and lower dimensional gauge theories having massive excitations  can be studied
 using other methods also (for example, see \cite{ms3,ms4}).

\section{Review of dimensional descent: 4+1 to 3+1 dimensions}
We begin our discussion of dimensional descent by noting that, the translational group $T(3)$ which generates gauge transformation
in 3+1 dimensional  $B\wedge F$ theory  is an invariant
 subgroup of $E(3)$.  Now, just as $E(2)$ is the generator of gauge transformation in 4-dimensional
Maxwell and massless KR theories, $E(3)$ generates gauge transformation in the 5-dimensional versions of these massless theories. This
indicates that the generators of gauge transformations in
$B\wedge F$ theory and  5-dimensional massless gauge theories  
are related. This relationship is explicitly demonstrated through the method
of dimensional descent \cite{bc2} which we will describe below. 

An element of Wigner's little group in 5 dimensions can be written as
\begin{equation}
W_5(p,q,r; \psi, \phi, \eta) =
\left( \begin{array}{ccccc}
1+ \frac{p^2 + q^2 + r^2}{2} & p & q & r & -\frac{p^2 + q^2 + r^2}{2} \\
p & & & & -p \\
q & &R(\psi, \phi, \eta) & & -q \\
r & & & & -r  \\
\frac{p^2 + q^2 + r^2}{2} &  p & q & r & 1 -\frac{p^2 + q^2 + r^2}{2}
\end{array}\right)
\label{888888}
\end{equation}
where $p,q,r$ are  any real numbers, while $R(\psi, \phi, \eta) \in
SO(3)$,  with $(\psi, \phi, \eta)$ being a triplet of Euler angles.
 The above result can be derived by following the standard treatment (see, for
example \cite{we}) adopted in appendix A for 3+1 dimensional
case. The element of the 
translational subgroup $T(3)$  of $W_5(p,q,r; \psi, \phi, \eta)$ can be
 trivially obtained by setting $R(\psi, \phi, \eta)$ to be the identity 
matrix and
will be denoted by $W(p,q,r)$;
\begin{equation}
W(p,q,r) \equiv W_5(p,q,r;0) =
\left( \begin{array}{ccccc}
1+ \frac{p^2 + q^2 + r^2}{2} & p & q & r & -\frac{p^2 + q^2 + r^2}{2} \\
p & 1 & 0 &0 & -p \\
q & 0  & 1 & 0 & -q \\
r & 0  & 0 & 1 & -r  \\
\frac{p^2 + q^2 + r^2}{2} &  p & q & r & 1 -\frac{p^2 + q^2 + r^2}{2}
\end{array}\right).
\label{88}
\end{equation}

Let us now consider free Maxwell theory in 5-dimensions,
\begin{equation}
{\cal L} = -{\frac{1}{4}}{F^{xy}F_{xy}} ; \hskip 1.0cm x
,y = 0, 1, 2, 3, 4.
\label{ds89}
\end{equation}
For a photon of energy $\omega$ (in 5-dimensional space-time) propagating in the $x = 4$ direction,
the momentum 5-vector is given by
\begin{equation}
k^x = (\omega, 0, 0, 0, \omega)^T.
\label{ds90}
\end{equation}
By following the plane wave method and proceeding exactly as in section \ref{tm}, one can show that the
maximally reduced form of the polarization vector of the photon is
\begin{equation}
\varepsilon^{x} = (0, a,b,c , 0)^T
\label{ds91}
\end{equation}
where $a,b,c$ represent the three
transverse degrees of freedom
(since the polarization vector satisfies the $`$Lorentz gauge'
$\varepsilon^{x}k_x = 0$).
If we now suppress the last rows of the column matrices  $k^{x}$ (\ref{ds90}) and $\varepsilon^{x}$ (\ref{ds91}), we end up respectively with the energy-momentum 4-vector and the polarization vector
 of 3+1 dimensional Proca theory in the rest frame of the quanta if we make the identification $\omega = m$ ($m$ being the mass of the Proca particle), or equivalently, of $B\wedge F$ theory since
the gauge invariant sector of the latter  is equivalent to the  former.
This is equivalent to applying
the projection operator given by the matrix
\begin{equation}
{\cal P}= diag(1, 1, 1, 1, 0)
\label{ds92}
\end{equation}
to the momentum 5-vector (\ref{ds90}) and the polarization vector (\ref{ds91}).

Similarly, to reproduce the polarization tensor of $B\wedge F$ or equivalently, massive Kalb-Ramond
theory in 3+1
dimensions let us consider free massless KR model in 5-dimensions,
\begin{equation}
{\cal L} = {\frac{1}{12}}{H^{xyz}H_{xyz}}
\label{krl-5}
\end{equation}
Analogous to (\ref{3-21}) (for massless KR theory in 3+1 dimensions), the 
maximally reduced form of the polarization tensor $\varepsilon^{xy}$ of the
4+1 dimensional massless KR model can be obtained by plane wave method as, 
\begin{equation}
 \{\varepsilon^{xy}\} = \left(
\begin{array}{ccccc}
0 & 0 & 0 & 0 & 0 \\
0 & 0 & \varepsilon^{12} & \varepsilon^{13} & 0 \\
0 & -\varepsilon^{12} & 0 & \varepsilon^{23} & 0 \\
0 & -\varepsilon^{13} & -\varepsilon^{23} & 0 & 0 \\
0 & 0 & 0 & 0 & 0
\end{array}\right)
\label{ex-5}
\end{equation}
Again deleting the last row and column, one gets the polarization 
tensor$({\varepsilon}^{ab})$ in (\ref{d14})  of the $B\wedge F$ model or of
 massive KR theory. This is equivalent to applying the projection operator as ${\cal P}{\varepsilon}{\cal P}$. Thus the polarization vector and  tensor of the Proca and  massive KR models  respectively, or $B\wedge F$ theory  have been
reproduced.
This is quite natural since, as mentioned in section 
\ref{bf},
the gauge invariant physical sector of $B\wedge F$ theory can be considered
equivalent either to Proca theory or to massive KR theory in 3+1 dimensions.

Now coming to the gauge transformation properties of polarization vector (\ref{ds91}) and polarization tensor (\ref{ex-5}) under the translational subgroup
$T(3)$, let $W (p, q, r)$ act on these objects one by one. First, acting on
$\varepsilon^{x} (\ref{ds91})$, one gets
\begin{equation}
\varepsilon^{x} \rightarrow {\varepsilon^{\prime x}} =
{{W (p, q, r)}^{x}}_{y} \varepsilon^{y} =\varepsilon^{x} +
\delta\varepsilon^{x} = \varepsilon^{x} + ( pa + qb + rc)
\frac{k^x}{\omega}
\label{ex-6}
\end{equation}
which is indeed a gauge transformation in (4+1) dimensional Maxwell theory.
Applying the projection operator ${\cal P}$  (\ref{ds92}) on
(\ref{ex-6}) yields
\begin{equation}
\delta {\varepsilon}^{a} = {\cal P}\delta\varepsilon^{x} 
= \frac{1}{\omega}( pa + qb + rc)k^{a}
\label{ex-8}
\end{equation}
where ${\varepsilon}^{a} = (0, a, b, c)^T$ and, modulus the
$i$-factor, corresponds to the expression in (\ref{d14}) of $B \wedge F$ theory.
This is precisely how the polarization vector in
 $B \wedge F$ theory transforms under gauge transformation \cite{bc1}.
In fact we can write
\begin{equation}
\delta {\varepsilon}^{a} = {D^{a}}_{b}(p, q, r){\varepsilon}^{b} - {\varepsilon}^{a} = \frac{1}{\omega}( pa + qb + rc)k^{a}
\label{ex1-8}
\end{equation}
where $D(p, q, r)$ is a representation of the translational group $T(3)$ given in (\ref{d15}),
\begin{equation}
D(p, q, r) = \left(
\begin{array}{cccc}
1 & p & q & r \\
0 & 1 & 0 & 0 \\
0 & 0 & 1 & 0 \\
0 & 0 & 0 & 1
\end{array}
\right).
\label{ex-13}
\end{equation}
In terms of the generators (\ref{mutually}) of $T(3)$, the change in the 
polarization vector (\ref{ex1-8}) can be expressed as the action of a Lie
algebra element:
\begin{equation}
\delta {\varepsilon}^{a} = (pT_1 + qT_2 + rT_3)\frac{{\varepsilon}^{a}}{\omega}.
\label{eex-13}
\end{equation}
Coming next to the polarization matrix $ \{ \varepsilon^{xy}\}$,
its transformation law is given by,
$$\{\varepsilon^{xy}\} \rightarrow\{ \varepsilon^{\prime xy }\} = W(p, q, r)\{\varepsilon^{xy}\} W^T(p, q, r) =\{\varepsilon^{xy}\} +\{ \delta \varepsilon^{xy}\}$$
where,
\begin{equation}
\{\delta \varepsilon^{xy} \} = \left(
\begin{array}{ccccc}
0 & \alpha_1 & \alpha_2 & \alpha_3 & 0 \\
-\alpha_1 & 0 & 0 & 0 & -\alpha_1 \\
- \alpha_2 & 0 & 0 & 0 & - \alpha_2 \\
-\alpha_3 & 0 & 0 & 0 & -\alpha_3 \\
0 & \alpha_1 & \alpha_2 & \alpha_3 & 0
\end{array}
\right)
\label{ex-10}
\end{equation}
and
$\alpha_1 = -(qc + rb),  ~~\alpha_2  = (pc - ra)),  ~~\alpha_3 = (pb + qa)$.
Again this can be easily recognized as a gauge transformation in (4+1)
dimensional KR theory involving massless quanta, as $\delta \varepsilon^{xy}$ can be expressed as $\approx (k^{x} f^{y}(k) - k^{y}f^{x}(k))$
with suitable choice for $f^{x}(k)$, where $k^x$ is of the form (\ref{ds90}).
 Now applying the projection operator
${\cal P}$ (\ref{ds92}) on (\ref{ex-10}), we get
the change in the 3+1 dimensional polarization matrix
$ \{ {\varepsilon}^{ab}\}$, by the formula,
$\{\delta \varepsilon^{ab}\} = {\cal P}\{\delta \varepsilon^{xy}\}{\cal P}^T$.
 This simply
amounts to a deletion of the last row and column of $\{\delta\varepsilon^{xy}\}$.
The result
can be expressed more compactly as
\begin{equation}
\{\delta {\varepsilon}^{ab}\} = (D(p,q,r)\{\varepsilon^{ab}\}D^T((p,q,r) - \{\varepsilon^{ab}\}).
\label{ex-12}
\end{equation}
Again this has the precise form of gauge transformation of the
 polarization matrix
of  $B \wedge F$ model, since it can be cast in the form 
\begin{equation}
\{\delta \varepsilon^{ab}\} = i(k^{a} f^{b}(k) - k^{b}f^{a}(k))
\label{ex-12-1}
\end{equation}
for a suitable $f^{a}(k)$, where $k^{a} = (m, 0, 0,0)^T$.

Thus   we have shown that the gauge generation representation $D(p,q,r)$ of
 $T(3)$ for topologically massive $B \wedge F$ gauge theory in 3+1 dimensions
 can be connected to the Wigner's little group for
massless particle in 4+1 dimensions through dimensional descent. This involved
 appropriate projections in the
intermediate steps, where the massless particles moving in  4+1 dimensions 
can be associated with a massive particle at rest in 3+1 dimensions.
Similarly the polarization vector and tensor respectively  of $B \wedge F$ theory in 
3+1 dimensions can be associated with polarization vector and polarization
 tensor of free Maxwell and KR theories in 4+1 dimensions. 
\label{dbf}

\section{Dimensional descent: 3+1 to 2+1 dimensions}
We have seen from the previous discussion that using dimensional descent
 one can obtain several properties regarding the gauge transformation in 3+1
dimensional topologically massive gauge theory (the $B \wedge F$ theory)
by starting from higher dimensional massless gauge theories. We have also seen
earlier that the MCS and ECS theories are topologically massive theories in 2+1
dimensions. Therefore one is naturally led to the question as to what would
be the role of dimensional descent in these 2+1 dimensional topologically 
massive gauge theories with respect to the massless gauge theories (Maxwell
and linearized gravity) in 3+1 dimensions.
This section is devoted to  a detailed discussion of this issue and the application of dimensional descent from 3+1 dimensions to 2+1 dimensions \cite{sc,bc2}.

\subsection{Proca theory and doublet of Maxwell-Chern-Simons theories}
\label{p-mcs}
Here we describe the method of dimensional descent from 3+1 to 2+1 dimensions
for vector theories \cite{bc2}.
Let us  recapitulate certain properties of
free Maxwell theory in 3+1 dimensions from section \ref{tm}, which are essential in the present context. The 3+1 dimensional Maxwell theory has 
two transverse degrees of freedom. Correspondingly the polarization vector
$\varepsilon^{a} $ takes the maximally reduced 
form $\varepsilon^{a} = (0, ,\varepsilon^1,\varepsilon^2, 0)^T$  (\ref{3-4}), if the 4-momentum is  $k^{a}
 = (\omega, 0, 0, \omega)^T$. We have seen earlier (section \ref{tm},  (\ref{3-5})) that 
the generator of gauge transformation in this case is $T(2)$, which is a
subgroup of $E(2)$:
\begin{equation}
\delta \varepsilon^{a}  =  {W^{a}}_{b}(p, q) \varepsilon^{b} - \varepsilon^{a}  =
 \left( p\varepsilon^1 + q\varepsilon^2\right)\frac{k^{a}}{\omega}.
\label{g3-5}
\end{equation}
 We now proceed forth to discuss   dimensional descent  from 3+1 dimensions to 2+1  dimensions. By  applying the projection operator ${\cal P} = diag (1, 1, 1, 0)$ on $\varepsilon^{a} $ and $k^{a}$, and then suppressing   the last rows 
of $\varepsilon^{a} $ and $k^{a}$, we obtain the descended objects $ {\varepsilon}^\mu =
(0, a, b)^T $ (with $\varepsilon^1 =a$ and $\varepsilon^2 = b$)    and $k^\mu = (\omega, 0, 0)^T $. These 
can be considered to be the polarization vector and momentum 3-vector in the
 rest frame of  2+1 dimensional Proca theory described by the Lagrangian 
\begin{equation}
{\cal L} = -{\frac{1}{4}}{F^{\mu\nu}F_{\mu\nu}} + {\frac{m^2}{2}}A^{\mu}A_{\mu}
\label{p26}
\end{equation}
 provided  we make
the identification $\omega = m$, the mass of Proca excitations. Analogous to 
(\ref{ex-8}), the projection
operator ${\cal P} = diag (1, 1, 1, 0)$ when applied on the gauge transformation
 (\ref{g3-5}) yields
\begin{equation}
\delta {\varepsilon}^{\mu } = {\cal P}\delta\varepsilon^{a}
= \frac{1}{\omega}( pa + qb )k^{\mu} 
\label{p33}
\end{equation}
Since  Proca theory is not a gauge theory, (\ref{p33}) cannot be considered as
 a gauge transformation. We have, however,  seen earlier that
Proca theory in 2+1 dimensions
is actually a doublet of
Maxwell-Chern-Simons theories \cite{bcs, bw, deser}
\begin{equation}
{\cal L} = {\cal L}_+ \oplus {\cal L}_-
\label{p29}
\end{equation}
where
\begin{equation}
{\cal L}_\pm = -{\frac{1}{4}}{F^{\mu\nu}F_{\mu\nu}} \pm {\frac{{\vartheta
}}{2}}{\epsilon}^{\mu\nu\lambda}A_{\mu}{\partial}_{\nu}A_{\lambda}
\label{p30}
\end{equation}
with $\vartheta > 0$ and each of ${\cal L}_+$ or ${\cal L}_-$ being a topologically massive gauge
 theory.  The mass of the MCS quanta is $m = \vartheta$, where $m$
 is the parameter entering in (\ref{p26}).  We can therefore study the gauge transformation generated in this
doublet. For this purpose, analogous to (\ref{ex-13}),   it is essential to provide a $3 \times 3$ representation of $T(2)$ (denoted by
${D}(p, q)$),
\begin{equation}
{D}(p, q) = \left(
\begin{array}{ccc}
1 & p & q   \\
0 & 1 & 0   \\
0 & 0 & 1   \\
\end{array}
\right).
\label{p27}
\end{equation}
The corresponding generators are given by,
\begin{equation}
{T}_1 = \frac{\partial {D(p,q)}}{\partial p} = \left(
\begin{array}{ccc}
0 & 1 & 0 \\
0 & 0 & 0 \\
0 & 0 & 0
\end{array}
\right); \hskip 1.0cm {T}_2 = \frac{\partial {D(p,q)}}{\partial q} =
\left( \begin{array}{ccc}
0 & 0 & 1 \\
0 & 0 & 0 \\
0 & 0 & 0
\end{array}
\right)
\label{p28}
\end{equation}
In analogy with (\ref{eex-13}), here also one can rewrite (\ref{p33}) as,
\begin{equation}\delta {\varepsilon}^{\mu} = {(p{T}_1 + q{T}_2)^\mu}_\nu {\varepsilon}^{\nu}
\label{p32}
\end{equation}
so that the change in the polarization vector ${\varepsilon}^{\mu}$ is obtained as the actoin of a Lie algebra element.
 Proca polarization vector ${\varepsilon}^{\mu}$ is just a linear
combination of the two real orthonormal canonical vectors $\varepsilon^1$
and $ \varepsilon^2$ where,
\begin{equation}
{\varepsilon}^{\mu} = a \varepsilon^{(1)} + b \varepsilon^{(2)};~~ \hskip 1.0 cm
\varepsilon^{(1)} = (0,1,0)^T,
\varepsilon^{(2)} = (0,0,1)^T.
\label{p34}
\end{equation}
Correspondingly the generators $T_1$ and $T_2$  (\ref{p28}),
form an
orthonormal basis as they satisfy  $tr({T}^{\dagger}_i{T}_j) = \delta_{ij}$. Furthermore,
\begin{equation}
{T}_1\varepsilon^{(1)} = {T}_2\varepsilon^{(2)} = (1, 0, 0)^T
= \frac{k^\mu}{m}, \hskip 1.0 cm {T}_1\varepsilon^{(2)} = {T}_2\varepsilon^{(1)} = 0
\label{p35}
\end{equation}

On the other hand, the rest frame  polarization vectors for ${\cal L}_\pm $, with only one degree of freedom for each of ${\cal L}_+$ and
${\cal L}_-$ (section \ref{mcs1}), as given by 
\begin{equation}
{\xi}^{\mu}_\pm = \frac{1}{\sqrt 2} \left( \begin{array}{c}
0 \\
1 \\
\mp i
\end{array}
\right)
\label{p31}
\end{equation}
also provide an orthonormal basis (complex) in the plane as
\begin{equation}
({\xi}^{\mu}_+)^{\dagger}({\xi}^{\mu}_-) = 0;
\hskip 1.0cm ({\xi}^{\mu}_+)^{\dagger}({\xi}^{\mu}_+) =
({\xi}^{\mu}_-)^{\dagger}({\xi}^{\mu}_-) = 1.
\label{p36}
\end{equation}
Here we note that spatial part $\vec{\xi}_\pm$ of ${\xi}_\pm$
 can be obtained from the space part of the  above mentioned canonical ones by appropriate
$SU(2)$ transformation. That is, $U= \frac{1}{\sqrt{2}}\left( \begin{array}{cc}
1 & i \\
i & 1 \end{array} \right)
\in $ SU(2) when acts on ${\vec{\varepsilon}}^{(1)} = \left( \begin{array}{c} 1 \\ 0 \end{array} \right)$ 
and ${\vec{\varepsilon}}^{(2)} = \left( \begin{array}{c} 0 \\ 1 \end{array} \right)$
yields respectively the vectors $\vec{\xi}_- = \frac{1}{\sqrt{2}}\left( \begin{array}{c} 1 \\ i \end{array} \right)$ and  $\vec{\xi}_+ = \frac{1}{\sqrt{2}}\left(
\begin{array}{c} 1 \\ -i \end{array} \right)$
(up to an irrelevant factor of $i$)\footnote{This ambiguity of $i$ factor is 
related to the U(1) phase arbitrariness of the polarization vector discussed in section \ref{mcs1}.}:
\begin{equation} 
i\vec{\xi}_+ = U {\vec{\varepsilon}}^{(2)}, ~~~~\vec{\xi}_- = U {\vec{\varepsilon}}^{(1)}.
\label{p37-1}
\end{equation}
 This suggests that we consider the following
 orthonormal basis for the Lie algebra of $T(2)$:
\begin{equation}
{T}_\pm = \frac{1}{\sqrt{2}}({T}_1 \pm i{T}_2) = \frac{1}{\sqrt{2}}\left( \begin{array}{ccc}
0 & 1 & \pm i \\
0 & 0 & 0 \\
0 & 0 & 0 \end{array} \right)
\label{p37}
\end{equation}
instead of ${T}_1$ and ${T}_2$. Note that they also satisfy relations similar to
the ($T_1-T_2)$ basis,
\begin{equation}
tr({T}_+^{\dagger}{T}_+) = tr({T}_-^{\dagger}{T}_-) = 1;~~~ tr({T}_+^{\dagger}{T}_-) = 0
\label{p38}
\end{equation}
One can now easily see that
\begin{equation}
{T}_+{\xi}_+ = {T}_-\xi_- = \frac{k^\mu}{m}, \hskip 1.0cm {T}_+{\xi}_- = {T}_-\xi_+ = 0
\label{p39}
\end{equation}
analogous to (\ref{p35}). Furthermore,
\begin{equation}
\delta {\xi}^{\mu}_\pm = p_{\pm}{T}_{\pm} {\xi}^{\mu}_\pm =
 \frac{p_{\pm} }{m}k^\mu.
\label{p40}
\end{equation}
This indicates that ${T}_\pm$ - the generators of the Lie algebra of $T(2)$
in the
rotated (complex) basis -
generate independent  gauge transformations in ${\cal L}_{\pm}$ respectively. One
therefore can understand how the appropriate representation
 of the generator of gauge
transformation in the doublet of MCS theory can be obtained from higher
3+1 dimensional Wigner's group through dimensional descent.
A finite gauge transformation is obtained by integrating (\ref{p40}) i.e., exponentiating the corresponding Lie algebra element. This gives two
representations of Wigner's little group for massless particles in $2+1$
dimensions, which is isomorphic to ${\cal R}\times {\cal Z}_2$, although here 
we are just considering the component which is connected to the identity,
\begin{equation}
D_\pm (p_\pm) = e^{p_\pm {T}_\pm} = 1 + p_\pm {T}_\pm = \left(
\begin{array}{ccc}
1 & \frac{p_\pm}{\sqrt{2}} & \pm i \frac{p_\pm}{\sqrt{2}} \\
0 & 1 & 0 \\
0 & 0 & 1
\end{array}
\right)
\label{p41}
\end{equation}
Note that $D_\pm (p_\pm) $ generates gauge transformation in the                 doublet ${\cal L}_{\pm}$,  
\begin{equation}
{D_\pm^\mu}_\nu {\xi}^{\nu}_\pm =  {\xi}^{\mu}_\pm + 
\frac{p_\pm}{|\vartheta | } k^\mu
\label{p42}
\end{equation}
and are related by complex conjugation.
This
complex conjugation is also a symmetry of the doublet as we saw in section \ref{appl}.

Therefore, it is clear that the gauge generating representation of little group
for MCS and ECS theories can be obtained by the method of dimensional descent.

\subsection{EPF theory and doublet of Einstein-Chern-Simons theories}
Now we discuss dimensional descent from 3+1 dimensions to 2+1 dimensions for
theories with symmetric second rank tensor fields \cite{sc}. For this, it
 is essential to discuss the relevant aspects 
of 2+1 dimensional Einstein-Pauli-Fierz (EPF) theory \cite{djt,fierz,fierz1} whose action  is given by
\begin{equation}
I^{EPF} = \int d^3 x \left(- \sqrt{g} R - \frac{\mu^2}{4}(h_{\mu\nu}^2 - h^2)
 \right). 
\label{p79}
\end{equation}
Note that the usual sign in front of Einstein action has been restored to avoid
ghosts and tachyons, as has been observed recently by Deser and Tekin 
\cite{tekin}. As noted in \cite{tekin}, both the relative  and overall signs
of the two terms in (\ref{p79}) have to be of conventional Einstein and Pauli-Feirz mass terms in order to have a physically meaningful theory. On the other hand in the ECS theory, sign of the Einstein term has to be 
opposite to that of the conventional one for the theory to be viable. Therefore,
if one attempts to couple the ECS theory with a Pauli-Fierz term, one is faced
 with an unavoidable conflict of signs. Upon linearization, (\ref{p79}) reduces to
\begin{equation}
{\cal L}_L^{EPF} = \frac{1}{2}h_{\mu\nu} \left[ R^{\mu\nu}_L -
\frac{1}{2} \eta^{\mu\nu} R_L\right] - \frac{\mu^2}{2}(h_{\mu\nu}^2 - h^2).
\label{p80}
\end{equation}
Analogous to the doublet structure of Proca theory discussed above,
the EPF theory
is a doublet, as was suggested in \cite{djt},  comprising of a pair of ECS theories having opposite helicities.
And just like the Proca theory,
EPF theory does not possess any gauge symmetry. The equation of motion following from the EPF Lagrangian is
given by 
\begin{equation}
- \Box h^{\mu\nu} + \partial^\mu \partial_\gamma  h^{\gamma\nu} + \partial^\nu\partial_\gamma  h^{\gamma\mu}
 - \partial^\mu \partial^\nu h + \eta^{\mu\nu}(\Box h -\partial_\gamma \partial_\delta h^{\gamma\delta}) - \mu^2(h^{\mu\nu} - \eta^{\mu\nu} h) = 0.
\label{p81}
\end{equation}
With the ansatz $ h^{\mu\nu} = {\chi}^{\mu\nu} e^{ik.x}$, where ${\chi}^{\mu\nu}$ is the polarization tensor in this case,  this equation can
be written as
\begin{equation}
k^2{\chi}^{\mu\nu} - k^\mu k_\gamma {\chi}^{\gamma \nu} -
k^\nu k_\gamma {\chi}^{\gamma\mu } +  k^\mu k^\nu {\chi}
+\eta^{\mu\nu}(-k^2 {\chi} +k_\gamma k_\delta {\chi}^{\gamma\delta}) -
\mu^2({\chi }^{\mu\nu} - \eta^{\mu\nu}{\chi}) = 0
\label{p82}
\end{equation}
We now proceed along the same lines as was done in the previous cases
to arrive at the physical polarization tensor of EPF theory.  By a heuristic argument we can easily 
see that the EPF theory does not possess any massless excitations. (It  can also be seen from the propagator of the $ h^{\mu\nu}$ field that the EPF theory has massive excitations \cite{fierz,fierz1}.)   If we  choose $k^2 =0$, the
equation of motion (\ref{p82}) leads, upon contraction with $k_\mu$, to the 
condition
\begin{equation}
k_\mu {\chi}^{\mu\nu} = k^\nu{\chi}. \label{p821}
\end{equation} 
On the other hand, the  contraction of (\ref{p821}) with $k_{\nu}$ leads to
\begin{equation}
k_\mu k_\nu {\chi}^{\mu\nu} = 0.\label{p822}
\end{equation}
A solution of the above pair of  equations (\ref{p821}, \ref{p822}) is given by
$  {\chi}^{\mu\nu} = k^af^b(k) + k^b f^a(k)$ where $f$'s are arbitrary functions of 
$k$ which satisfies the condition $k.f=0$. However,
such a solution is compatible with  the equation of motion (\ref{p82}) if and 
only if the $f$'s vanish identically thus demonstrating that EPF theory does not have
massless excitations. 

Now for $k^2 \neq 0$, in the rest frame $k^{\mu} = (m, 0, 0)^T$ and so the $(00)$ component of the equation of motion yields
$\mu^2({\chi}^{00} - {\chi}) = 0$ which in turn gives
\begin{equation}
{\chi}^{11} + {\chi}^{22} = 0.
\label{p82+1}
\end{equation}
Therefore, one is free to arbitrarily choose either ${\chi}^{11}$ or ${\chi}^{22}$. 
Similarly, the  $(0i)$ component in the rest frame becomes $\mu^2  {\chi}^{0i} = 0$  implying
\begin{equation}
{\chi}^{0i} = 0.
\label{p82+2}
\end{equation}
The space part ($(ij)$-components) of the equation of motion with $i=j=1$ 
and $i=j=2$, respectively yields in the rest frame,
\begin{equation}
-\mu^2 ({{\chi}^0}_0 + {{\chi}^2}_2) + m^2 {\chi}^{11}  = 0
\label{puthu}
\end{equation}
\begin{equation}
-\mu^2 ({{\chi}^0}_0 + {{\chi}^1}_1) + m^2 {\chi}^{22}  = 0.
\end{equation}
Adding the above two equation gives,
\begin{equation}
{\chi}^{00} = 0.
\label{p82+3}
\end{equation}
Substituting (\ref{p82+3}) back in (\ref{puthu}) and using (\ref{p82+1}) one arrives at 
\begin{equation}
(m^2 - \mu^2){{\chi}^1}_1 = 0.
\label{p82+333}
\end{equation}
Finally, the spatial component $(\mu =i,~\nu =j)$  of  (\ref{p82}) for $i\ne j$
in the rest frame becomes
\begin{equation}
(m^2 - \mu^2){{\chi}^1}_2 = 0.
\label{newp82+3}
\end{equation}
The equations (\ref{p82+333}) and (\ref{newp82+3}) can be satisfied if either $(m^2 - \mu^2) = 0$  or   ${{\chi}^1}_1 = {{\chi}^1}_2 =  0.$ On account of
(\ref{p82+1}), (\ref{p82+2}) and (\ref{p82+3}), the latter choice will lead to
a null theory and can be ruled out. Therefore, we must have
\begin{equation}
m^2 = \mu^2
\label{p82+4}
\end{equation}
thus establishing that the mass of the EPF excitation to be $|\mu |$. This
choice leaves  the two components $\chi^{11}$ and $\chi^{12}$  arbitrary
 representing the two physical degrees of freedom in the theory. 

On the other hand, if the Einstein term in (\ref{p79}) were +ve (i.e., the same as 
the Einstein term in (\ref{57+1})), instead of
(\ref{newp82+3}) we would have had
\begin{equation}
(m^2 + \mu^2){{\chi}^1}_2 = 0.
\label{newp82+4}
\end{equation}
This in term would have meant $ m^2  =- \mu^2$ leading to an unphysical
tachyonic mode for theory. Thus, in EPF theory, the sign of the Einstein
term must necessarily be negative, i.e., the conventional one. Therefore, our analysis also shows the unviability
of coupling both Chern-Simons and Pauli-Fierz terms to Einstein gravity
in 2+1 dimensions since such a coupling puts conflicting demands
on the sign on the Einstein term.

Therefore, it is obvious that we can write the polarization tensor of EPF
model in the rest frame as
\begin{equation}
\{\chi^{\mu \nu}\} 
 = \left(
\begin{array}{ccc}
0 & 0 & 0   \\
0 & a & b \\
0 & b & -a
\end{array}
\right).
\label{p83}
\end{equation}
where ${\chi}^{11} = 
-{\chi}^{22}=a $ and $ {\chi}^{12} =  {\chi}^{21}= b $.
With the aid of the expressions (\ref{p83}) for $\chi^{\mu \nu}$ of EPF theory  
and (\ref{76}) for  $\{\chi_\pm\}$ of a pair of ECS theories, we now embark on a discussion of
dimensional descent for the case of second rank symmetric tensor gauge fields emphasizing
the near exact parallel with the  case of vector gauge fields discussed in the previous subsection.
One can obtain the 
momentum 3-vector $k^\mu$ and polarization tensor $\{\chi^{\mu\nu}\}$ (\ref{p83}) of EPF model in 2+1 dimensions
from those of linearized gravity in 3+1 dimensions as follows.  
By applying the projection operator ${\cal P} = $diag$(1, 1, 1, 0)$
on momentum 4-vector
 $k^a = (\omega, 0, 0, \omega )^T $ of a massless graviton moving in the $z$-direction of 3+1 dimensional linearized Einstein gravity and subsequently
deleting the last row of the resulting vector, one get momentum 3-vector $k^\mu$ of 2+1 dimensional
EPF quanta at rest. By a similar application of ${\cal P}$ on ${\{\chi^{ab}\}}$ (\ref{32-16}) and deleting
the last row and column, one gets the polarization tensor $\{\chi^{\mu\nu}\}$ (\ref{p83})  in the rest frame of the EPF quanta.
Next we notice that, just like the way Proca polarization vector ${{\varepsilon}}^\mu$ is written as a linear combination of two orthonormal canonical vectors (\ref{p34}), one can write the EPF
polarization tensor $\{\chi^{\mu\nu}\}$ as 
\begin{equation}
\{\chi^{\mu\nu}\} = a {\cal X}_1 + b {\cal X}_2 =  a \left(
\begin{array}{ccc}
0 & 0 & 0   \\
0 & 1 & 0 \\
0 & 0 &-1
\end{array}
\right) + b \left(\begin{array}{ccc}
0 & 0 & 0   \\
0 & 0 & 1 \\
0 & 1 & 0
\end{array}
\right).
\label{p84}
\end{equation}
We may consider the above equation to be the EPF analogue of (\ref{p34}) in the case
Proca theory.
Notice that the space parts of the matrices appearing in the above linear
combination are nothing but the Pauli matrices
\begin{equation}
\sigma_1 = \left(\begin{array}{cc}
 0 & 1 \\
1 & 0
\end{array}
\right),  ~~~~ \sigma_3 = \left(\begin{array}{cc}
 1 & 0 \\
0 & -1
\end{array}
\right).
\end{equation}
Clearly, the space part $\{{\chi_\pm}^{ij}\}$ of the ECS polarization tensors $\{\chi_\pm^{\mu\nu}\}$ (\ref{76}) can be expressed in terms of $ \sigma_1$ and $\sigma_3$
as follows:
\begin{equation}
\{{\chi_\pm}^{ij}\} = \frac{1}{\sqrt{2}}(\frac{1}{\sqrt{2}}\sigma_3 \mp \frac{i}{\sqrt{2}} \sigma_1 )~.
\label{p85-1}
\end{equation}
This amounts to an SU(2) transformation in the 2-dimensional subspace (of the
SU(2) Lie algebra in an orthonormal basis) spanned by  
$ \Sigma_1 = \frac{1}{\sqrt{2}}\sigma_1$ and $\Sigma_2 =   \frac{1}{\sqrt{2}} \sigma_3$ as can be seen from the following\footnote{Note
that the $\{\chi^{ij}_-\}$ obtained in (\ref{p85-1}) differs from the one 
obtained by SU(2) rotation by an irrelevant $i$ factor just as in the vector case.}:
\begin{equation}
\frac{1}{\sqrt{2}}\left( \begin{array}{cc}
1 & i \\
i & 1 \end{array} \right) \left( \begin{array}{c}\Sigma_1 \\ \Sigma_2 \end{array} \right) =  \left( \begin{array}{c} i\{{\chi_+}^{ij}\} \\ \{{\chi_-}^{ij}\}
\end{array} \right).   
\label{p85-1-0}
\end{equation}
It should be noticed that (\ref{p85-1}) is the analogue of
  (\ref{p37-1}) in the case of Proca and MCS theories. In the case of vector
  (Proca and MCS) field theories, the basis vectors $\varepsilon^{(1)}$ and
$\varepsilon^{(2)}$ (\ref{p34}) of the Proca polarization vector, when transformed 
by a suitable SU(2) transformation yield the polarization vectors
$\xi_\pm$
of a pair of MCS theories. We can see an exact analogy of this in the EPF
 theory as follows. 
 Similarly, in the case of tensor (EPF and ECS) field
 theories,  the same SU(2) transformation when acted on the ${\cal X}_1$ and 
${\cal X}_2$ provides the polarization tensors $\{\chi_\pm^{\mu\nu}\}$ of a doublet of
ECS theories with opposite helicities just as  Proca theory is a doublet
of MCS theories having  opposite spins.
This corroborates the proposition that EPF theory is consisted of 
a doublet of ECS theories with opposite spins at least at the level of
polarization tensor. Moreover, as we have discussed earlier, 
the polarization tensor and momentum vector of (2+1 dimensional) EPF theory can be obtained
from those of linearized Einstein gravity (in 3+1 dimensions) by applying suitable projection operator. This relationship between EPF and ECS theories resemble the one between Proca and MCS theories. Therefore we expect that 
the procedure of dimensional descent to be valid here as well. As described in section \ref{p-mcs},
the generator of the  representations of  $T(1)$, obtained by dimensional descent,  that generate 
gauge transformation in a pair of MCS theories with opposite helicities are
given by
${T}_\pm$  (\ref{p37}).
Also, the ECS polarization tensors $\chi_\pm$ can be made to 
satisfy the orthonormality relations
(\ref{75}) similar to (\ref{p36}) for MCS case owing to the fact that the 
former is a tensor product of MCS polarization  vectors.
Hence it is natural to expect that the $T(1)$ group representation $D_\pm (p_\pm)$ (\ref{p41}) obtained by exponentiation of ${T}_\pm$ generates gauge 
transformation in  ECS doublet, which in fact it does, as we have shown in
(\ref{79+0}).
Therefore, it is evident that by a dimensional descent from 3+1 dimensional linearized gravity
one could obtain the representations of $T(1)$ that generate gauge transformations in the doublet of
topologically massive ECS theories in 2+1 dimensions. This is similar to  the dimensional descent
from 3+1 dimensional Maxwell theory to 2+1 dimensional MCS theory discussed in 
the subsection \ref{p-mcs}.

\section{Summary}
We presented a  review of how the gauge transformations in the 3+1 dimensional $B\wedge F$ 
theory is related through the method of dimensional descent to the gauge
transformations in Maxwell and massless Kalb-Ramond theories in 4+1 dimensions.
In the same fashion, dimensional descent relates the gauge transformations
in 3+1 dimensional Maxwell theory to those in the doublet of  Maxwell-Chern-Simons theories which is the equivalent to a Proca theory in 2+1 dimensions.
There exists an analogous relationship between the gauge transformation in
3+1 dimensional linearized gravity and in a doublet of Einstein-Chern-Simons
theories in 2+1 dimensions. Analysis of the polarization tensors of the
Einstein-Pauli-Fierz theory and  of a doublet
of Einstein-Chern-Simons theories with opposite helicities suggests that
the EPF theory is the doublet of ECS theories just like the Proca theory is a 
doublet of MCS theories. However, the analogy between EPF and Proca
theories with their respective doublet structures breaks down if one considers
the fact that sign of the Einstein term flips from EPF to ECS theories in
contrast to Proca theory where the sign of the Maxwell term remains unchanged
irrespective of whether it is coupled to a Chern-Simons term or a usual mass
term.
Therefore, further investigations are necessary in order to rigorously
establish the doublet structure, if any, of EPF theory beyond the level of polarization
tensors.
 
\chapter{Massive Gauge theories}
In the preceding chapters we have seen that various translational groups in 
their appropriate representations act as generators of gauge transformations
 in a variety of Abelian gauge theories. We have demonstrated this for
 theories which  have vector, symmetric and antisymmetric second rank tensors as
the underlying gauge fields.  Such theories considered so far are either 
massless or topologically massive gauge theories.   However,
there are other types of gauge theories which are
obtained by converting second class
constrained systems(in the  language of Dirac's theory of constraint dynamics) to first 
class (gauge) systems using the generalized
prescription of Batalin, Fradkin and Tyutin \cite{bft,bft1,bt,berg,bneto}. By such a prescription, one can obtain 
from the massive  Proca theory, the St\"uckelberg  model for vector field  that
is massive while possessing a gauge invariance \cite{bb}.
Similarly one can obtain the St\"uckelberg  extended versions of massive KR and
EPF theories. The discussion in this chapter is confined to 3+1 dimensions
since the St\"uckelberg  embedded theories are usually studied in that space-time.
Nevertheless, with suitable modifications,  all the methods as well as results of this chapter are equally
valid in 2+1 dimensions also.
 This chapter consists of the study of
the gauge transformations in such theories and their relationship with translational groups. Here we show that the same representation of the translational
group $T(3)$
that generates gauge transformation in  the topologically massive $B\wedge F$ theory also generates gauge
transformation in the St\"uckelberg  extended first class version of Proca, massive KR and EPF theories in 3+1 dimensions.

\section{Massive vector gauge theory}
\label{maxwell}
One can convert the 4-dimensional Proca theory (which does not possess any  gauge symmetry) into a gauge theory
by  St\"uckelberg  mechanism with the introduction of a new scalar field  $\alpha
(x)$
as follows;
\begin{equation}
{\cal{L}} = -\frac{1}{4}F_{a b }F^{a b } + \frac{m^2}{2} (A_a + \partial_a \alpha)
(A^a + \partial^a \alpha)
\label{554}
\end{equation}
The Lagrangian remains invariant under the transformations
\begin{equation}
A_{a}(x) \rightarrow
A^{\prime}_{a}(x) =  A_{a}(x) + \partial_{a} \Lambda (x), ~~~~~~\alpha (x)  \rightarrow  \alpha '(x) = \alpha (x) -\Lambda (x)
\label{54}
\end{equation}
where $\Lambda (x)$ is an arbitrary scalar function.
The equations of motion for the theory are
\begin{equation}
-\partial_b F^{a b } + m^2 (A^a + \partial^a \alpha)  = 0,~~~~~ \partial^a (A_a + \partial_a \alpha) = 0.
\label{55}
\end{equation}
One must notice that  by
operating $\partial_a $ on the first equation in (\ref{55}) one yields the second. Hence the
latter is consequence of the former. This implies that the gauge transformation
 of the $\alpha $-field can be deduced by knowing that of the $A^a$-field. 
Similar to (\ref{3-4-4}), here we adopt  the ansatz $A^{a}(x) = \varepsilon^a \exp (ik.x)$ and $\alpha (x)
= \tilde{\alpha}(k) \exp (ik.x)$. As before, $\varepsilon^a (k)$ is the polarization vector of the field $A^a(x)$ and $\tilde{\alpha}(k)$ is a particular
Fourier component of $\alpha (x)$ . In terms
of the polarization vector $\varepsilon^a (k)$, the equations of motion become respectively,
\begin{equation}
k_b(k^a \varepsilon^b - k^b \varepsilon^a ) + m^2(\varepsilon^a + ik^a \tilde{\alpha}) = 0, ~~~
ik^b(\varepsilon_b + ik_b \tilde{\alpha}) = 0.
\label{566}
\end{equation}
Analogous to (\ref{55}), the second equation in (\ref{566}) is a consequence of the first one.
For massless excitations $k^2 = 0$, the second equation of the above pair of
equations gives the Lorentz condition  $ k_b \varepsilon^b = 0$ which
when substituted in the first gives,
\begin{equation}
\varepsilon^a  =- i k^a \tilde{\alpha}.
\label{57}
\end{equation}
Since this is a solution proportional to the 4-momentum $k^a$, it can be gauged away by an appropriate
choice of the gauge. Thus, massless excitations are gauge artefacts. For $k^2 = M^2 $ (massive excitations),
the equations of motion (\ref{566}) becomes,
\begin{equation}
(m^2 - M^2) \varepsilon^a + k^a  k_b \varepsilon^b + im^2  k^a \tilde{\alpha}  = 0, ~~~ \tilde{\alpha} = \frac{ik_b \varepsilon^b}{M^2}.
\label{5i}
\end{equation}
Substituting  the second equation in (\ref{5i}) in  the first yields,
\begin{equation}
(m^2 - M^2)\left[ \eta^{ab} - \frac{k^a k^b}{M^2}\right] \varepsilon_b =0
\label{5j}   
\end{equation}
In this equation, the expression inside the parenthesis is a projection 
operator which projects
out the transverse component of $\varepsilon_b$. Since longitudinal component
can be gauged away, transverse part of $\varepsilon_b$ should be nonvanishing
in order to avoid
having no physical excitations. Therefore we must have $m^2 - M^2 =0$.
Therefore, one can conclude that the mass of the excitation is given by 
$m$ itself and  the rest frame momentum 4-vector of the
theory can be written as $ k^b = (m, 0, 0, 0)$. In the rest frame, the second
equation in (\ref{566}) gives
\begin{equation}
\varepsilon^0 = -i m \tilde{\alpha}.
\label{6i}
\end{equation}
Therefore, the polarization vector of $A^a (x)$ field of (\ref{554}) can be written as
\begin{equation}
\varepsilon^{a} = (-i m \tilde{\alpha}, \varepsilon^1, \varepsilon^2, \varepsilon^3)^T
\label{6j}
\end{equation}
The maximally reduced form of the polarization vector can be obtained from
(\ref{6j}), by a gauge transformation with the choice $\Lambda (x) = \alpha (x)$ and is given by
\begin{equation}
\varepsilon^{a} = (0, a, b, c)^T
\label{62i}
\end{equation}
with the free components $ a= \varepsilon^1, b=\varepsilon^2, c=\varepsilon^3 $
representing the three physical degrees of freedom of the theory.
One must note  that (\ref{62i}) is of the same form as that of the $B\wedge F$ theory polarization vector (section \ref{bf}).  Therefore, just as in the case of
$B\wedge F$ theory, the action of representation $D(p,q,r)$ (\ref{d15}) of 
$T(3)$ on the polarization vector  (\ref{62i}) amounts to a gauge transformation
 in St\"uckelberg  extended Proca theory:
\begin{equation}
\varepsilon^{a} \rightarrow \varepsilon^{\prime a} = {D^{a}}_{b}(p,q,r)\varepsilon^{b} = \varepsilon^{a} + \frac{i}{m}(pa +
q b + r  c)k^{a}
\label{622}
\end{equation}
The above transformation can be cast in the form of the momentum space 
gauge transformation
\begin{equation}
\varepsilon^{a}\rightarrow \varepsilon^{a} +
ik^a \lambda (k)
\label{6221}
\end{equation}
 (where $\Lambda(x)= \lambda (k) e^{ik.x}$) corresponding
to the field $A(x)$, by choosing the field $ \Lambda(x)$ such that 
\begin{equation}
\lambda (k)= \frac{(pa +
q b + r  c)}{m}.
\label{6222}
\end{equation}
As mentioned before, it is possible to obtain the gauge transformation property of
$\alpha$-field from that of the $A^a$-field for which we now proceed as follows.
\newpage
Consider the second relation in (\ref{5i}); 
\begin{equation}
\tilde{\alpha} = \frac{ik_a \varepsilon^a}{m^2}.
\label{6223}
\end{equation}
and let $\varepsilon^a$ undergo the gauge transformation (\ref{6221}) which has the effect
of making a corresponding transformation in $\alpha$-field as  
\begin{equation}
\tilde{\alpha} \rightarrow \tilde{\alpha}'  = \frac{ik_a (\varepsilon^a + ik^a \lambda)}{m^2} =\frac{ik_a \varepsilon^a}{m^2} - \lambda = \tilde{\alpha} - \lambda  .
\label{62233}
\end{equation}
Here $\lambda$ is given by (\ref{6222}) corresponding to the gauge transformation generated by the
translational group $T(3)$ in the $A^a(x)$ field. Notice that the above equation (\ref{62233})
corresponds to the second equation in (\ref{54}). We have thus obtained the gauge
transformation generated in the $\alpha$ field by $T(3)$ from that in the $A^a (x)$-field. It follows therefore that $\alpha$-field can be gauged away
completely by a suitable gauge fixing condition (unitary gauge) and it does not
appear in the physical spectrum of the theory.  
\label{sp}
Hence it is obvious that the representation $D(p,q,r)$  of $T(3)$ generates 
gauge transformation in the massive vector gauge theory governed
by (\ref{554}).

We noticed that the maximally  reduced polarization vector $\varepsilon^a $ 
(\ref{62i}) of the St\"uckelberg  embedded Proca model takes the same form as that of $B\wedge F$ theory  (\ref{d14}). Also both are massive gauge  theories
and the gauge transformations in both cases are generated by the representation
$D(p,q,r)$ of $T(3)$. Therefore, by starting from 4+1 dimensional Maxwell
theory, the dimensional descent can be employed to
study gauge transformations in St\"uckelberg  embedded Proca model just as the
gauge transformation of the vector field in  $B\wedge F$ theory is studied
using dimensional descent in section \ref{dbf}.

\section{Massive symmetric tensor gauge theory}
\label{sepf}
Consider the massive and non-gauge  Einstein-Pauli-Fierz (EPF) theory
in 3+1 dimension as given by the Lagrangian,
\begin{equation}
{\cal L}_L^{EPF} = \frac{1}{2}h_{ab} \left[ R^{ab}_L - 
\frac{1}{2} \eta^{ab} R_L\right] - \frac{\mu^2}{2}\left((h_{ab})^2
 - h^2 \right).
\label{63}
\end{equation}
The EPF theory does not posses any gauge invariance.
Just as the Proca theory was elevated to a  gauge theory (section \ref{maxwell}) by St\"uckelberg 
mechanism, the linearized EPF theory also can be provided with a gauge symmetry  by introducing the an additional vector field $A^a$ as follows:
\begin{equation}
{\cal L}_L^{EPF} = \frac{1}{2}h_{ab} \left[ R^{ab}_L -
\frac{1}{2} \eta^{ab} R_L\right] - \frac{\mu^2}{2}\left(\left(h_{ab} +
\partial_a A_b + \partial_b A_a          \right)^2 - \left(h +
2 \partial \cdot A\right)^2\right).
\label{64i}
\end{equation}
The  transformations
\begin{equation}
h_{ab} \rightarrow h'_{ab} = h_{ab} + \partial_a \Lambda_b
+ \partial_b \Lambda_a
\label{65i}
\end{equation}
\begin{equation}
A_a (x)\rightarrow A'_a (x) = A_a (x) - \Lambda_a (x)
\label{66i}
\end{equation}
represent the gauge symmetry of the theory described by (\ref{64i}).
The equation of motion for $h_{ab}$ is
$$- \Box h^{ab} + \partial^a \partial_c  h^{c b} + \partial^b\partial_c  h^{c a}
 - \partial^a \partial^b h + \eta^{ab}(\Box h -\partial_c
\partial_d h^{cd})$$
\begin{equation}
 - \mu^2 \left[ (h^{ab} +
\partial^a A^b + \partial^b A^a) - \eta^{ab}(h + 2 \partial \cdot A)\right] = 0
\label{67i}
\end{equation}
and that for $A_a$ is \begin{equation}
\Box A^a + \partial_b h^{ba} -  \partial^a h -\partial^a (\partial \cdot A)  = 0.
\label{68i}
\end{equation}
Analogous to the case of massive vector gauge theory discussed before, the equation of motion
(\ref{68i}) for $A_a$ can be obtained from (\ref{67i}) by applying the operator $\partial_b$.
Therefore, gauge transformation of $A^a$  is obtainable by knowing the gauge transformation of
the $h^{ab}$ field. 
As in the previous cases, we employ the plane wave method to obtain the
maximally reduced polarization
tensor $\chi_{a b}$  and vector $\varepsilon_a$  involved in $h_{a b}$ and $A_a$ respectively. Considering only the negative frequency part of a single  mode in the corresponding mode expansions, we write,
\begin{equation}
h_{a b}(x) = \chi_{a b}(k) e^{ik.x}
\label{69i}
\end{equation}
\begin{equation}
A_a(x) = \varepsilon_a (k) e^{ik.x}.
\label{70i}
\end{equation}
In terms of the polarization tensor $\chi_{a b}$ and vector $\varepsilon_a$  the gauge transformations (\ref{65i}) and (\ref{66i})  respectively can be written as
\begin{equation}
\chi_{ab} \rightarrow \chi_{ab}' = \chi_{ab} + ik_a \zeta_b
+ ik_b \zeta_a
\label{70+1}
\end{equation}
\begin{equation}
\varepsilon_a \rightarrow \varepsilon_a' = \varepsilon_a - \zeta_a
\label{70+2}
\end{equation}
where $\Lambda_a (x) = \zeta_a (k) \exp (ik.x)$.
Substituting (\ref{69i}) and (\ref{70i}) in (\ref{67i}), one gets
$$k^2\chi^{ab} - k^a k_c \chi^{cb} -
k^b k_c \chi^{ca }
+  k^a k^b \chi
+\eta^{ab}(-k^2 \chi +k_c k_d \chi^{cd})$$
\begin{equation}
-\mu^2 \left[ \chi^{ab} + ik^a\varepsilon^b + ik^b\varepsilon^a
-\eta^{ab}(\chi + 2ik_c\varepsilon^c) \right] = 0.
\label{71i}
\end{equation}
Contracting with $\eta_{ab}$ and considering only massless ($k^2 = 0$)
excitations the above equation reduces to
\begin{equation}
2k_ak_b\chi^{ab} + \mu^2\left[3(\chi + 2ik_a\varepsilon^a)\right] =
0.
\label{72i}
\end{equation}
The solution of the above equation is
\begin{equation}
\chi^{ab} = -i(k^a\varepsilon^b + k^b\varepsilon^a).
\label{73i}
\end{equation}
Hence it is also the solution of (\ref{71i}) with $k^2 = 0$.
It is obvious that this solution is a gauge artefact since one can choose the
arbitrary vector field $\Lambda_a = A_a$ in (\ref{66i}) so as to make this solution vanish.

Next we consider the massive case ($k^2 = M^2, M \neq 0$) and consider the $(00)$
component of the equation of motion (\ref{71i})  which, by a straightforward algebra,
can be reduced to
\begin{equation}
{\chi^1}_1 + {\chi^2}_2 + {\chi^3}_3 = 0
\label{74i}
\end{equation}
Similarly the ($0i$) component of (\ref{71i}) gives
\begin{equation}
{\chi_{0i}} = -iM \varepsilon_i
\label{75i}
\end{equation}
Now, the ($ij$) component of (\ref{71i}) is given by
\begin{equation}
k^2\chi_{ij} - \eta_{ij}M^2(\chi - \chi^{00}) - \mu^2[\chi_{ij} - \eta_{ij}
(\chi + 2iM\varepsilon^0)] =0
\label{76i}
\end{equation}
Using (\ref{74i}), the above equation can be reduced to
\begin{equation}
M^2\chi_{ij} - \mu^2[\chi_{ij} - \eta_{ij}
(\chi + 2iM\varepsilon^0)] =0
\label{77i}
\end{equation}
For $i= j = 1, 2, 3$ respectively in (\ref{77i}), we have  the following set of equations;
$$M^2\chi_{11} - \mu^2[\chi_{00} - \chi_{22} - \chi_{33}]
- 2iM\varepsilon_0 =0, $$
$$M^2\chi_{22} - \mu^2[\chi_{00} - \chi_{11} - \chi_{33}]
- 2iM\varepsilon_0 =0, $$
$$M^2\chi_{33} - \mu^2[\chi_{00} - \chi_{11} -\chi_{22}]
- 2iM\varepsilon_0 =0. $$
Adding the above three equations together and subsequently using (\ref{74i}), we
arrive at
\begin{equation}
\chi_{00} = -2 iM\varepsilon_0
\label{78i}
\end{equation}
When $i\neq j$, the equation (\ref{77i}) reduces to
\begin{equation}
( \mu^2 - M^2) \chi_{ij} = 0.
\label{79}
\end{equation}
At this juncture notice that only two of the three components $\chi_{ii}, i= 1,
2, 3 $ are
independent on account of the equation (\ref{74i}). Also, the $\chi_{00}$ and
$\chi_{0i}$ components can be set equal to zero by choosing the arbitrary 
field
$\Lambda_a$ to be $A_a$. Therefore, if $\chi_{ij} =0$
(for $ i\neq j$) in the above equation (\ref{79}), the number of independent components of $\chi_{a b}$ will  be only two. Since this is not the case, we can satisfy the equation (\ref{79})
only if $\mu^2 = M^2 $. Thus we see that the parameter $\mu$  represents the mass of the physical excitations of the field $h_{a b}$ and that its polarization
tensor is
\begin{equation}
\{ \chi_{a b} \} = \left( \begin{array} {cccc}
-2 i\mu\varepsilon_0 & -i\mu\varepsilon_1 & -i\mu\varepsilon_2 & -i\mu\varepsilon_3 \\
-i\mu\varepsilon_1   & \chi_{11}        &   \chi_{12}      &  \chi_{13}  \\
-i\mu\varepsilon_2   & \chi_{12}        &   \chi_{22}      &  \chi_{23}  \\
-i\mu\varepsilon_3   & \chi_{13}        &  \chi_{23}       &  \chi_{33} \end{array} \right)
\label{80}
\end{equation}
where $\chi_{11} + \chi_{22} + \chi_{33} = 0$ (see (\ref{74i})).
As mentioned before, by choosing the  field $\Lambda_a$ to be $A_a$
and making a gauge transformation, the above form of the polarization tensor can be converted
 to its maximally reduced form given by
\begin{equation}
\{ \chi_{a b} \} = \left( \begin{array} {cccc}
0 & 0 & 0 & 0 \\
0 &  \chi_{11}        &   \chi_{12}      &  \chi_{13}  \\
0 &  \chi_{12}        &   \chi_{22}      &  \chi_{23}  \\
0 &  \chi_{13}        &  \chi_{23}       &  \chi_{33} \end{array} \right);~~~ \chi_{11} + \chi_{22} + \chi_{33} = 0.
\label{81}
\end{equation}

Our next task is to show explicitly that it is possible to obtain the gauge transformation of $A^a$
from that of $h^{ab}$. For this purpose we 
consider now the equation of motion (\ref{68i}) corresponding to the vector field $A_a$ and
the associated polarization tensor $\varepsilon_a$. Substituting (\ref{69i}) and (\ref{70i}) in
 (\ref{68i})(or by contracting (\ref{71i}) with $k_b$ ) we get,
\begin{equation}
k^2\varepsilon^a -  k^a k_b\varepsilon^b - ik_b \chi^{b a} + ik^a \chi = 0.
\label{82}
\end{equation}
On making a gauge transformation (\ref{70+1}) in the polarization tensor $\chi_{ab}$, the polarization
vector $\varepsilon_a$ in (\ref{82}) automatically undergoes a gauge transformation.
\begin{equation}
k^2\varepsilon'_a -  k_a k_b\varepsilon'^b =  ik_b ({\chi^b}_ a + ik_a \zeta^b
+ ik^b \zeta_a)
 - ik_a (\chi + 2ik^b \zeta_b).
\label{82+1}
\end{equation}
This implies 
\begin{equation}
k^2\varepsilon'_a -  k_a k_b\varepsilon'^b =  [ik_b {\chi^b}_a -ik_a \chi ] -k^2 \zeta_a + k_a (k.\zeta)
\label{82+2}
\end{equation} 
From  equation (\ref{82}), substitute for the expression inside the square bracket in (\ref{82+2}) to obtain
\begin{equation}
[k^2\varepsilon'_a -  k_a k_b\varepsilon'^b ] -[k^2\varepsilon_a -  k_a k_b\varepsilon^b ] = -k^2 \zeta_a + k_a (k.\zeta)
\label{82+3}
\end{equation}
It is now easy to see that this relation can be satisfied only if $\varepsilon'_a -\varepsilon_a = - \zeta_a $ in agreement with
the previous relation (\ref{70+2}).
Therefore a knowledge of the gauge transformation of $h_{ab}$ is enough to deduce the gauge transformation 
property of $A_a$-field. Like the $\alpha$-field in St\"uckelberg  extended
Proca theory, this $A_a$-field too disappears from the physical spectrum.

Now we study the gauge transformation properties of the field $h_{ab}$
under the action of translational group $T(3)$. It is easy to see that, similar to the case of $B\wedge F$ theory,
the representation $D(p, q, r)$ (\ref{d15}) of $T(3)$ generates gauge
transformation of the massive field ($h_{ab}$). The action of $D(p, q, r)$ on the polarization tensor
$\{\chi_{ab}\}$ (\ref{81}) is given by,
$$\{\chi_{ab}\} \rightarrow  \{\chi_{a}\}' =  D(p, q, r) \{\chi_{ab}\} D^T(p, q, r) =\{\chi_{ab}\} $$
\begin{equation}
 + \left( \begin{array}{cccc}
\left(\begin{array}{c}p(p\chi_{11} + q\chi_{12} + r\chi_{13}) \\
+ q(p\chi_{12} + q\chi_{22} + r\chi_{23}) \\
+ r( p\chi_{13} + q\chi_{23} + r\chi_{33})  \end{array}\right)
&\left(\begin{array}{c} p\chi_{11} + q\chi_{12} \\+ r\chi_{13} \end{array}\right) &
\left(\begin{array}{c}  p\chi_{12} + q\chi_{22} \\ + r\chi_{23} \end{array} \right)&
\left(\begin{array}{c}  p\chi_{13} + q\chi_{23} \\
+ r\chi_{33}  \end{array}\right)  \\
p\chi_{11} + q\chi_{12} + r\chi_{13} & 0 & 0 & 0 \\
p\chi_{12} + q\chi_{22} + r\chi_{23} & 0 & 0 & 0 \\
 p\chi_{13} + q\chi_{23} + r\chi_{33} & 0 & 0 & 0 \\
\end{array} \right)
\label{87+1}
\end{equation}
By choosing
$$ \zeta_0 = \frac{1}{2} (p\zeta_1 + q\zeta_2 + r\zeta_3) $$
$$ \zeta_1 = \frac{1}{\mu} (p\chi_{11} + q\chi_{12} + r\chi_{13}) $$
$$ \zeta_2 = \frac{1}{\mu} (p\chi_{12} + q\chi_{22} + r\chi_{23}) $$
$$ \zeta_3 = \frac{1}{\mu}(p\chi_{13} + q\chi_{23} + r \chi_{33});~~~~\chi_{11} + \chi_{22} + \chi_{33} =0 $$
it is straightforward to see that (\ref{87+1}) is of the form (\ref{70+1})
which is the gauge transformation of $\chi_{ab}$. Notice that when one 
makes the above choices for  the components $\zeta_1, \zeta_2, \zeta_3$ in terms of
the parameters $p,q,r$ of the translational group $T(3)$, the component 
$\zeta_0$ gets automatically fixed. Therefore, in  the gauge transformation
(\ref{87+1}) generated by the representation $D(p,q,r)$ of $T(3)$ only the
three space components of the field $\zeta_a$ remain arbitrary.
 However, in the complete set of gauge transformations (\ref{70+1}) all the 
four components of $\zeta_a$ should be chosen independent of one another.
Hence, the above gauge transformations (\ref{87+1})  generated by the 
translational group $D(p,q,r)$ does not exhaust the complete set of gauge
transformations
available to the massive symmetric tensor gauge theory (\ref{64i}). This  is because
of the fact that in order to generate the entire  gauge equivalence 
class of the maximally reduced polarization tensor (\ref{81}) we require
four independent variables (corresponding to the four components of the 
arbitrary vector function $\zeta_a (k)$ which represents the gauge freedom) whereas the translational group $T(3)$
provides only three independent parameters.

Therefore,  in the present case (\ref{64i}) of massive tensor gauge theory,
a partial set of  gauge transformations are generated  the representation
$D(p,q, r)$ (\ref{d15}) of the translational group $T(3)$. The gauge 
transformation
of the $A_a$-field can be obtained from that of the $h_{ab}$-field, though
 the former one does not appear in the physical spectrum of the theory.

By considering the  linearized gravity in 4+1 dimensions,
one may obtain  
through the method of dimensional descent, the above discussed results
concerning the gauge transformation properties of St\"uckelberg  extended version
of EPF theory. The maximally reduced form of the polarization tensor for
linearized gravity in 4+1 dimension can be obtained using plane wave method
by proceeding exactly as in the 3+1 dimensional  case (section \ref{tlg}) and  is given by
\begin{equation}
\{ \chi^{xy} \} = \left( \begin{array} {ccccc}
0 & 0 & 0 & 0 & 0 \\
0 &  \chi^{11}        &   \chi^{12}      &  \chi^{13}  & 0 \\
0 &  \chi^{12}        &   \chi^{22}      &  \chi^{23}   & 0 \\
0 &  \chi^{13}        &  \chi^{23}       &  \chi^{33}   & 0 \\
0 & 0 & 0 & 0 & 0 \end{array} \right); ~~~~ \chi^{11} + \chi^{22} + \chi^{33} =0.
\label{9494}
\end{equation}
Notice now that by suitably applying the projection operator ${\cal P} =$diagonal$ (1,1,1,1,0) $ on this polarization tensor, one can get the maximally reduced polarization tensor (\ref{81}). Also by a similar application of this ${\cal P}$
on the energy-momentum vector $k^x =(\omega, 0,0,0,\omega)$  of a five 
dimensional particle propagating along the $x=4$ direction yields the rest frame
momentum 4-vector of particle belonging to St\"uckelberg  extended version
of EPF theory. Therefore it is clear that the dimensional descent connects
these two theories also.

\section{Massive antisymmetric tensor gauge theory}

Here we show that the translational group $T(3)$ generates the
full range of gauge transformations in the St\"uckelberg  extended
massive Kalb-Ramond theory. Similar to the massless KR theory discussed in
section \ref{tkr}, the gauge transformation in St\"uckelberg  extended massive KR theory
is also reducible.  Though the 
analysis in this case closely resembles that of St\"uckelberg  extended EPF theory
detailed before, here the reducibility  of the  
gauge transformation  is manifested in the  gauge
generation by $T(3)$. 

The Lagrangian of the St\"uckelberg  extended
massive KR theory is
\begin{equation}
{\cal L} = \frac{1}{12}H_{abc}H^{abc} - \frac{m^2}{4} (B_{ab} + \partial_a A_b -
\partial_b A_a ) (B^{ab} + \partial^a A^b -
\partial^b A^a )
\label{n1}
\end{equation}
with $B_{ab} = -B_{ba}$ and $~H_{abc} =
\partial_a B_{bc} + \partial_b B_{ca} +
\partial_c B_{ab}$. It can be easily verified that the above Lagrangian
is invariant under the joint gauge transformations
\begin{equation}
B_{ab}(x) \rightarrow B_{ab}(x) + \partial_a F_b
(x) - \partial_b F_a (x)
\label{n2}
\end{equation}
and
\begin{equation}
A_a (x)\rightarrow A_a (x) - F_a (x).
\label{n3}
\end{equation}
Here we must notice that the transformation (\ref{n2}) is  reducible exactly
as  in the case of massless KR theory; i.e.,
the transformation (\ref{n2}) remains invariant if we make the change 
$F_a (x) \rightarrow F_a (x) + \partial_a \beta (x)$. Therefore, there
exist a $`$gauge invariance of gauge transformation' in the theory described by
(\ref{n1}) also.
The equation of motion corresponding to $ B^{bc}(x)$ is given by
\begin{equation}
\partial_a H^{abc} + m^2 (B^{bc} + \partial^b A^c
- \partial^c  A^b ) = 0
\label{n4}
\end{equation}
and that corresponding to $A^b$ is 
\begin{equation}
\partial_c (B^{cb} + \partial^c A^b - \partial^b A^c ) = 0.
\label{n5}
\end{equation}
As in the case of the St\"uckelberg  extended massive theories considered
previously in sections \ref{sp} and \ref{sepf}, the equation of motion 
(\ref{n5}) for
$A^b$ can be obtained from the equation (\ref{n4})  by the 
 application of the operator $\partial_c$ upon
the latter equation. Hence,  one can  obtain the gauge 
transformation property (\ref{n3}) of the $A^b$-field from that of the
$ B^{bc}$-field. (This can be easily achieved by a straightforward
procedure  similar to the ones adopted before in the cases
 of St\"uckelberg  extended Proca and EPF theories for the same purpose 
and hence is not elaborated here again.)

We now proceed to obtain the maximally reduced polarization tensor
$\varepsilon^{ab}(k)$ corresponding to the antisymmetric field 
$B^{ab}(x)$
so that the role of $T(3)$ as a generator of gauge transformations in 
(\ref{n1}) can be studied. For this purpose, as usual we use the ansatz
\begin{equation} 
B^{ab}(x) = \varepsilon^{ab}(k) e^{ik\cdot x},~~~~A^{a}(x) = \varepsilon^{a}(k)  e^{ik\cdot x},~~~~F^a(x) = f^{a}(k)e^{ik\cdot x}
\label{n6}
\end{equation}
and employ the plane wave method. 
The momentum space gauge transformation corresponding to (\ref{n2})
now has the same form as (\ref{3-13});
\begin{equation}
\varepsilon_{a b}(k) \rightarrow  \varepsilon_{a b}(k) + i(k_{a}f_{b}(k) - k_{b}f_{a}(k))
\label{nn}
\end{equation}
Then the equation of motion
 (\ref{n4}) can be written (in the momentum space) as 
\begin{equation}
-k^2 \varepsilon^{bc} - k^b k_a \varepsilon^{ca} - k^c k_a \varepsilon^{a b} + m^2 (\varepsilon^{bc} + ik^b \varepsilon^{c} - ik^c \varepsilon^{b}) = 0.
\label{n7}
\end{equation}
If $k^2 = 0$ (massless excitations), the above equations reduces to
\begin{equation}
- k^b k_a \varepsilon^{ca} - k^c k_a \varepsilon^{a b} + m^2 (\varepsilon^{bc} + ik^b \varepsilon^{c} - ik^c \varepsilon^{b}) = 0
\label{n8}
\end{equation}
whose solution must be of the form 
\begin{equation}
\varepsilon^{ bc}(k) = C(ik^b \varepsilon^{c} - ik^c 
\varepsilon^{b}) + D (\epsilon^{bcde}
k_d\varepsilon_e )
\label{n9}
\end{equation}
where $C$ and $D$ are constants to be fixed.
Substituting (\ref{n9}) in (\ref{n8}), we can easily see that
$C = -1$ and $D = 0$.
Therefore, the solution to (\ref{n7}) corresponding to  massless excitations
is
\begin{equation}
\varepsilon^{ bc}(k) = -ik^b \varepsilon^{c} + ik^c
\varepsilon^{b}.
\label{n10}
\end{equation} 
However, such solutions can be gauged away by choosing the arbitrary field $F^a (x)  = A^a (x)$ in (\ref{n3}), which shows that massless excitations are gauge artefacts.

Next we consider the massive case, $k^2 = M^2,~(M\neq 0)$ where it is
possible to go to the rest frame and one has the momentum 4-vector 
$k^a = (M, 0,0,0)^T$.  In the rest frame,
the equation of motion (\ref{n7}) 
reduces to
\begin{equation}
(m^2 -M^2) \varepsilon^{bc} -M(k^b \varepsilon^{c 0} + k^c \varepsilon^{ 0b }) + m^2 (ik^b \varepsilon^{c} - ik^c
\varepsilon^{b}) = 0.
\label{n11}
\end{equation}
Note that, since the polarization tensor $\varepsilon^{ bc}$ is antisymmetric,
all its diagonal entries are automatically zero. Considering the components
of (\ref{n11}) for which $(b =0, c = i)$, we have
\begin{equation}
 \varepsilon^{ i0} = iM \varepsilon^{ i}
\label{n12}
\end{equation}
For $(b =i, c = j)$ with $i\neq j$, the equation (\ref{n11}) gives 
\begin{equation}
(m^2 -M^2) \varepsilon^{ij} = 0.
\label{n13}
\end{equation}
This leads to two possibilities; either $\varepsilon^{ij} = 0$ or $ M^2 = m^2$.
The former possibility can be ruled out by the following reasoning.
Since (\ref{n1}) is the first class version (obtained by a St\"uckelberg
extension mechanism)
of massive KR theory possessing three physical degrees of
freedom, the former too must inherit the same number of degrees of freedom. 
However, the $\varepsilon^{ i0}$ elements can all be made to vanish by the gauge choice $F_a = A_a$. Therefore the possibility
 $\varepsilon^{ij} = 0$ leads to a null theory and hence should be discounted. 
Therefore, we have $M^2 = m^2$ which  is also consistent with the degrees of
freedom counting. Finally, analogous to (\ref{81}), the maximally reduced 
form of the polarization tensor corresponding to (\ref{n1}) is given by,
\begin{equation}
\{\varepsilon^{ab}\} = \left( \begin{array} {cccc}
0 & 0 & 0 & 0 \\
0 &  0        &   \varepsilon^{12}      &  \varepsilon^{13}  \\
0 &     -\varepsilon^{12}     &     0    &  \varepsilon^{23}  \\
0 &   -\varepsilon^{13}      &  -\varepsilon^{23}       &  0 \end{array} \right).
\label{n14}
\end{equation}
As in the case of St\"uckelberg  extended EPF theory, the $A_a$-field  disappears from the physical spectrum in this case also.
Here it must be emphasized that the maximally reduced form of polarization tensor
of $B\wedge F$ theory also has the same form (\ref{n14}). This is not 
surprising since the physical sector of $B\wedge F$ theory is equivalent to
massive KR theory whose first class version is the theory (\ref{n1}) under
consideration now.
It is now straight forward to see that the translational group $T(3)$ in the 
representation $D(p, q, r)$ (\ref{d15}) generates the full set  of gauge
transformations in the theory described by (\ref{n1}) also. The action of 
$D(p, q, r)$ on (\ref{n14}) is given by,
$$\{\varepsilon^{ab}\}\rightarrow \{\varepsilon'^{ab}\} = D(p, q, r) \{\varepsilon^{ab}\} D^T(p, q, r)$$
\begin{equation}
=\{\varepsilon^{ab}\} + \left( \begin{array} {cccc}
0 & -q\varepsilon^{12} - r \varepsilon^{13} & p\varepsilon^{12} - r \varepsilon^{23} & p\varepsilon^{13} + q \varepsilon^{23} \\
q\varepsilon^{12} + r \varepsilon^{13} & 0 & 0& 0 \\
-p\varepsilon^{12} + r \varepsilon^{23}& 0 & 0& 0 \\
- p\varepsilon^{13} - q \varepsilon^{23}& 0 & 0& 0 \\
\end{array} \right)
\label{n15}
\end{equation}
Since in the rest frame $k^a = (M,0,0,0)^T$, (\ref{n15})
 can be considered to be the gauge
transformations of the form (\ref{nn}) if 
\begin{equation}
f^1 =\frac{1}{im}(q\varepsilon^{12} + r \varepsilon^{13}),~~~
f^2 =  \frac{1}{im}(-p\varepsilon^{12} + r \varepsilon^{23}), ~~~
f^3 = \frac{-1}{im}( p\varepsilon^{13} + q \varepsilon^{23})
\label{n16}
\end{equation}
Here one must clearly note that the component $f^0$ remains completely
undetermined and does not depend at all either on the parameters $p,q,r$ of 
$T(3)$ or on the components of  maximally reduced polarization tensor of the theory
whereas the other components $f^1, f^2, f^3$ are determined by these
parameters and the elements of the polarization tensor.
Interestingly, it is exactly in the same fashion as in the present case (of 
St\"uckelberg   extended massive KR theory) that gauge transformations
of $B\wedge F$ theory are generated by the translational group $D(p,q,r)$ (section \ref{bf}). Hence, analogous to the gauge transformation
generated by $W(p,q)$ in massless KR theory, for any given set of ($f^1, f^2, f^3$)
we have a continuum of values for $f^0$, representing the reducibility of
the gauge transformation in the underlying 2-form field  both in  St\"uckelberg
extended first class version of massive KR theory and in the $B\wedge F$ theory.
Therefore, the complete independence of the time-component of $f^a$ on the
maximally reduced polarization tensor or on the parameters of the group 
$D(p,q,r)$ is a consequence of the reducibility of the gauge transformations
of these theories.

Dimensional descent relates the gauge transformations in 5+1 dimensional
 massless KR theory to those in the presently discussed St\"uckelberg  extended
massive KR theory just the same way it is related to the gauge transformation of the antisymmetric field of $B\wedge F$ theory.

\section{summary}
We   showed that, just like in the case of topologically massive  $B\wedge F$
 theory, translational group $T(3)$ acts as generators of gauge transformations in gauge theories obtained by St\"uckelberg  embeddings of  massive  theories having no gauge symmetry.
We illustrate these with the examples of
St\"uckelberg  extended first class versions of Proca, Einstein-Pauli-Fierz and
massive Kalb-Ramond theories  in 3+1 dimensions. In each of these cases,
we have shown that the representation $D(p,q,r)$ of $T(3)$ generates gauge transformation when acted suitably on the corresponding maximally reduced 
polarization tensor. The reducibility of the  gauge transformations
transformation in St\"uckelberg  extended  massive Kalb-Ramond theory
 is manifested clearly in the gauge generation by the translational group
$T(3)$. With suitable 4+1 dimensional  theories as the respective starting 
points,  dimensional  descent can be applied  consistently in all the
 St\"uckelberg  embedded models considered here.

\chapter{Conclusions}

We have studied several aspects of a variety of planar field theories with
particular emphasis  on topologically massive gauge theories and the
role of translational groups in generating gauge transformations. Attention is
also focused   on the interrelationship between the different
theories. We  considered  Abelian models with vector, symmetric and
antisymmetric second rank  tensors as underlying fields.  Our methodology 
relied crucially on the maximally reduced form (i.e., representing just the
physical sector) of the polarization vectors  and tensors of these theories.
The maximally reduced polarization vector/tensor is usually derived by
considering a plane wave solution corresponding to a single component in 
the mode expansion of the basic field of the respective  theory.
Hence this method of derivation is named the  $`$plane wave method. 
In this method,
the plane wave solution is substituted in the equation of motion of the
theory and confining to a particular reference frame  we obtain the explicit
form of  the maximally reduced  polarization vector/tensor corresponding to that frame, using
mainly  the structure of equation motion itself and finally making a gauge transformation.  The plane wave method is used for the derivation of the maximally
reduced form of the polarization vector/tensor of every  model considered
in this work.

To begin with we obtained, using the plane wave method, the polarization vectors of Maxwell-Chern-Simons and Maxwell-Chern-Simons-Proca models in both Lagrangian  Hamiltonian frameworks with compatible results.  In Lagrangian formalism
the polarization vector of MCS theory is calculated in the Lorentz gauge while
in the Hamiltonian formalism, the calculation is done in a variety of covariant gauges and it is found that the maximally reduced polarization vector has the
same form in all these gauges. 
The structure of the polarization vectors  explicitly display the doublet
structure of Maxwell-Chern-Simons-Proca theory consisting of a pair of 
Maxwell-Chern-Simons models  with opposite helicities and different mass
parameters. The polarization vector of self-dual model is  of exactly the same
form as the maximally reduced polarization vector of a Maxwell-Chern-Simons
theory with positive helicity whereas there is a similar correspondence between
 antiself-dual model and a Maxwell-Chern-Simons theory with negative helicity. This is in agreement with the well known equivalence
of self and antiself-dual models with a pair of Maxwell-Chern-Simons theories with
opposite parities.

We have also made a comparison of the polarization vectors of Maxwell and
 Maxwell-Chern-Simons theories  in different gauge choices and in different
reference frames. The time component of the Coulomb gauge polarization
vector of Maxwell-Chern-Simons theory is undefined in the rest frame.
Like Maxwell theory, basic field in  Maxwell-Chern-Simons model can be made
to satisfy the spatial transversality condition ${\bf k \cdot A } =0$  
 in a boosted frame provided it undergoes a suitable gauge
 transformation. However, unlike Maxwell theory,  the temporal gauge condition
$ A^0 =0$ cannot be satisfied simultaneously by Maxwell-Chern-Simons field
except in the ultra-relativistic limit, whereas it blows up in the rest frame.
In comparison, the Maxwell-Chern-Simons polarization vector satisfying Lorentz gauge condition
 is well defined in all reference frames including the rest frame.
Notice that rest frame is not available in Maxwell theory.

We  reviewed, with necessary details, the role of translational subgroup
 $T(2)$ of Wigner's little group for massless particles in generating
gauge transformations in 3+1 
dimensional massless gauge theories - Maxwell, massless Kalb-Ramond
and  linearized gravity theories. In all these cases, it is demonstrated that
the  transformation of the respective maximally reduced  polarization
 vector/tensor under the
 representation of $T(2)$ inherited from defining representation of the
little group amounts to a gauge transformation in momentum space.  Our analysis,
however showed that, the gauge transformations generated by the translational
group in massless Kalb-Ramond and linearized gravity constitute only certain subsets of the 
full spectra of gauge transformations available in these two theories.
This is attributed to the fact that the gauge freedoms in these second rank tensor
theories are represented by arbitrary vector fields with four
components while the translational group provides only two parameters.
Because of this deficit in the number of parameters  the range of gauge
transformations  generated by $T(2)$ become restricted. We also reviewed the
gauge generation by the translational group $T(3)$ in the 3+1 dimensional
topologically massive $B\wedge F$ gauge theory. Both massless Kalb-Ramond and 
$B\wedge F$ theories have reducible gauge transformations. We have shown that
the reducibility of gauge transformations is manifested clearly in the
gauge generation by translational groups.

Just like $T(2)$ generates gauge transformations in 3+1 dimensional massless
theories, translational group $T(1)$ acts as generator of gauge transformations
in 2+1 dimensional Maxwell theory. (Massless Kalb-Ramond and linearized gravity
possess no physical degrees of freedom in 2+1 dimensions.) We show that 
 $T(1)$ generate gauge transformations in the topologically massive planar
gauge  theories namely Maxwell-Chern-Simons and Einstein-Chern-Simons theories.
A suitable representation of $T(1)$  when acted on the respective
maximally reduced polarization vector and tensor  generate momentum-space
 gauge transformations in these theories.  The polarization tensor of
Einstein-Chern-Simons theory is found to be the direct product of the 
polarization vector of a Maxwell-Chern-Simons theory with itself.

It has been known that the gauge transformations in a topologically massive
theory living in a certain space-time dimensions can be related to those
of massless gauge theories in a space-time of one higher dimension through
the method of dimensional descent.  We presented a short review of dimensional
descent from 4+1 dimensional Maxwell and massless Kalb-Ramond theories  to
3+1 dimensional $B\wedge F$ theory and derived the gauge generating 
representation of $T(3)$  with the help of the gauge transformation properties
of these higher dimensional theories. Subsequently we have derived, using 
dimensional descent, the previously mentioned gauge generating representation of
 $T(1)$ for Maxwell-Chern-Simons and Einstein-Chern-Simons theories in 2+1
dimensions by considering the gauge transformation properties of 3+1
dimensional Maxwell and linearized gravity theories respectively. 
In this process,
we have also considered the massive   non-gauge Einstein-Pauli-Fierz theory 
in 2+1 dimensions whose basic field transforms like a second rank symmetric
tensor. A comparison of the polarization tensors of Einstein-Pauli-Fierz theory
with those of a pair of Einstein-Chern-Simons theories having opposite
helicities suggested a doublet structure for Einstein-Pauli-Fierz theory. 
This is analogous to the vector case in which Proca theory in 2+1 dimensions
 is equivalent to a doublet of Maxwell-Chern-Simons theories with opposite helicities. In  connection with this analogy, it should  be noted that the relative 
and overall signs of the Einstein term as well as the Pauli-Fierz term in  the
Einstein-Pauli-Fierz theory should be  the same as the conventional ones for 
the theory to be meaningful. However, the sign of the Einstein-term in
 Einstein-Chern-Simons theory should be opposite to that the conventional one.
Therefore, a theory with all three terms present simultaneously (the tensor
analogue of Maxwell-Chern-Simons-Proca theory) is not possible
because of the unavoidable conflict of requirements on the signs of the Einstein term. Therefore, the suggestion that Einstein-Pauli-Fierz theory is a doublet
of Einstein-Chern-Simons theories needs further investigation. 

Next we have considered, the role of translational group in generating 
gauge transformations in some massive  gauge theories obtained by elevating
massive non-gauge theories to their first class versions by  St\"uckelberg  
embedding mechanism. Such theories considered in this work are the St\"uckelberg
extended versions of Proca, Einstein-Pauli-Fierz and massive Kalb-Ramond 
theories which have vector, symmetric and antisymmetric second rank tensors respectively
as the basic fields. The investigation of these theories is done in 3+1
dimensional space-time in which such models are usually studied.
However, with suitable    modifications the methods and results of our
study of these theories are applicable to the corresponding 2+1 dimensional
theories as well. We have shown that the representation of $T(3)$ that 
generates  gauge transformation in $B\wedge F$ theory also generates gauge
 transformations in the above mentioned massive gauge theories in 3+1 
dimensions. Analogous to the case of linearized gravity, the gauge
transformations generated in the St\"uckelberg  extended Einstein-Pauli-Fierz 
theory by $T(3)$  constitute only a subgroup of the full set of gauge
 transformations in the theory. The gauge transformations in the St\"uckelberg  
extended massive Kalb-Ramond theory are reducible and this reducibility is
clearly exhibited in the gauge generation by the translational group.
\newpage
\addcontentsline{toc}{chapter}{Appendix}
\appendix
\chapter{Wigner's Little group in 3+1 dimensions}
Wigner's little group preserves the energy-momentum vector of a particle and  is a subgroup of homogeneous Lorentz group.
Here we give a 
derivation of the explicit expression for the little group for a
3+1 dimensional  massless particle \cite{we}.  Let  
\begin{equation}
W = \left(\begin{array}{cccc}
 m_1 & m_2 & m_3 & m_4 \\
 p_1 & p_2 & p_3 & p_4 \\
q_1 & q_2 & q_3 & q_4 \\
r_1 & r_2 & r_3 & r_4 \end{array}\right) 
\label{aa1}
\end{equation}
be a representation of an arbitrary  element of the little group
that preserves the  momentum 4-vector of a massless
particle of energy $\omega$ propagating along the $z$-direction:
\begin{equation}
{W^a}_b k^b = k^a, ~~~~~~~k^a =\omega(1,0,0,1)^T.
\label{aa2}
\end{equation}
With $l^a = (1,0,0,1)^T$, we obviously have ${W^a}_b l^b = l^a$ also. 
Consider a time-like vector $\alpha^a = (1,0,0,0)^T$ which has the
following obvious Lorentz covariance properties,
\begin{eqnarray}
(W\alpha)^a(Wl)_a = (W\alpha)^al_a = \alpha^a l_a  =  1,
\label{aa3}\\
 (W\alpha)^a(W\alpha)_a  =\alpha^a \alpha_a =  1.
\label{aa4}
\end{eqnarray}
Since $(W\alpha)^a = (m_1, p_1, q_1, r_1)^T $ (which is nothing but the first column of $W$),  one can see
using (\ref{aa3}) that $r_1 = m_1 -1$. Therefore, 
 the form of $(W\alpha)^a$ that satisfies the relation (\ref{aa4}) is given by
\begin{equation}
(W\alpha)^a = \left( 1+ \frac{ p_1^2 + q_1^2}{2},  p_1, q_1, \frac{ p_1^2 + q_1^2}{2}\right)^T.
\label{aa5}
\end{equation}
Now,  $W$  acting on the space-like unit vector 
$\beta^a = (0,0,0,1)^T$ yields $(W\beta)^a = (m_2, p_2, q_2, r_2)^T $.
Analogous to (\ref{aa3}) and (\ref{aa3}) in this case we have
\begin{eqnarray}
(W\beta)^a (Wl)_a = (W\beta)^al_a =\beta^al_a = -1,
\label{aa6}\\
(W\beta)^a(W\beta)_a = \beta^a\beta_a = -1.
\label{aa7}
\end{eqnarray}
It is easy to see that (\ref{aa6}) and (\ref{aa7}) restrict the form of
$(W\beta)^a$ to \\
$\left(-\frac{ p_4^2 + q_4^2}{2},p_4,q_4, 
1-\frac{ p_4^2 + q_4^2}{2}\right)^T$ which, along with
(\ref{aa6}), when substituted in the relation
\begin{equation}
(W\alpha)^a(W\beta)_a = \alpha^a \beta_a = 0
\label{aa8}
\end{equation}
 yields $(p_1 + p_4)^2 + (q_1 +q_4)^2 =0$. This implies that
$ p_1 = -p_4$ and $q_1 =-q_4$. Thus  the fourth column of $W$ (\ref{aa1})is
given by
\begin{equation}
(W\beta)^a = \left( -\frac{ p_1^2 + q_1^2}{2}, - p_1,- q_1, 1- \frac{ p_1^2 + q_1^2}{2}\right)^T.
\label{aa9}
\end{equation}
In a similar fashion, by considering another space-like unit vector $\gamma ^a = (0,0,1,0)^T$ we can obtain $(W\gamma)^a =(m_3, p_3, q_3, r_3)^T   $,  the third column of $W$. 
A simple algebra using  the properties
\begin{equation}
(W\gamma)^a(Wl)_a = (W\gamma)^al_a = \gamma ^al_a = 0, ~~~
(W\gamma)^a(W\gamma)_a = \gamma ^a\gamma_a = -1 
\label{aa10}
\end{equation}
leads to the condition $ p_3^2 + q_3^2 = 1$ which enables one to make the parametrization
$ p_3 = \sin\phi,~~q_3 = \cos\phi$. Hence we have $(W\gamma)^a =(m_3, \sin\phi,
\cos\phi,m_3)^T$. Furthermore, since
\begin{equation}
(W\beta)^a(W\gamma)_a = \beta^a\gamma_a =0
\label{aa11}
\end{equation}
we obtain the relation $m_3 = p_1\sin\phi +q_1\cos\phi$ wherein we used (\ref{aa9}) and the form $(W\gamma)^a$ mentioned above.
Therefore
\begin{equation} 
(W\gamma)^a =(p_1\sin\phi +q_1\cos\phi,\sin\phi, \cos\phi,p_1\sin\phi +q_1\cos\phi)^T.
\label{aa12}
\end{equation}
Lastly,  we introduce the spacelike vector $\delta^a =(0,1,0,0)^T$ so that \\
$(W\delta)^a = (m_2, p_2, q_2, r_2)^T$ is the second column of $W$ (\ref{aa1}). Then, just as in the previous case,  using the covariance properties
$$
(W\delta)^a(Wl)_a = (W\delta)^a l_a = \delta^al_a = 0,$$
$$(W\delta)^a(W\delta)_a = \delta^a\delta_a = -1,~~
(W\delta)^a(W\alpha)_a = \delta^a \alpha_a = 0$$
we get $(W\delta)^a = ( p_1\cos\theta + q_1\sin\theta, \cos\theta, \sin\theta,
 p_1\cos\theta + q_1\sin\theta)^T$. Contracting with $(W\gamma)_a$ and using the relation 
\begin{equation}
(W\delta)^a(W\gamma)_a =\delta^a \gamma_a = 0
\label{aa14}
\end{equation}
we deduce that $\theta = -\phi$. Therefore we have
\begin{equation}
(W\delta)^a = (p_1\cos \phi - q_1\sin\phi,~~ \cos \phi, ~~-\sin\phi,~~p_1\cos \phi - q_1\sin\phi)^T.
\label{aa15}
\end{equation}
From (\ref{aa5}), (\ref{aa9}), (\ref{aa12}) and (\ref{aa15})  
we then obtain  the representation of the little group
for a massless particle in 3+1 dimensions as
\begin{equation}
W(p, q;  \phi) =
\left( \begin{array}{cccc}
1+ \frac{p^2 + q^2 }{2} & p\cos \phi - q \sin \phi & p\sin \phi + q \cos \phi  & -\frac{p^2 + q^2 }{2} \\
p & \cos \phi & \sin \phi  & -p \\
q & -\sin \phi & \cos \phi  & -q \\
\frac{p^2 + q^2 }{2} &  p\cos \phi -q \sin \phi & p\sin \phi + q \cos \phi  & 1
-\frac{p^2 + q^2 }{2}
\end{array}\right)
\label{aa16}
\end{equation}
dropping the subscripts on  $p,q$ which  are  any real numbers.
This little group can be written as
\begin{equation}
W(p, q;  \phi) = W(p, q) R(\phi)
\label{aa17}
\end{equation}
where
\begin{equation}
W(p, q) \equiv W(p, q;  0) = \left( \begin{array}{cccc}
1+ \frac{p^2 + q^2 }{2} & p & q  & -\frac{p^2 + q^2 }{2} \\
p & 1 & 0 & -p \\
q & 0 & 1 & -q \\
\frac{p^2 + q^2 }{2} &  p & q  & 1 -\frac{p^2 + q^2 }{2}
\end{array}\right)
\label{aa18}
\end{equation}
is a particular representation of $T(2)$ - the group of  translations in a plane and $ R(\phi)$ represents a $SO(2)$ rotation in the plane. 
It is clear that $W(p, q)$ and  are Abelian subgroups
of the little group $W(p, q;  \phi)$:
\begin{eqnarray}
 W(p, q)W(\bar{p},\bar{q}) = W(p+\bar{p}, q+\bar{q})\\
\label{aa19}
 R(\phi) R(\bar{\phi}) = R(\phi +\bar{\phi})
\label{aa20}
\end{eqnarray}
Moreover, the subgroup $W(p, q)$ is invariant since
\begin{equation}
 R(\phi) W(p, q)R^{-1}(\phi) =W(p\cos \phi + q\sin \phi, -p\sin \phi + q\cos \phi).
\label{aaa20}
\end{equation}
Therefore, the little group is not semi-simple.  
The group generators are given by
\begin{eqnarray}
A =-i \frac{\partial W(p, 0;  0)}{\partial p}\mid_{p=0} = -i\left(\begin{array}{cccc}
0 & 1 & 0 & 0 \\
1 & 0  & 0 & -1 \\
0 & 0  & 0 & 0\\
0 & 1  & 0 & 0
\end{array}\right), 
\label{aa211} \\
B =-i \frac{\partial W(0, q;  0)}{\partial q}\mid_{q=0} = -i\left(\begin{array}{cccc}
0 & 0 & 1 & 0 \\
0 & 0  & 0 & 0 \\
1 & 0  & 0 & -1\\
0 & 0  & 1 & 0 
\end{array}\right),
\label{aa21} \\
J_3 =-i\frac{\partial W(0, 0; \phi)}{\partial \phi }\mid_{\phi = 0} = -i\left(\begin{array}{cccc}
0 & 0 & 0 & 0 \\
0 & 0  & 1 & 0 \\
0 & -1  & 0 & 0\\
0 & 0  & 0 & 0
\end{array}\right).
\label{aa22}
\end{eqnarray}
It is important to note that the generators $A$ and $B$ can be expressed as a
combination of the generators of boosts and rotations\footnote{We denote by 
$J_i~~(i=1,2,3)$  the  generator of rotation about the  $j$-th axis.
Similarly, the generator of  boost
 along $x_i$-direction is denoted by  $K_i$. In this notation, the algebga
of the homogeneous Lorentz group is given by $[J_i,J_j] = i\epsilon_{ijk}J_k,~~
[J_i, K_j] = i\epsilon_{ijk}K_k, ~~ [K_i,K_j] = -i\epsilon_{ijk}J_k$.}:
\begin{equation}
A = J_2 + K_1,~~~ B= -J_1 + K_2.
\label{aaab22} 
\end{equation}
The Lie algebra of the little group is given by
\begin{equation}
\left[A, B\right] = 0,~~~ \left[J_3, A\right] = iB,~~~ \left[J_3, B\right] = -iA.
\label{aaa22}
\end{equation}
This is identical to the algebra of $E(2)$ - the Euclidean group in two dimensions, comprising of two mutually commuting translation generators (corresponding
to $A$ and $B$) and a generator of rotation in the plane ($J_3$). Thus, the 
algebra of the 3+1 dimensional little group for massless particles
is isomorphic to the $E(2)$ algebra \cite{wkt}.

\end{document}